\documentclass[fleqn,usenatbib]{mnras}

\usepackage{newtxtext,newtxmath}

\usepackage[T1]{fontenc}
\usepackage{natbib}

\DeclareRobustCommand{\VAN}[3]{#2}
\let\VANthebibliography\thebibliography
\def\thebibliography{\DeclareRobustCommand{\VAN}[3]{##3}\VANthebibliography}

\usepackage{graphicx}	
\usepackage{amsmath}	
\usepackage{xspace}
\usepackage{verbatim}
\usepackage[normalem]{ulem}

\usepackage[dvipsnames]{xcolor}

\newcommand{\decode}{\textsc{decode}\xspace}
\newcommand{\galics}{\textsc{GalICS}\xspace}
\newcommand{\paperI}{Paper I\xspace}

\defcitealias{fu_2022}{Paper I}

\title[\decode and galaxy star formation rates]{Unveiling the (in)consistencies among the galaxy stellar mass function, star formation histories, satellite abundances and intracluster light from a semi-empirical perspective}

\author[H. Fu et al.]{
	Hao Fu$^{1}$\thanks{E-mail: \href{mailto:h.fu@soton.ac.uk}{h.fu@soton.ac.uk}},
	Francesco Shankar$^{1}$\thanks{E-mail: \href{mailto:f.shankar@soton.ac.uk}{f.shankar@soton.ac.uk}},
        Mohammadreza Ayromlou$^{2}$,
        Ioanna Koutsouridou$^{3,4}$,
        Andrea Cattaneo$^{5}$,
        \newauthor
        Caroline Bertemes$^{2}$,
        Sabine Bellstedt$^{6}$,
        Ignacio Mart\'in-Navarro$^{7,8}$,
        Joel Leja$^{9}$,
        Viola Allevato$^{10}$,
        Mariangela \newauthor Bernardi$^{11}$,
        Lumen Boco$^{12}$,
        Paola Dimauro$^{13}$,
        Carlotta Gruppioni$^{14}$,
        Andrea Lapi$^{12}$,
        Nicola Menci$^{13}$,
        Iv\'an 
        \newauthor
        Mu\~noz Rodríguez$^{15,1}$,
        Annagrazia Puglisi$^{1}$,
        Alba V. Alonso-Tetilla$^{1}$
	\\\\
	$^{1}$School of Physics and Astronomy, University of Southampton, Highfield SO17 1BJ, Southampton, UK\\
        $^{2}$Universit{\"a}t Heidelberg, Zentrum f{\"u}r Astronomie, Institut f{\"u}r theoretische Astrophysik, Albert-Ueberle-Str. 2, 69120 Heidelberg, Germany\\
	$^{3}$Dipartimento di Fisica e Astronomia, Universita degli Studi di Firenze, Via G. Sansone 1, I-50019 Sesto Fiorentino, Italy\\
	$^{4}$INAF/Osservatorio Astrofisico di Arcetri, Largo E. Fermi 5, I-50125 Firenze, Italy\\
	$^{5}$Observatoire de Paris/LERMA, PSL University, 61 av. de l’Observatoire, 75014 Paris, France\\
        $^{6}$ICRAR, The University of Western Australia, 7 Fairway, Crawley, WA 6009, Australia\\
        $^{7}$Instituto de Astrof\'{i}sica de Canarias, V\'{i}a L\'{a}ctea s/n, E-38205 La Laguna, Tenerife, Spain\\
        $^{8}$Departamento de Astrof\'{i}sica, Universidad de La Laguna, E-38205 La Laguna, Tenerife, Spain\\
        $^{9}$Department of Astronomy \& Astrophysics, The Pennsylvania State University, University Park, PA 16802, USA\\
        $^{10}$INAF – Osservatorio Astronomico di Capodimonte, Via Moiariello 16, 30131 Napoli, Italy\\
        $^{11}$Department of Physics and Astronomy, University of Pennsylvania, Philadelphia, PA 19104, USA\\
        $^{12}$SISSA, Via Bonomea 265, 34135 Trieste, Italy\\
        $^{13}$INAF - Osservatorio Astronomico di Roma, via di Frascati 33, 00078 Monte Porzio Catone, Italy\\
        $^{14}$INAF - Osservatorio di Astrofisica e Scienza dello Spazio di Bologna via Gobetti 93/3, 40129, Bologna, Italy\\
        $^{15}$Institute for Astronomy and Astrophysics, National Observatory of Athens, V. Paulou \& I. Metaxa 11532, Greece
}

\date{Accepted XXX. Received YYY; in original form ZZZ}

\pubyear{2023}

\begin{document}
\label{firstpage}
\pagerange{\pageref{firstpage}--\pageref{lastpage}}
\maketitle

\begin{abstract}
    In a hierarchical, dark matter-dominated Universe, stellar mass functions (SMFs), galaxy merger rates, star formation histories (SFHs), satellite abundances, and intracluster light, should all be intimately connected observables. However, the systematics affecting observations still prevent universal and uniform measurements of, for example, the SMF and the SFHs, inevitably preventing theoretical models to compare with multiple data sets robustly and simultaneously. We here present our holistic semi-empirical model \decode (Discrete statistical sEmi-empiriCal mODEl) that converts via abundance matching dark matter merger trees into galaxy assembly histories, using different SMFs in input and predicting all other observables in output in a fully data-driven and self-consistent fashion with minimal assumptions. We find that: 1) weakly evolving or nearly constant SMFs below the knee ($M_\star \lesssim 10^{11} \, M_\odot$) are the best suited to generate star formation histories aligned with those inferred from MaNGA, SDSS, GAMA, and, more recently, JWST; 2) the evolution of satellites after infall only affects the satellite abundances and star formation histories of massive central galaxies but not their merger histories; 3) the resulting SFR-$M_\star$ relation is lower in normalization by a factor of $\sim 2$ with respect to observations, with a flattening at high masses more pronounced in the presence of mergers; 4) the latest data on intracluster light can be reproduced if mass loss from mergers is included in the models. Our findings are pivotal in acting as pathfinder to test the self-consistency of the high-quality data from, e.g., JWST and Euclid.
\end{abstract}




\section{Introduction}

In a $\Lambda$CDM scenario galaxies form within the potential well of dark matter haloes. They are believed to grow their stellar mass via both \textit{in-situ} and \textit{ex-situ} processes. \textit{In-situ} star formation originates from cooling of the infalling gas, while \textit{ex-situ} growth is predominantly driven by mergers with other galaxies. Several works have shown that most of the galaxies with $M_\star<10^{11}\,M_\odot$ grow primarily via star formation, with mergers becoming increasingly relevant in more massive galaxies (e.g., \citealt{van_dokkum_2010, leitner_2012, shankar_2015, buchan_2016, groenewald_2017, matharu_2019}; \citetalias{fu_2022}; \citealt{eisert_2023}).

In traditional models of galaxy evolution, galaxies are evolved following the hierarchical growth of their host dark matter haloes, with a tendency to build up via mergers more massive systems at later epochs (e.g., \citealt{guo_2008, oser_2010, cattaneo_2011, lackner_2012, rodriguez_gomez_2016, pillepich_2018, monachesi_2019, grylls_paper2}; \citetalias{fu_2022}). However, in stellar archaeological findings and star formation histories analysis via spectral energy distribution (SED) fitting \citep[e.g.,][]{thomas_2019, sanchez_2019, bellstedt_2020}, a different pattern emerges. Galaxies are observed to follow a downsizing trend, wherein more massive galaxies form at earlier times, possibly triggered by large bursts of star formation, while less massive galaxies form at later times with progressively lower star formation rates probably induced by an overall ``cosmic starvation'' caused by the general reduction in available cold gas available to feed galaxies \citep[e.g.,][]{larson_1980, gunn_1972, cowie_1977}. Many galaxies, especially those with a prominent bulge component \citep[e.g.,][]{bell_2008, bluck_2014, lang_2014, bluck_2022, dimauro_2022}, tend to show signs of a more rapid quenching of their star formation. The primary physical processes responsible for rapid quenching are still unclear, with leading theories proposing a combination of internal and external processes, such as stellar and AGN feedback, halo quenching, morphological quenching,  mergers, ram-pressure stripping \citep[e.g.,][]{granato_2004, dekel_2006, shankar_2006, dekel_2008, martig_2009, lilly_2013, schawinski_2014, lapi_2018, xu_2021}.

Many theoretical models and techniques have been developed in the last decades to study galaxy formation and evolution at different levels of detail, most notably, hydrodynamical simulations, semi-analytical models (SAMs) and semi-empirical models (SEMs). Hydrodynamical simulations, as a first example, are an extremely useful tool for studying galactic evolution, as they evolve dark matter and stellar particles simultaneously, allowing for a comprehensive overview of their co-evolution \citep[][]{vogelsberger_2014, nelson_2019}. However, these simulations generally require high computational power and are affected by mass/volume resolution limitations. Semi-analytic models and semi-empirical models, as complementary models, require a lower computational cost. Semi-analytic models typically initialize gas at early epochs and subsequently apply physical assumptions and parameterization for the latter \citep[][]{de_lucia_2006, gonzalez_2011, guo_2011, cattaneo_2020, ayromlou2021, koutsouridou_2022}, but could be susceptible to the degeneracy resulting from the high number of parameters. Semi-empirical models, instead, initialize galaxy stellar mass by matching the galactic properties (such as the stellar mass, star formation rate or luminosity) to the dark matter halo properties (such as the halo mass, accretion rate or peak velocity) via their relative abundances or via analytic parameterizations \citep[e.g.,][]{hopkins_2010a, shankar_2006, moster_2013, shankar_2014, tollet_2017, moster_2018, behroozi_2019, grylls_paper1, boco_2023}, and evolve galaxies ensuring that at each redshift the main statistical observational properties of galaxies (e.g., star formation, number densities, clustering) are reproduced. By design, semi-empirical models strongly rely on observational data as input and, therefore, having a robust determination of the statistical properties of galaxies and scaling relations is fundamental for building a successful data-driven model.

The need for well calibrated, uniform, and self-consistent observational data sets is indeed of paramount importance for all types of theoretical galaxy evolution models, not just semi-empirical ones. Besides the quality of the data, it is fundamental to decipher the consistency of the different data sets used as term of comparison. Failures in simultaneously fitting distinct observational probes may be certainly ascribed to shortcomings in the underlying modelling, but they could also arise from disagreements in different data sets. For instance, the galaxy stellar mass function (SMF) contains essential information on the galaxy stellar mass growth. However, stellar masses are commonly inferred via SED fitting, which can introduce potential biases due to the SED fitting algorithm, observational bands and cosmic variance. Indeed, several studies have reported different SMFs, suggesting various slopes at the bright end and evolution with redshift \citep[e.g.,][]{bernardi_2013, shankar_2014, bernardi_2016, bernardi_2017, davidzdon_2017, Kawinwanichakij_2020, weaver_2023}. Furthermore, observed star formation rates (SFRs) might not align consistently with the observed stellar masses. Typically, SFRs predicted by theoretical models and simulations tend to be lower than the observationally estimated values, leading to discrepancies in the observed stellar mass growths and mass functions \citep[see e.g.,][]{bernardi_2010, leja_2015, lapi_2017, grylls_paper2, leja_2022}. For example, \citet{leja_2015} demonstrated that the observed SMF could not be reconciled with the observed main sequence. \citet{grylls_paper1} showed that by using the observed SFRs \citep[from, e.g.,][]{tomczak_2016} the predicted satellites abundances are not compatible with those inferred from observations \citep[e.g.,][]{bernardi_2017}, and also showed that the predicted merger rates are significantly different when adopting different SMFs as inputs in semi-empirical models \citep[see also, e.g.,][]{hopkins_2010b, oleary_2021}.

Semi-empirical models, by empirically linking different observables, prove to be extremely useful for testing possible inconsistencies among distinct data sets, which can often arise due to significant systematics in, e.g., measurements of stellar masses or star formation rates. For example, studying how galaxies build up their stellar mass across cosmic time will yield valuable insights into the properties and systematics of the SMF. In the previous paper of this series \citep[][hereafter referred to as Paper I]{fu_2022} we presented \decode, the Discrete statistical sEmi-empiriCal mODEl and we showed its suitability in addressing these tasks. In this work, we use \decode to test the systematic inconsistencies that may be present among distinct observables, namely the SMF, the star formation histories (SFHs), the satellite abundances, and the intracluster light (ICL). To achieve this goal we will use the most up to date estimates of the SFHs in galaxies from SED fitting, as well as the latest determinations of the SMF, of the satellite abundances and the ICL. We will show that there are still significant systematic inconsistencies among these data sets, which must be seriously considered when attempting to provide a holistic model of galaxy evolution.

This paper is structured as follows. In Section \ref{sec_data}, we describe the data sets that we use in this work. In Section \ref{sec_decode}, we provide the details of our methodology and test the self-consistency of \decode in predicting SFHs against hydrodynamical simulations. In Section \ref{sec_results}, we present \decode's predictions for the SFHs of central galaxies, main sequence, satellite abundances and intracluster light. In Sections \ref{sec_discuss} and \ref{sec_conclu}, finally, we discuss the results that we found and draw our conclusions. In what follows we adopt the $\Lambda$CDM cosmological model with parameters from \cite{planck_2018_cosmo_params} best fit values, i.e. $(\Omega_{\rm m}, \Omega_{\Lambda}, \Omega_{\rm b}, h, n_{\rm S}, \sigma_8) = (0.31, 0.69, 0.049, 0.68, 0.97, 0.81)$, and we use a \citet{chabrier_2003} stellar initial mass function.

\section{Data}\label{sec_data}

In this work, we use (1) simulated data from the TNG100 simulation to validate our methodology and test the performance of \decode for predicting SFHs; (2) the \galics semi-analytic model to compare \decode to an ab initio galaxy formation model; (3) observational data from SDSS (Sloan Digital Sky Survey), MaNGA (Mapping Nearby Galaxies at Apache Point Observatory) and GAMA (Galaxy And Mass Assembly) to test \decode's predictions for SFHs and satellite abundances. Below we provide more details on each of these simulated and observed data sets.

    \subsection{The TNG simulation}\label{sec_data_tng}

    In the present study, we employ the TNG100 simulation, a component of the IllustrisTNG hydrodynamical simulation suite \citep[TNG henceforth;][]{nelson18a,pillepich_2018,springel2018first,marinacci2018first,naiman2018first}. The TNG simulations are performed utilizing the moving-mesh \textsc{AREPO} code \citep[][]{springel2010pur}, which provides solutions for a combination of gravity and magnetohydrodynamical equations \citep[][]{pakmor2011magnetohydrodynamics,pakmor2013simulations}. TNG facilitates the simulation of galaxies via subgrid modelling of critical galaxy formation processes, such as gas cooling, star formation, stellar evolution, feedback from stars, and supermassive black hole associated processes, including seeding, merging, and feedback (see \citealt{pillepich2018Simulating,weinberger17} for a comprehensive model description).
    
    The TNG simulations encompass varying box sizes and mass resolutions. In this paper, we utilize the TNG100 simulation, executed within an approximately 100 Mpc box, which is the most suitable simulation for the objectives of our study. This approach enables us to circumvent any potential non-trivial resolution effects (see \citealt{pillepich2018Simulating} for further details).
    
    For the identification of haloes, the TNG simulation employs the Friends of Friends algorithm \citep[FOF, ][]{Davis1985TheEvolution}, which locates a group of dark matter particles whose mutual distance is less than a linking length $b=0.2$. Subsequently, the subhaloes are identified as gravitationally bound groups of particles, by executing the SUBFIND algorithm \citep[][]{springel2001populating}. Each FOF halo possesses a central subhalo, typically the most massive subhalo of the halo, with all other subhaloes identified as satellite subhaloes. Galaxies correspond to subhaloes with a non-zero stellar mass. FOF haloes generally lack a well-defined shape. However, it is customary to consider $R_{200}$ ($R_{500}$), the radius within which the mean density is 200 (500) times the critical density of the Universe, as the halo radius.\footnote{$R_{200}$ is also referred to as the virial radius, $R_{\rm vir}$, in the literature, although they are not exactly the same.} Given that FOF haloes can extend beyond the $R_{200}$, satellites of a FOF halo can exist and evolve both within and beyond the halo boundary (e.g. see \citealt{Ayromlou2021Comparing,Rohr2023Jellyfish}).
        
    In order to evaluate our abundance matching technique against the TNG simulation, we calculate the SMHM relation from the TNG data, by measuring the ratio between the total stellar mass of all galaxies (centrals and satellites) and the total mass within $R_{200}$, for all haloes in the simulation.
    
    We then employ subhalo merger trees, created using the \textsc{SubLink} algorithm \citep{rodriguez_gomez_2015}, to trace the evolution of individual galaxies over time. For each galaxy, we start from the present time ($z=0$, snapshot=99) and trace back the evolution of its main progenitor through the 100 snapshots/redshifts of the TNG100 simulation. We continue this process until reaching the first appearance of the galaxy in the simulation. This approach enables us to extract several properties of all individual galaxies across cosmic time, including stellar mass growths and star formation histories. We also segregate the stellar masses of galaxies into \textit{in-situ}, formed within the main progenitor branch of a galaxy, and \textit{ex-situ}, originating from minor and major mergers of the galaxy with other objects (see \citealt{rodriguez_gomez_2016}). This allows the careful study of the evolution of galaxies, as presented in the subsequent sections.

    \begin{figure*}
        \includegraphics[width=\textwidth]{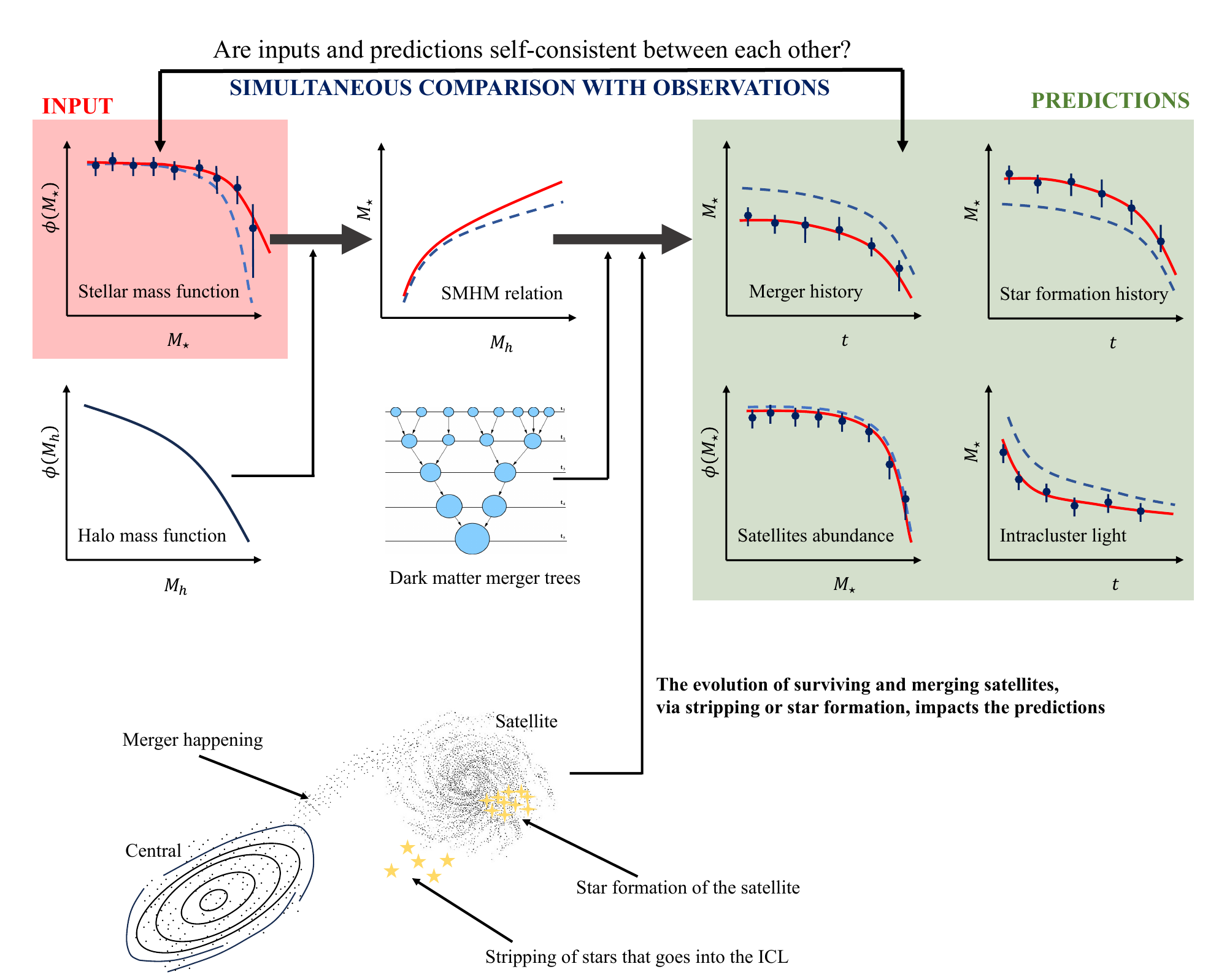}
        \caption{Cartoon of \decode's conception as described in Section \ref{sec_decode}. The galaxy stellar mass function and dark matter halo mass function are taken as inputs in \decode to calculate the SMHM relation via abundance matching (Section \ref{sec_decode_smhm}). The SMHM relation, along with the halo merger trees, is used in \decode to predict galaxy merger histories and abundances of satellites, as described in \paperI. The star formation histories are computed from the difference between the total mass growth and the merger history (Section \ref{sec_decode_SF}). Finally, the intracluster light assumed to be originating from satellites stripping and/or from stellar mass loss during mergers. Different photometries, or different input stellar mass functions, will produce different SMHM relations, merger histories and star formation histories, as shown by the example red solid and blue dashed lines in the cartoon. The red lines in the cartoon are shown to line up with all data. However, in reality, the data sets in the green panel are highly heterogeneous, derived using distinct methods and assumptions, and possibly susceptible to a number of diverse systematic errors.}
        \label{fg_decode_cartoon}
    \end{figure*}

    \subsection{GALICS}\label{sec_data_galics}

    We make use of \galics to compare the stellar mass assemblies and star formation histories predicted by \decode. \galics \citep[Galaxies In Cosmological Simulations;][]{hatton_2003} is a semi-analytic model which follows the evolution of baryons in halo merger trees extracted from cosmological N-body simulations. In this work, we consider the \galics 2.0 framework which includes information on dark matter substructures within each halo \citep[][]{cattaneo_2017}. Central galaxies are at the center of their host haloes. Satellite galaxies are associated with subhaloes. A formula based on hydrodynamic simulations by \citet{jiang_2008} is used to determine how long a satellite galaxy should take to merge with the central galaxy of its host system. As soon as a galaxy becomes a satellite, a merging countdown timer starts ticking.

    In \galics 2.1 \citep[][]{cattaneo_2020} there was introduced a physical criterion to determine when the gas that accretes onto a halo streams onto the central galaxy in cold flows and when it is shock-heated. In absence of feedback from active galactic nuclei (AGN), the shock-heated gas in a halo or subhalo can cool onto the associated galaxy. Its cooling provides a second mechanism for gas accretion. Environmental effects such as ram pressure and tidal stripping reduce the importance of cooling in satellite galaxies \citep[][for details]{koutsouridou_2019, koutsouridou_2022}.

    Gas accreted through cold accretion and cooling settles into discs, the sizes of which are determined by the conservation of angular momentum, and forms stars on a timescale equal to 25 orbital times at one exponential scale-length from the galactic centre \citep[][]{cattaneo_2017}. The conversion of gas into stars is much faster in merger-driven starbursts. Mergers also cause galaxies to grow bulges. The modelling of the effects of galaxy mergers is based on hydrodynamic simulations by \citet{kannan_2015} and described in \citet{koutsouridou_2022} (\galics 2.2). At each merger, a stellar mass equal to $20\%$ of the stellar mass transferred to the bulge is scattered into the halo. We note that this assumption is very similar to the one made in \decode for major mergers, which transfer most of the stars to the bulge component.

    The main new feature of the \galics 2.2 version is AGN feedback, described with the empirical model of \citet{chen_2020}. Black holes grow and deposit feedback energy into the surrounding gas, until this feedback energy is larger than a fraction or multiple of its gravitational binding energy, at which point the gas is unbound or, more realistically, heated to high entropy, so that its cooling time becomes long. By doing so, AGN feedback quenches star formation and prevents its reactivation. The shutdown of star formation induced by AGN quenching is assumed to be permanent because when cooling restarts, it is self-limited. As soon as some gas cools, it reactivates the central AGN,  which shuts down cooling again (maintenance mode).
    
    \subsection{Sloan Digital Sky Survey}\label{sec_data_sdss}
    
    To test the stellar mass function of satellites predicted by \decode at $z\sim0.1$, we use the photometric catalogue of \citet{meert_2015, meert_2016} which is based on the  Sloan Digital Sky Survey Data Release 7 \citep[DR7;][]{abazajian_2009} images. As also done in \paperI (Section 2.4 therein), we use the Meert et al. catalogue of stellar masses derived from the best-fitting \texttt{S\'ersic-Exponential} / \texttt{S\'ersic} photometry on r-band observations, with mass-to-light ratios by \citet{mendel_2014}. Furthermore, we adopt the truncation of the light profile as prescribed in \citet{fischer_2017}. The Meert et al. catalogues are matched with the \citet{yang_2007, yang_2012} group catalogues, which allow us to identify the satellite galaxies.

    We also compare the stellar mass growths from the \citet{mendel_2014} sample with an average redshift of $z\sim0.1$, selecting 100 galaxies in each stellar mass bin of interest. To calculate the star formation histories, we combine the MILES stellar population synthesis models of \citet{vazdekis_2010} with the Penalized PiXel-Fitting (pPXF) inversion algorithm. MILES models provide single stellar population (SSP) predictions in the optical range ($3540 - 7410$ \AA) for a wide range of ages (0.03 to 14 Gyr) and metallicities ($-2.27 < [M/H] < 0.26$) at a $2.51$ \AA \,resolution based on the BaSTI isochrones \citep[][]{pietrinferni_2004, pietrinferni_2006} and assuming a Milky Way-like IMF. We then feed the pPXF with these models, finding the best-fitting linear combination of MILES SSPs in order to reproduce the observed SDSS spectra. The output of pPXF provides the weight to each SSP model with given age and metallicity. Then, star formation histories are computed as the sum of the weights over the metallicities as a function of age (see \citealt{cappellari_2017} for more details).

    \subsection{MaNGA galaxies and their star formation histories}\label{sec_data_manga}
    
    Mapping Nearby Galaxies at Apache Point Observatory (MaNGA) survey \citep{Bundy2015} is part of the fourth generation of SDSS (Section \ref{sec_data_sdss}; see also \citealt{Blanton2017_SDSS-IV}). The data set provides optical-IFU spectroscopy for $\sim 10,000$ galaxies at low redshift ($0.01<z<0.15$). We use the observationally-inferred star formation histories of star-forming MaNGA galaxies from \citet{Bertemes2022}, which were derived based on a full spectral fitting procedure following the stellar population synthesis (SPS) method. 
    In more detail, the integrated MaNGA spectra were fit in the optical wavelength range of $3700 - 7400$ \AA\ simultaneously with associated photometry from the NASA-Sloan Atlas \citep{Blanton2011} using the Bagpipes code \citep{Carnall2018}. Bagpipes is based on the 2016 version of the \citet{Bruzual2003} models, and assumes a \citet{Kroupa2002} IMF. 

    In this paper we use piecewise constant SFHs, which consist of $7$ segments of constant SFR in $7$ age bins, and thus leave a significant amount of freedom to the evolution of the SFR in time. To favour smooth SFHs in the absence of constraining information, the jumps in SFR between the age bins are drawn from a prior corresponding to a Student's t-distribution \citep{Leja2019a}. A \citet{Calzetti2000} extinction law was assumed (with the dust attenuation $A_{V}$ being a free parameter), and all stars were assumed to share a single metallicity, which is left to vary between $0.1 \, Z_\odot$ and $2 \, Z_\odot$. All emission lines were subtracted for the fitting process by using the emission line only (EMLINE) model cube produced by the MaNGA DAP \citep{Westfall2019}. Further, spectra were scaled to a $\rm S/N$ of $30$ prior to fitting to allow the procedure to sufficiently explore the parameter space. 

    \subsection{GAMA}\label{sec_data_gama}
    
    The Galaxy And Mass Assembly (GAMA) survey \citep{driver_2011, liske_2015} is a spectroscopic survey which provides redshifts for $\sim$300,000 galaxies over 5 regions with a total sky area of 230 square degrees. The spectroscopic data are accompanied by 20-band photometry from the ultraviolet to the infrared, providing a vast multi-wavelength data set which allows us to measure SFHs, SFRs and stellar masses of the detected galaxies via SED fitting. As of the final data release DR4\footnote{\url{http://www.gama-survey.org/dr4/}} \citep{driver_2022}, which is based on a new derivation of the underlying photometry \citep{bellstedt_2020a} using the source-extracting software \textsc{ProFound} \citep{robotham_2018}, the survey is 95\% complete down to an $r$-band magnitude of 19.65 mag.
    
    We use the SFHs from \citet{bellstedt_2020}, who studied a sample of 6,688 galaxies with redshift $z<0.06$ and $r<19.5 \, {\rm mag}$ in the three equatorial fields. The 20-band photometry were processed with the SED fitting code \textsc{ProSpect} \citep{robotham_2020} to derive the star formation histories, using the parametric \texttt{massfunc\_snorm\_trunc} function to describe the SFH (corresponding to a skewed Normal function with a truncation in the early Universe), and an evolving metallicity implementation, where the final metallicity is fitted as a free parameter. For further details on the SED fitting implementation, we refer the reader to \citet{bellstedt_2020}.

\section{The DECODE modelling}\label{sec_decode}

To probe the consistency among the different data sets, we make use of the Discrete statistical sEmi-empiriCal mODEl, \decode, presented in \paperI. \decode relies on statistical merger trees generated stochastically from analytical halo mass functions (HMF) and subhalo mass funtions (SHMF). Halo accretion tracks and merger histories are converted transparently into stellar mass growths via input SMHM relations.

In this Section, we describe our recipes to model the evolution of satellite galaxies after the time of infall, such as star formation and stellar stripping, and to predict the mean SFH of central galaxies. Our aim in this work is to model the evolution of central and satellite galaxies with minimal assumptions and input parameters, to be guided as much as possible by the data in extracting constraints on the evolution of galaxies in a cosmological context.

The main steps that we follow to compute SFHs can be summarized as follows:
\begin{itemize}
    \item generation of halo merger trees via \decode (\paperI);
    \item assignment of galaxies to dark matter haloes (Section \ref{sec_decode_smhm});
    \item evolution of satellite galaxies after infall (Section \ref{sec_decode_sat_evo});
    \item prediction of merger histories (\paperI);
    \item computing star formation histories (Section \ref{sec_decode_SF}).
\end{itemize}

Figure \ref{fg_decode_cartoon} shows a cartoon sketching the idea behind \decode, giving a basic description of how different shapes and evolution of the input SMF (therefore SMHM relation) can impact the output SFHs. In brief, our methodology is based on the dark matter merger trees generated following the recipe presented in \paperI, and converted into galaxy mergers to compute the mean cumulative mass assembly history of mergers. The total stellar mass growth of a galaxy is computed by converting the total halo mass assembly via the SMHM relation. Then, the stellar mass growth history from star formation is derived by computing the difference between the total mass growth and the merger history, as we will further detail in Section \ref{sec_decode_SF}. In this work, we include the evolution of satellites after their infall, unlike in \paperI where we focused only on a frozen model where the mass of satellites is assumed to be constant. While a frozen model represents a good first-order approximation, it is not realistic to ignore any further evolutionary process affecting individual satellites after infall, although we will see that in some instances, a frozen model provides a very good approximation, as in the comparison with the TNG100 simulation (see Section \ref{sec_decode_valid_tng}). In this new version of \decode, we therefore also allow for the possibility for the satellites to form stars, quench their star formation and to undergo stellar tidal stripping. The mass stripped from satellites or the mass lost during mergers will contribute to the formation of the ICL. Depending on the evolution of the merging and surviving satellites, the merger and star formation history of central galaxies will also change consequently. Throughout, we assume that $20\%$ of the stellar mass is lost during each merger, following the results of several semi-analytical models and hydrodynamical simulations \citep[e.g.,][]{contini_2014, fontanot_2015, kannan_2015, koutsouridou_2022}. We note that by increasing such mass loss fraction up to $\sim 40\%$, which is among the highest values adopted in cosmological models in the literature (e.g., \citealt{moster_2018}), the resulting galaxy merger rate is reduced by a factor of $0.1$ dex, even for the most massive galaxies with stellar mass $M_\star \sim 10^{12}\, M_\odot$, but the rest of the relevant observables is barely impacted, including the satellite mass function and the SFHs.

Ideally, semi-empirical models, developed on top of a dark matter-dominated hierarchical Universe, are designed to be self-consistent between their input and output observables. This means, in our case, that the input observed SMF should produce a SMHM relation which predicts SFHs, morphologies, ICL, satellite abundances and so on, self-consistent among each other when compared simultaneously with observational data. Therefore, simultaneously matching robust and homogeneous data sets via semi-empirical hierarchical models like ours, will not only be able to test the input assumptions on, e.g., infall timescales and amount of stellar stripping, but also will probe the degree of reliability of a $\Lambda$CDM-based galaxy evolution model in reproducing the real Universe. However, these are challenging goals for a number of reasons. First of all, observations are not fully consistent among themselves, for instance integrated SFHs have often resulted in the literature to overproduce the mass locked up in galaxies when compared to the integrated SMFs at any given epoch, as further discussed below. Furthermore, several observables predicted by \decode depend not only on the input SMF, but also on the way we model satellites evolution, which can be affected by a number of, sometime degenerate, physical processes (e.g., star formation, quenching, stripping). Nevertheless, we will discuss below that some observables are more dependent than others on post-infall satellite evolution and it is thus possible to pin down the effectiveness of a given SMF and SMHM relation to reproduce different data sets.

As shown in Figure \ref{fg_decode_cartoon}, the working flow of \decode, from the left to the right, starts by applying abundance matching between the SMF and central halo mass function to define a redshift and mass dependent SMHM relation. The SMHM relation is in turn used to translate halo assembly histories into galaxy growth histories, which are then decomposed into mergers and star formation histories as detailed below.

    \subsection{Abundance matching}\label{sec_decode_smhm}
    
    The connection between the galaxy stellar mass and the host dark matter halo mass is one of the main ingredients of \decode. Our SMHM relation is computed via the abundance matching technique presented in \paperI (Equation 8) according to the formalism of \citet{aversa_2015}
    \begin{equation}\label{eq_aversa_AM}
    \begin{split}
        \int_{\log M_*}^{+\infty} \phi(M_*', z) & \mathrm d \log M_*' = \\ 
        \int_{-\infty}^{+\infty} & \frac{1}{2} \mathrm{erfc} \Bigg\{ \frac{\log M_h (M_*) - \log M_h' }{\sqrt{2} \Tilde{\sigma}_{\log M_*}} \Bigg\} \\
        & \cdot \phi(M_h', z) \mathrm d \log M_h' \; ,
    \end{split}
    \end{equation}
    where $\Tilde{\sigma}_{\log M_*} = \sigma_{\log M_*} / \mu $, with $\sigma_{\log M_*}$ being the Gaussian scatter in stellar mass at given halo mass and $\mu = \mathrm d \log M_* / \mathrm d \log M_h$ the derivative of the SMHM relation. This method allows to compute the mean stellar mass at fixed halo mass taking as input the dark matter HMF and galaxy SMF, and an input scatter in stellar mass. We make use of the total HMF which includes also surviving unstripped subhaloes, which is computed via the correction to the parent HMF as described in Appendix B in \paperI. For the SMF, we employ several models which are described in Section \ref{sec_res_smhm_models}. We then make use of the output SMHM relation to populate parent haloes and infalling unstripped subhaloes with galaxies. We remind the reader that although \decode generates a statistical ensemble of merger trees, it only robustly predicts mean galaxy assembly, merger, and star formation histories, as detailed in Appendix C of \paperI. We checked that all the sanity checks carried out in \paperI still hold even after inclusion of different satellite evolutionary histories.

    \begin{figure*}
        \includegraphics[width=0.85\textwidth]{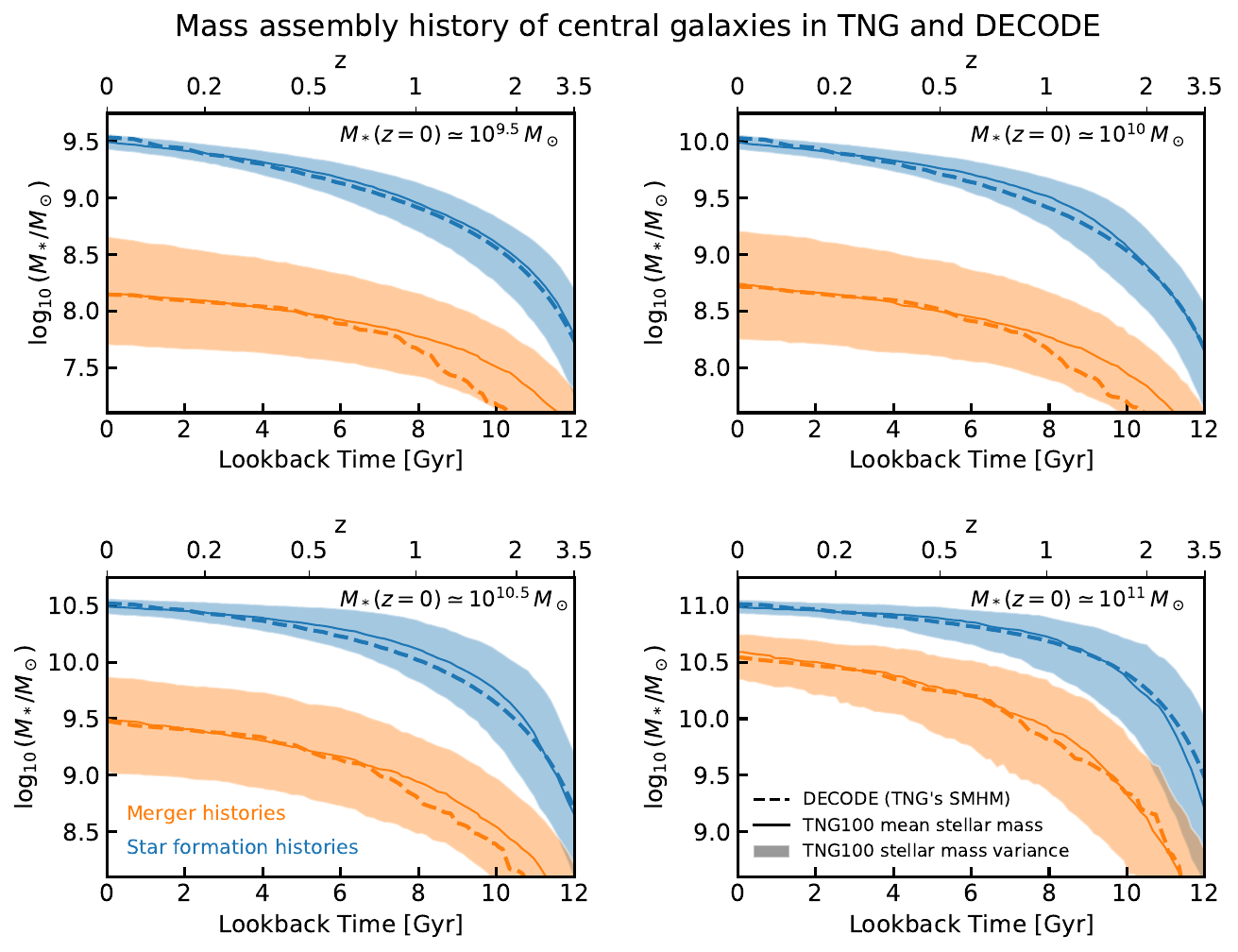}
        \caption{Stellar mass from \textit{in-situ} star formation (blue lines) and \textit{ex-situ} mergers (orange lines) as a function of redshift for central galaxies of four different mass bins at $z=0$ ($M_\star \simeq 10^{9.5}, \, 10^{10}, \, 10^{10.5} \, {\rm and} \, 10^{11} \, M_\odot$), as labelled. The solid lines and shaded areas show the mean stellar mass growths and standard deviation from the TNG simulation, respectively. The dashed lines show the predictions from \decode with the TNG's SMHM relation as input. The error on the mean, $\sigma_{\rm std}/\sqrt{N_{\rm gal}}$, is of the order of $0.01$ dex in all the stellar mass bins for both merger and star formation histories.}
        \label{fg_sfh_comp_tng}
    \end{figure*}
    
    \subsection{Satellite evolution}\label{sec_decode_sat_evo}
    
    In \paperI we assumed that the mass of satellite galaxies remains mostly constant after infall. This assumption allows to rapidly predict the galaxy merger rates, satellite abundances and star formation rates, with minimal computational time. Although the mass of each individual satellite is expected to evolve after infall via several physical processes such as stripping, star formation and quenching \citep[e.g.,][]{cattaneo_2011, wetzel_2013, fillingham_2016, smith_2016, shi_2020, wright_2022, engler_2023}, we will show in Section \ref{sec_res_sat_abun} that a frozen model is sufficiently accurate for our purposes. In this work, we show how these processes impact the prediction of the satellite abundances, merger rates and SFH of central galaxies. Below, we provide a brief description of how we model these additional key physical processes.
    
        \subsubsection{Stellar stripping}\label{sec_decode_strip}

        We apply the recipe for stellar stripping according to the results from the N-body simulations by \citet{smith_2016}
        \begin{equation}\label{eq_stripping_smith}
            f_{\rm str} = \exp (1 - 14.2 f_{\rm DM}) \; ,
        \end{equation}
        where $f_{\rm DM}$ is the ratio between the mass of the subhalo at given redshift $z$ and its peak mass, which we assume to be in good approximation equal to the mass of the subhalo at infall, and $f_{\rm str}$ is the fraction of the satellite's stellar mass at redshift $z$ after stripping.
        
        \subsubsection{Satellites star formation and quenching}\label{sec_decode_sat_sf}

        For each model rendition we first generate the mean main sequence SFR-stellar mass relation (see Section \ref{sec_res_sfhs}), which we then use to assign a SFR to the infalling satellites. We assume that the satellites have enough gas reservoirs to sustain this SFR until a given quenching time which we define below. We stress that in \decode the SFR for centrals is a prediction, while for the satellites it is an input assumption. Obviously, the prediction of the former slightly depends on the latter according to the prescription of Section \ref{sec_decode_SF}, since the SFHs of centrals depend on the merger histories, and therefore on the way we treat satellites evolution. We also allow for the possibility to include a scatter in the SFR-$M_\star$ relation as well. However, we found that by adding a scatter up to $0.2$ dex, the predicted mean quantities of interest here do not vary appreciably.
        
        We quench the star formation of satellites following the recipe from \citet{wetzel_2013} of the ``delayed-then-rapid'' quenching, according to which satellite galaxies continue to form stars for a period equal to the delay timescale $\tau_{\rm delay}$ (see Section 4.3 therein), and after which they undergo a rapid quenching where the SFR is truncated exponentially.
        
    \subsection{Star formation histories of centrals}\label{sec_decode_SF}
    
    Star formation is an additional process, complementary to mergers, significantly contributing to the mass assembly history of a galaxy. In order to compute the SFH of central galaxies, we simply assume that the main components of the stellar mass growth of a galaxy are in first approximation 1) the stellar mass coming from all mergers with other satellite galaxies and 2) the \textit{in-situ} formation of stars. The stellar mass formed \textit{in-situ} is derived by the difference between the total stellar mass growth history, which is inferred by converting the halo mass evolutionary track to stellar mass via the SMHM relation, and that from the cumulative merger history of the galaxy
    \begin{equation}
        M_\star^{\rm SF} (t) = M_\star^{\rm tot} (t) - M_\star^{\rm mer} (t) \; .
    \end{equation}
    $M_*^{\rm SF}$ indicates the stellar mass formed via star formation including also 1) any star formation triggered by wet mergers \citep[e.g.,][]{hopkins_2008}, 2) any star formation from secular process. Our SFRs, when integrated to yield the fractional stellar mass originating from star formation, have to then be corrected at each time step by the cumulative amount of stellar mass returned in gas to the interstellar medium (ISM), described by the global gas mass loss rate (GMLR). In this paper, we make use of the recipe presented in Section 3.2 of \citet{leitner_2011} (see also discussion therein) to account for the gas loss component to derive the SFRs from stellar masses, based on the equation
    \begin{equation}
        {\rm GMLR}(t) = \int_{t_0}^t \dot M_\star^{\rm SF} (t') \dot f_{\rm ML} (t-t') \mathrm dt' \; ,
    \end{equation}
    where $\dot M_\star^{\rm SF} = \mathrm d M_\star^{\rm SF} / \mathrm d t$ is the uncorrected SFR, and $f_{\rm ML}$ is the fraction of mass loss given by Equation (1) of \citet{leitner_2011}.

    \subsection{Validating DECODE's self-consistency}\label{sec_decode_valid_tng}

    Before progressing into generating all \decode's predictions for the mass assembly histories of galaxies, SFHs, merger rates, ICL fractions, using observed SMFs as input, we proceed by testing the inner self-consistency of \decode by comparing with the TNG100 simulation. We specifically aim to test if, by starting from the SMF generated by the TNG100 simulation, \decode is able to predict mass assembly histories, SFHs and merger rates consistent with those in the TNG100 simulation. We stress that the aim here is not to use the TNG data to calibrate our modelling, but to test the self-consistency and performance of \decode in predicting the aforementioned quantities by taking TNG's inputs, i.e., SMHM relation and scatter in stellar mass.
    
    First of all, we test our abundance matching prescription described in Section \ref{sec_decode_smhm}, which takes as inputs the galaxy SMF and the HMF extracted from the TNG simulation. We find that the SMHM relation implied by Equation (\ref{eq_aversa_AM}) is in good agreement with the one directly computed from the simulation itself at all redshifts, demonstrating the robustness of our abundance matching procedure in determining the right mapping between central galaxies and host haloes. We show the comparison in Figure \ref{fg_smhm_TNG_AM_comp} in Appendix \ref{app_abun_match_check} for redshifts $z=0, 1 \, {\rm and} \, 2$, where the upper panels show the SMHM relation from our abundance matching compared to those from the TNG, and the lower panels show the logarithmic difference between the two of them, with a small residual of less than $0.1$ dex.
    
    Secondly, we test the self-consistency of our methodology in calculating the SFH of galaxies with stellar mass today of $M_\star (z=0) \lesssim 10^{11} \, M_\odot$. In Figure \ref{fg_sfh_comp_tng} we show the mean stellar mass accreted \textit{in-situ} (blue lines) and \textit{ex-situ} (orange lines) for four different mass bins of central galaxies at redshift $z=0$, as labelled. The SFHs and merger histories predicted from \decode using TNG's SMHM relation in input are in good agreement with those extracted from the simulation itself. This test proves the validity of our methodology which we aim to extend to observationally driven SMHM relations in Section \ref{sec_res_sfhs}.

    \begin{figure}
        \includegraphics[width=\columnwidth]{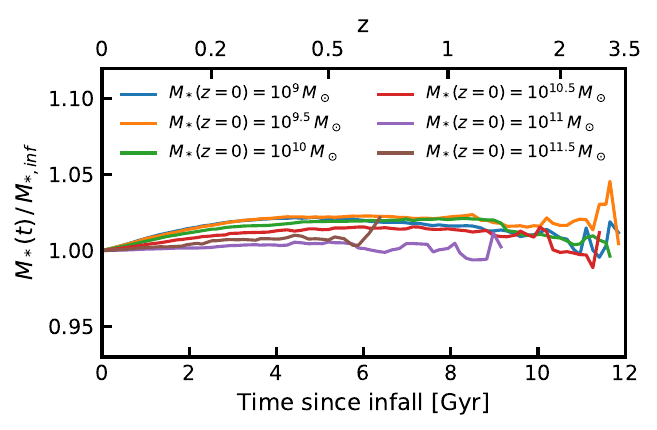}
        \includegraphics[width=\columnwidth]{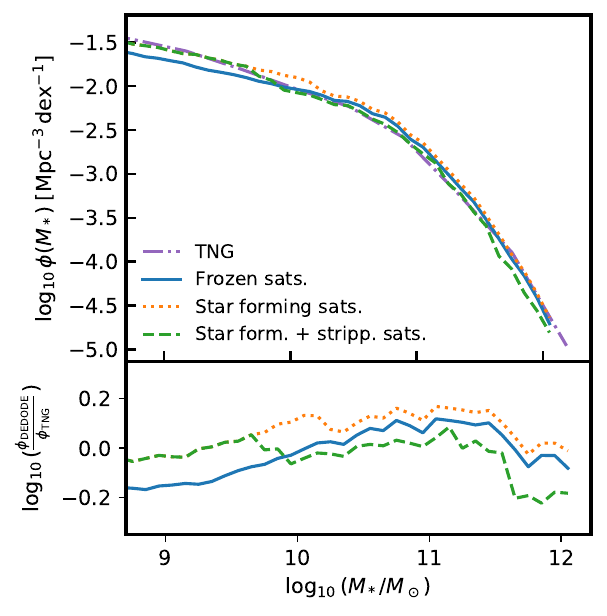}
        \caption{Upper panel: fractional stellar mass growth of satellite galaxies in the TNG100 simulation as a function of time from their infall, for satellites selected in different stellar mass bins at $z=0$. The solid lines and dashed areas show the mean mass growth histories and their standard deviation, respectively. Lower panel: satellites stellar mass function from the TNG100 simulation (purple dash-dotted line), compared to that predicted by \decode using the simulation's SMHM relation in input for the frozen, star-forming and stripped satellites models (blue solid, orange dotted and green dashed lines, respectively), along with the logarithmic difference between \decode's and TNG's satellite stellar mass functions.}
        \label{fg_TNG_sats_Mgrowth}
    \end{figure}
    
    We now turn to test the applicability of our methodology to self-consistently predict the SFHs, or at least the mass assembly histories, of more massive galaxies. For galaxies of stellar mass above $M_\star (z=0) \gtrsim 10^{11} \, M_\odot$, the contribution from mergers becomes progressively more important, and it should be carefully accounted for when predicting SFHs. Since the total mass assembly history of galaxies in our modelling are given by the SMHM relation and the assembly histories of the host dark matter haloes, it is vital to carefully determine the contribution of mergers to derive reliable estimates of the implied SFHs (as described in Section \ref{sec_decode_SF}). In order to accurately predict galaxy merger histories, it is important to have a good understanding of the evolution of merging satellite galaxies, whose mass increase or decrease will lead to different merger rates, as already discussed in Section \ref{sec_decode}. The upper panel of Figure \ref{fg_TNG_sats_Mgrowth} shows the fractional mass evolution of satellite galaxies from the TNG100 simulation. The curves show the fraction of stellar mass growth of the TNG satellites as a function of their time since infall, for satellite galaxies of different masses selected at redshift $z=0$. The aim of the plot is to highlight the stellar mass evolution of the TNG satellites since infall, regardless of their different values of infall redshift. For stellar mass above $M_\star \gtrsim 10^{10} \, M_\odot$, some galaxies have growing mass due to star formation, some others have decreasing mass due to ram-pressure stripping. However, regardless of the physical mechanism that controls the evolution of the single satellite galaxy, we find that in the TNG most of the satellites have, on average, a roughly constant stellar mass growth history since their epoch of infall. Therefore, over a large sample, the frozen model is a good approximation for the evolution of satellites in the TNG simulation. Only the very least massive satellites continue to form stars after infall but are then rapidly quenched \citep[e.g.,][]{shi_2020, ding_2024}, resulting in an average increase in their final stellar mass of roughly $0.1-0.2$ dex since infall.

    \begin{figure*}
        \includegraphics[width=0.9\textwidth]{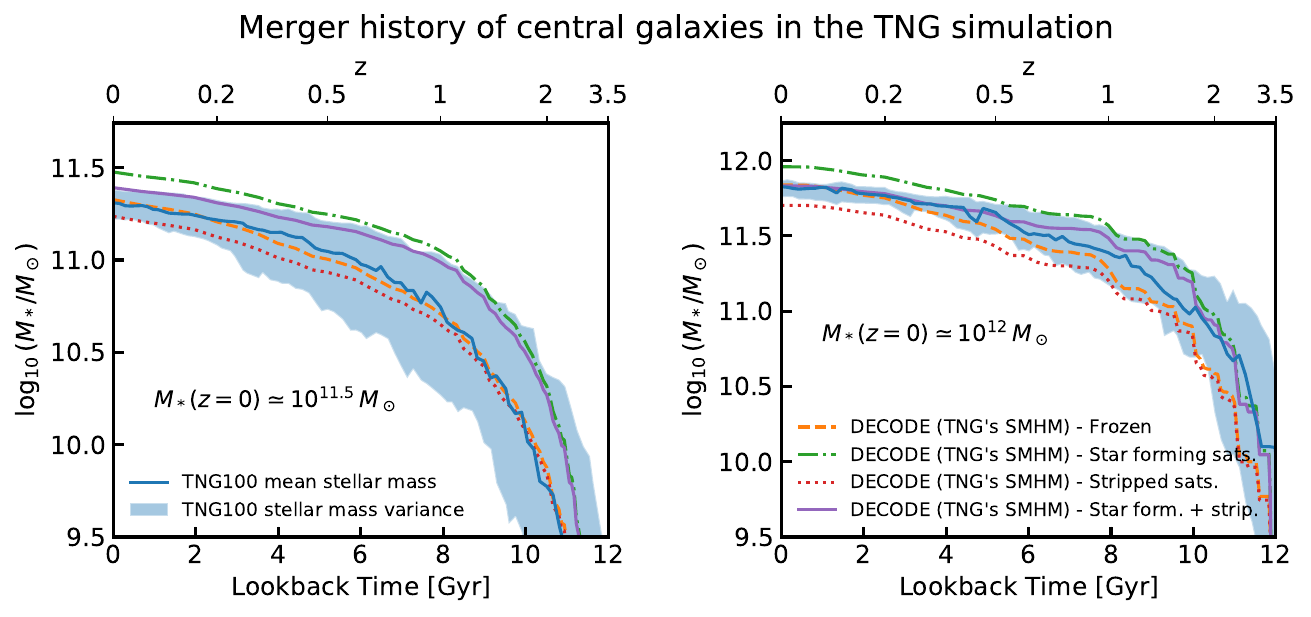}
        \includegraphics[width=0.9\textwidth]{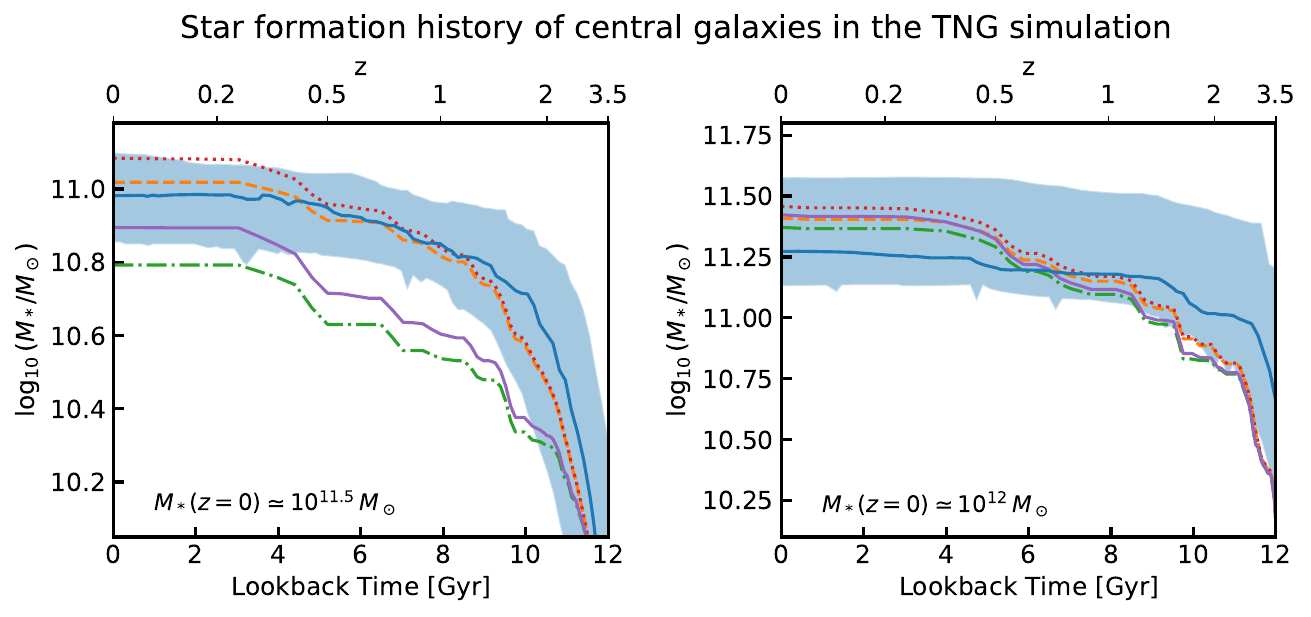}
        \caption{Upper panels: mean stellar mass accreted from \textit{ex-situ} mergers for two stellar mass bins at $z=0$ ($M_\star \simeq 10^{11.5}$ and $10^{12} \, M_\odot$). The blue solid lines and shaded areas show the mean and standard deviation from the TNG simulation, respectively. The orange dashed, green dash-dotted, red dotted and purple solid lines show the results from \decode with TNG's SMHM relation in input for the frozen, star-forming, stripped and star formation+stripping scenarios, respectively. Lower panels: same as upper panels, but for the mean stellar mass accreted via \textit{in-situ} star formation. The error on the mean, $\sigma_{\rm std}/\sqrt{N_{\rm gal}}$, is of the order of $0.01$ dex in both stellar mass bins for both merger and star formation histories.}
        \label{fg_TNG_merg_history}
    \end{figure*}
    
    The lower panel of Figure \ref{fg_TNG_sats_Mgrowth} reports the satellite SMF in the TNG simulation at $z=0$ (dash-dotted, purple line), compared to the predictions from \decode when adopting the TNG's SMHM relation in input, as well as the logarithmic difference between the two of them. We found that a frozen satellites model can relatively well reproduce the simulation's satellite abundances, except for the less massive satellites which continue to form some mass via star formation after their time of infall. When including both stellar stripping and satellite star formation after infall, \decode is able to reproduce the stellar mass function of satellite galaxies with high accuracy across almost the full stellar mass range considered here, further validating \decode's methodology in successfully, flexibly, and self-consistently reproducing the outputs of a comprehensive galaxy evolution model like the TNG, by only using its SMHM relation in input. It is interesting to note from the bottom panel of Figure \ref{fg_TNG_sats_Mgrowth} that the detailed evolutionary histories of satellites after infall, within the modelling explored in this work, have a relatively minor effect on the resulting SMF at $z=0$ of less than $\sim0.1$ dex, for satellites with stellar mass $M_\star \gtrsim 10^{9} \, M_\odot$, which is comparable or even less than the statistical uncertainties in observational estimates \citep[e.g.,][]{grylls_paper1}. The only appreciable difference would apply to lower mass satellites, which could be matched by allowing a fraction of galaxies to increase their stellar mass via star formation before their cold gas being removed via tidal stripping or ram-pressure stripping (see, e.g., \citealt{Rohr2023Jellyfish, ding_2024}).
    
    Having tested the accuracy of \decode in reproducing the observed local SMF of satellite galaxies, we are now in a position to proceed to the comparison between predicted and observed stellar mass assembly histories of massive, central galaxies. The upper panels of Figure \ref{fg_TNG_merg_history} show the cumulative stellar mass accreted from mergers for two mass bins of central galaxy ($M_\star (z=0) = 10^{11.5} \, {\rm and} \, 10^{12} \, M_\odot$), as labelled in the figure, computed via \decode using TNG's SMHM relation as input compared to the merger histories computed from the simulation directly. We found that \decode is capable of reproducing fairly well (within $\sim0.1$ dex) the merger histories of the TNG simulation when taking as input its SMHM relation and adopting a frozen model for the mass of satellites. In both stellar mass bins the scenario where satellites lose their mass through stripping tends to slightly underestimate the TNG's merger histories. On the other hand, a model, in which satellites evolve either via star formation only or both star formation and stellar stripping, tends to somewhat overestimate the TNG's merger histories. At the level in which it is modelled in \decode, the stellar stripping tends to have less impact than the SFR in shaping satellites after infall, and thus less influence on the merger histories.
    
    The lower panels of Figure \ref{fg_TNG_merg_history} show the SFHs for the same galaxy stellar mass bins. As expected, at fixed stellar mass bin, a satellite evolutionary model that produces a higher merger history predicts lower SFHs and vice versa. Similarly as for the merger histories, the frozen satellites model allows to reproduce the TNG's SFHs fairly in agreement. We note that, for the stellar mass bin of $M_\star (z=0) = 10^{12} \, M_\odot$, overall \decode is slightly less performant in reproducing TNG's star formation history for a couple of reasons. First of all, the limited number of clusters ($<10$) in the TNG100 simulation could introduce some bias in the sample. More crucially, the bright end of the SMHM relation computed from the simulation could be affected by a non-negligible numerical uncertainty, with tiny variations in the slope causing large discrepancies in the mass assembly histories in \decode.

\section{Results}\label{sec_results}

Having tested the efficacy of \decode in self-consistently reproducing the mean galaxy assembly histories of the TNG simulation in terms of star formation, mergers, and satellite abundances, by using in input only their SMHM relation, we are now ready to turn to real data. The aim of this Section is to use the latest multiple (but different) determinations of the SMFs at both low and high redshifts, extract via abundance matching the respective SMHM relations implied by these SMFs in a $\Lambda$CDM cosmological context, and then derive the expected SFHs, mergers, and satellite abundances against the corresponding, independently derived data sets. Our aim here is twofold, we aim 1) to test whether the currently available data sets are inherently self-consistent with each other and 2) to discern the best models capable of simultaneously reproducing most of the data sets considered here, including ICL and satellite abundances. This Section is structured as follows. We will first present the determinations of the expected SMHM derived from current data on the SMFs  in Section \ref{sec_res_smhm_models}. We will then show the implied SFHs for low-mass and high-mass central galaxies in Section \ref{sec_res_sfhs}, the corresponding populations of surviving satellites and the implied ICL in Sections \ref{sec_res_sat_abun} and \ref{sec_res_icl}, respectively.

    \subsection{Stellar mass-halo mass models}\label{sec_res_smhm_models}
    
    As already discussed by many works in the literature (e.g., \citealt{shankar_2006, guo_2011, moster_2010, moster_2013, grylls_paper2}; \citetalias{fu_2022}), the SMHM relation is directly determined by the input observed SMF, the shape and evolution of which are still a matter of hot debate, even at low redshifts. On one hand, some groups suggested a negligible time evolution of the SMF. For example, \citet{moustakas_2013}, \citet{bernardi_2013, bernardi_2016, bernardi_2017} and \citet{leja_2020} showed that there is no evolution in the high mass end of the SMF in PRIMUS, SDSS, CMASS and 3D-HST up to $z \simeq 1$. \citet{Kawinwanichakij_2020} suggested that there is no evolution in the SHELA survey's SMF even up to $z=1.5$. On the other hand, other groups found more substantial evolution \citep[e.g.,][]{tomczak_2014, davidzdon_2017, leja_2020, weaver_2023}. At any given redshift, even at $z=0$, the shape of the SMF is still far from being firmly established. \citet{bernardi_2010, bernardi_2013, bernardi_2016} extensively discussed how different choices of background, photometry, mass-to-light ratios, have a profound impact on the SMF, especially at high stellar masses. Other groups have derived very different shapes at both low and high masses and even around the knee of the SMF at all measured redshifts (e.g., \citealt{tomczak_2014, davidzdon_2017, Kawinwanichakij_2020}; \citealt{weibel_2024}).
    
\begin{figure}
        \includegraphics[width=0.85\columnwidth]{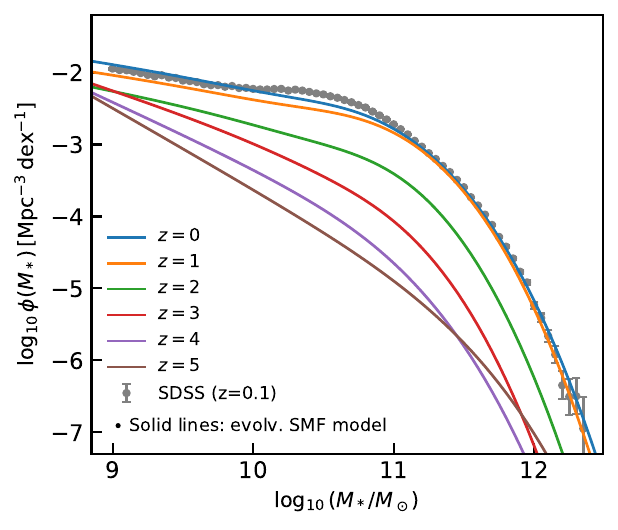}
        \includegraphics[width=\columnwidth]{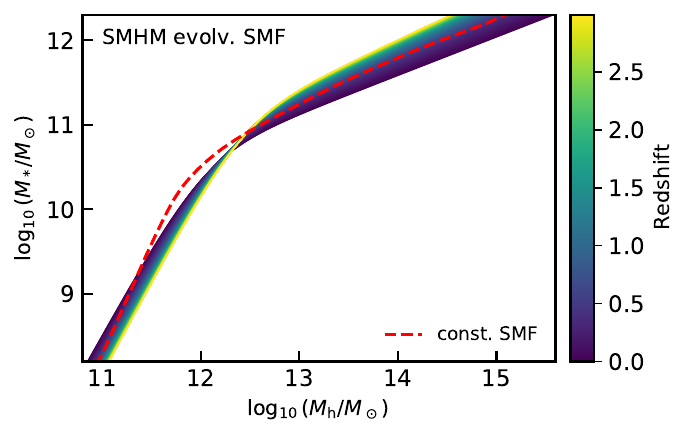}
        \caption{Upper panel: Evolving stellar mass function toy model as described in Section \ref{sec_res_smhm_models} (solid lines) at different redshifts, compared to those from SDSS (grey dots and error bars) at $z=0.1$ from \citet{bernardi_2017}. Lower panel: Stellar mass–halo mass relation computed via abundance matching for the evolving stellar mass function model in the range of redshift denoted by the colour code, compared to the relation at $z=1$ implied by the constant stellar mass function (red dashed line).}
        \label{fg_smf_smhm_ref}
    \end{figure}

    In \paperI we found that a SMHM relation implied by a SMF with flatter and evolving bright end is more suitable to describe the galaxy major merger rates, elliptical fractions and bulge-to-total distributions, compared to a model derived from a constant SMF up to $z=1.5$ at least for $M_\star \gtrsim 10^{11} \, M_\odot$. On the other hand, the constraints on the shape of the input SMHM relation become gradually significantly less tight at stellar masses below $10^{11} \, M_\odot$, for example, major mergers are relatively rare at lower stellar masses (e.g., \citealt{hopkins_2010a, hopkins_2010b, rodriguez_gomez_2015, oleary_2021, husko_2022}; \citetalias{fu_2022}). In what follows, we will underline the role of SFHs and their close link with the shape of the SMF in particular at low masses. Flatter SFHs are on average better mapped into slowly evolving SMFs, as expected also from basic continuity equation arguments \citep[e.g.,][]{leja_2015, grylls_paper2}. To highlight the connections between SFHs and SMFs, we will consider two models for the SMHM relation that broadly bracket the current measurements of the SMF at different redshifts: 

    \begin{itemize}
        \item \textit{constant SMF model}: SMF constant up to redshift $z \sim 1.5$ \citep{bernardi_2017} and gradually dropping in normalization above this redshift according to Equation (11) of \paperI, $\log_{10} \phi (M_\star(z)) \simeq f(z) \cdot \log_{10} \phi(M_\star (z=0.1))$, where $f(z)=1$ for $z\leq1.5$ and $f(z) = (0.99 + 0.13(z-1.5))$ for $z>1.5$.
        \item \textit{evolving SMF model}: SMF characterized by a weakly evolving low-mass end (below the knee) at low redshifts $z \lesssim 1.5$ following \citet{davidzdon_2017}, as plotted in the top panel of Figure \ref{fg_smf_smhm_ref}.
    \end{itemize}

    \begin{figure*}
        \includegraphics[width=0.95\textwidth]{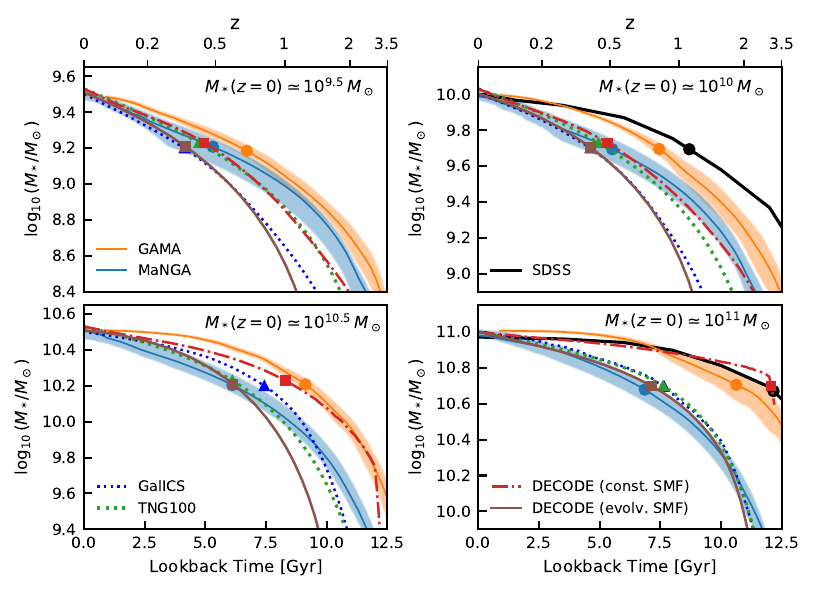}
        \caption{Mean integrated star formation history for four stellar mass bins ($M_\star = 10^{9.5} , \, 10^{10} , \, 10^{10.5} \; {\rm and} \; 10^{11} \, M_\odot$) at $z=0$, as predicted by \decode, compared to the data from the TNG100 simulation as well as SDSS, MaNGA and GAMA surveys. The red dash-dotted and brown solid lines show the predictions from models with constant and evolving stellar mass functions, respectively. The blue and green dotted lines show the results from \galics and the TNG simulation, respectively. The blue and orange lines and shaded areas show the mean SFHs from SEDs and uncertainties of MaNGA and GAMA, respectively. The colored dots, triangles and squares show the 50\% of the stellar mass formed today for the observations (MaNGA and GAMA), theoretical models (TNG and \galics) and \decode's SMHM relations, respectively.}
        \label{fg_sfhs_multiMstar}
    \end{figure*}

    \begin{figure*}
        \includegraphics[width=0.95\textwidth]{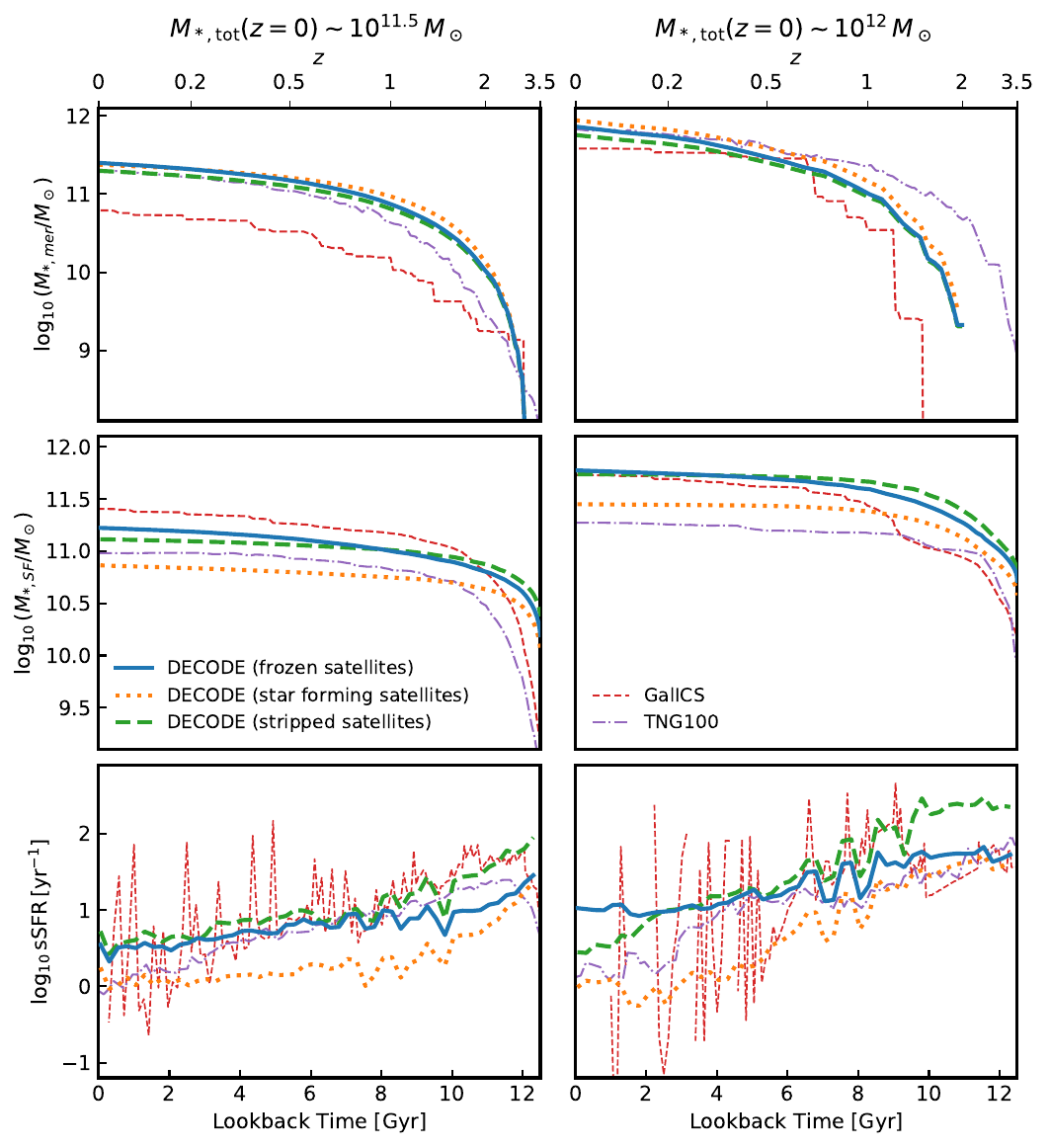}
        \caption{Upper panels: mean merger histories as a function of lookback time for central galaxies of stellar mass $M_\star = 10^{11.5} \; {\rm and} \; 10^{12} \, M_\odot$ at $z=0$, respectively, when inputting an evolving SMF model. The blue solid, yellow dotted and green dashed show the cases where merging satellite galaxies are assumed to not evolve their mass after infall, accrete mass via star formation and lose mass via tidal stripping, respectively. The results are compared with the mean growth histories from \galics (red dashed lines) and the TNG simulation (purple dash-dotted lines). Central panels: same as upper panels but for mean stellar mass accreted via star formation.  Lower panels: same as upper panels but for specific star formation rates.}
        \label{fg_sfhs_highMstar}
    \end{figure*}

    We note that \citet{davidzdon_2017}'s SMF predicts less abundant number of massive galaxies with respect to \citet{bernardi_2017}, which could be attributed to the choice of different mass-to-light ratios and/or different photometries. As we are here mainly interested in the differences in the evolution of the SMFs, and their impact in the SFHs, we have added a scatter of $0.3$ dex to the original fit of \citet{davidzdon_2017}'s SMF to align their fit to \citet{bernardi_2017}'s, but retained their original redshift evolution. We also note that \citet{davidzdon_2017}'s double power-law fit to the SMF predicts slightly less abundant numbers of galaxies around the knee with respect to \citet{bernardi_2017}'s estimate for the SMF (solid blue line and data in the upper panel of Figure \ref{fg_smf_smhm_ref}), but this relatively small difference has little impact on any of the results presented below. Furthermore, we notice that the SMF built in this way is well consistent in shape and normalization at high redshifts ($z\sim 4-5$) with the latest inferred JWST SMF from \citet{weibel_2024}, who reported a steep low-mass end of the SMF at those redshifts with moderate evolution.
    
    The upper panel of Figure \ref{fg_smf_smhm_ref} shows the two models of SMF described above and the lower panel shows the redshift evolution of the SMHM relation computed using the evolving SMF model in the redshift range $0 < z < 3$. The resulting SMHM relation from the evolving model is characterized by an increase in stellar mass at fixed halo mass of $0.29$ dex from $z=0$ to $z=1$ in the high mass range ($M_{\rm h} \sim 10^{14}\, M_\odot$), and by a decrease of $0.24$ dex in the low mass range ($M_{\rm h} \sim 10^{11.5}\, M_\odot$). For both models, we choose an input scatter in stellar mass at fixed halo mass of $0.15$ dex, noticing that varying this input parameter, within reason, does not affect any of our main conclusions. In \citetalias{fu_2022}, indeed, we showed that a redshift- and/or mass-dependent scatter does not significantly alter the merger rates and satellite abundances, mainly because varying the scatter mainly impacts the steepening of the high-mass end of SMHM relation, and thus all the observables below $10^{11}\, M_\odot$, which is the main focus of the present work, remain largely unaltered.

    \subsection{Star formation histories}\label{sec_res_sfhs}

        As already discussed above, the high-mass end slope of the SMHM relation has a profound imprint on the major merger histories of galaxies, steeper relations would preferentially correspond to less major mergers (e.g., \citealt{hopkins_2010a, grylls_paper2, oleary_2021}; \citetalias{fu_2022}). On the other hand, at fixed total stellar mass, a larger contribution of (major) mergers will inevitably correspond to less mass growth via star formation, and thus the shape of the input SMHM relation is expected to also have a significant impact on the SFH of a galaxy.  

        In Figure \ref{fg_sfhs_multiMstar} we compare \decode's predictions on the SFHs of galaxies of different stellar masses at $z=0$, with the SFHs extracted from SED fitting from different surveys and other theoretical models, as labelled. We here focus only on galaxies with stellar mass $M_\star \lesssim 10^{11}\, M_\odot$ at $z=0$. We restrict to this mass limit in this Figure because we want to isolate the full effect of SMF variations on the SFHs. Most theoretical models indeed support the view that galaxy growth is fully SF dominated below $10^{11}\, M_\odot$  \citep[e.g.,][]{oser_2010, lackner_2012, rodriguez_gomez_2016, pillepich_2018, davison_2020, connarozzo_2022, trujillo_gomez_2023}, and \decode indeed predicts the same behaviour (see also \paperI). More specifically, we compare \decode's outputs (red dash-dotted and brown solid lines) with the TNG100 hydrodynamical simulation (green dotted lines), \galics semi-analytic model (blue dotted lines) and the inferred (integrated) star formation histories from the MaNGA and GAMA surveys (orange and cyan solid lines and shaded areas). There are some differences between the MaNGA and GAMA SFHs, especially at stellar masses $M_\star > 10^{10} \, M_\odot$, which are possibly due to the differences between the two data sets themselves. First of all, the two surveys cover different wavelengths, with MaNGA focusing only in the optical regime, while GAMA covers a broad wavelength range from the UV to the IR. Furthermore, due to the smooth truncation of the skewed normal function shape at the beginning of the Universe, the GAMA SFHs might have higher SFRs at high redshifts with respect to MaNGA, where no truncation is used. Additionally, the usage of constant metallicity without modelling the chemical evolution in the MaNGA fit might introduce some slight bias towards younger galaxies compared to those from GAMA. Additionally, we note that there might be a tiny shift in redshift between the star formation histories from theoretical models and observations, due to the different mean redshifts of the samples from the latter, being $z\lesssim 0.1$ for SDSS and MaNGA, and $z<0.06$ for GAMA. We have checked that this effect would be negligible compared to the effects introduced by the systematics of the surveys.

        The mean SFHs in \decode are computed by selecting in the stellar bin of interest a relatively large number of galaxies, which are then evolved in mass following the recipes described in Section \ref{sec_decode_SF}. Here, we show the SFHs from \decode computed via the frozen stellar mass model for satellites. Indeed, similarly to the tests on the TNG in Section \ref{sec_decode_valid_tng}, we have checked that the SFHs and SFRs for central galaxies of $M_\star \lesssim 10^{11}\, M_\odot$ are not altered by the satellites evolution model, varying by less than $0.05$ dex (see Appendix \ref{app_sats_evo_effect}).
        
        The constant input SMF model generates SFHs that are broadly consistent with the SFHs from the TNG simulation, \galics and MaNGA at $M_\star (z=0) = 10^{9.5} \, {\rm and} \, 10^{10} \, M_\odot$, while it produces flatter SFHs for the other two mass bins, being more consistent with the GAMA and SDSS surveys. On the other hand, the SMHM relation implied by an evolving SMF produces mean SFHs steeper than those predicted by a constant SMF model, as expected, given that there is more stellar mass build up in the former model. The evolving SMF model produces SFHs that are steeper than all the data at $M_\star \lesssim 10^{10}\, M_\odot$ (top panels), while at larger masses $M_\star \sim 10^{10} - 10^{11} \, M_\odot$ they are consistent with what inferred from MaNGA, but significantly steeper than those from GAMA and SDSS (bottom panels). The SFHs from the evolving SMF model are broadly consistent with those predicted by the TNG100 at $M_\star \sim 10^{10} - 10^{11} \, M_\odot$, a trend expected as the SMF of the TNG100 has a weaker evolution below the knee but predicting a stronger evolution at higher stellar masses \citep[see, e.g.,][]{pillepich_2018}.
    
        While at lower masses the total stellar mass growth of a galaxy is in good approximation mostly contributed by star formation alone, for more massive galaxies the comparison with observations is less straightforward as mergers can provide a non-negligible contribution to the overall mass assembly histories of galaxies, especially for SMHM relations characterized by a flatter high-mass slope (e.g., \citealt{grylls_paper2}; \citetalias{fu_2022}), as discussed above. On the other hand, the SFHs retrieved from SED fitting record the mass formed in stars but cannot distinguish between the mass formed \textit{ex-situ} or \textit{in-situ}. For this reason, in what follows we do not compare directly with measured SFHs, but rather show the predicted mass grow histories of massive galaxies predicted by \decode using our constant and evolving SMF models, distinguishing between the mass growth via \textit{in-situ} star formation and via  mergers. In what follows we will mainly present the results using the evolving SMF model, but highlight the differences in our results when switching to a constant SMF model. In addition, for completeness, we also discuss the predicted mass accretion histories for different types of satellite evolution after infall.
        
        Figure \ref{fg_sfhs_highMstar} shows the mean merger histories (upper panels) and the integrated SFHs (central panels) for galaxies with stellar masses of $M_\star = 10^{11.5} \; {\rm and} \; 10^{12} \, M_\odot$ at $z=0$, as well as the specific SFRs as a function of time (lower panels). The results in the plot are shown for the evolving SMF as input, and we checked that the stellar mass growths change by less than $\sim0.15$ dex in normalization but maintain the same redshift evolution when using the constant SMF as input. We find that, all the merger histories predicted by \decode are similar to each other, irrespective of the chosen model for the satellite evolution (orange dotted, blue solid, green long dashed lines, as labelled), with only a slight marginal increase of $\sim10\%$ in stellar mass growth when satellites continue forming stars after infall in the most massive centrals (orange dotted lines, top right panel). In this respect, the shape of the SMHM relation has a larger impact than the satellite evolution in shaping the merger histories of central galaxies.

        The predicted stellar mass growth via \textit{in-situ} star formation (central panels) is relatively smaller than the one from mergers by a factor of $\sim2$ and $\sim3$ in the two stellar mass bins, respectively, for both the cases of evolving and constant SMF. It is interesting to note that the SFHs, being overall contributing less to the stellar mass growth in these massive galaxies, are proportionally more dependent on the satellite evolution, in particular the model with star-forming satellites tends to predict a SFH lower by a factor of $\sim2$ than that other two models (orange dotted versus blue solid and green long dashed lines). We stress that the total stellar mass growth remains the same in all models at fixed SMHM relation, i.e., at fixed SMF model, but only the relative proportions of merging and star formation change depending on the satellite evolution. Moreover, the stronger dependence of the \textit{in-situ} stellar mass growth on the satellite evolution with respect to the \textit{ex-situ} growth is an apparent visual effect, due to the high \textit{ex-situ} fraction with respect to the \textit{in-situ} fraction across cosmic time. Indeed, differences of the order of ${\rm few} \times 10^{-2} \, {\rm dex}$ in the \textit{ex-situ} growth will translate into differences of 10 times larger in the \textit{in-situ} growth, causing the apparent more sensible dependence on the satellite evolution. The specific star formation rates (sSFR) plotted in the bottom panels of Figure \ref{fg_sfhs_highMstar} all show, irrespective of the evolution of satellites, a clear decreasing trend with cosmic time, steepening in more massive galaxies, mimicking the overall gas starvation that all galaxies are expected to undergo given the strong fall off in the cosmic SFR below $z\sim1-2$ \citep[e.g.,][]{madau_2014}. It is interesting to note that in our data-driven model \decode, which does not contain any recipe for gas exhaustion or quenching, it is ultimately the shape of the halo accretion rate that drives the stellar mass growth and the sSFR histories, again lending further support to the capital importance of the underlying halo assembly histories in driving the growth of central galaxies \citep[e.g.,][]{neistein_2010, wechsler_2018, bose_2019, jiang_2021, boco_2023, lyu_2023}.

        \begin{figure}
            \includegraphics[width=\columnwidth]{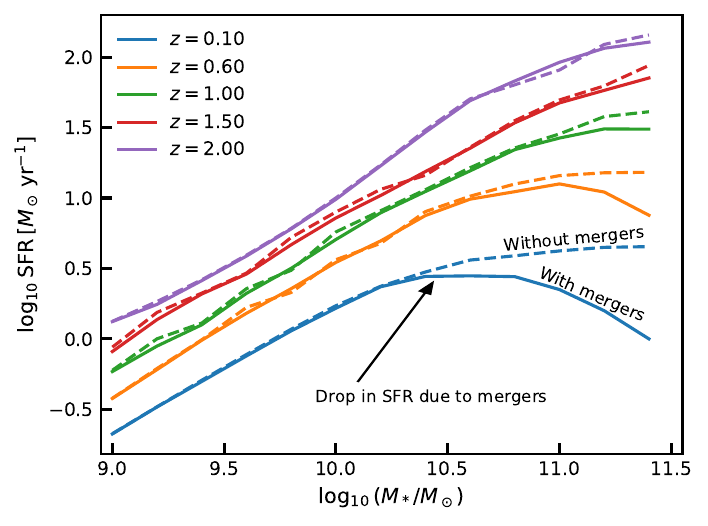}
            \includegraphics[width=\columnwidth]{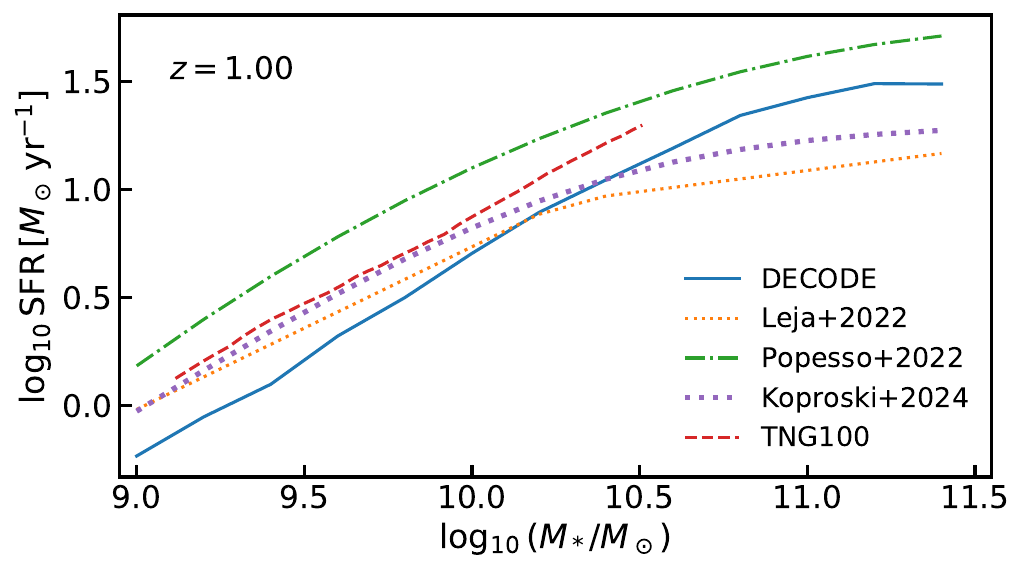}
            \includegraphics[width=\columnwidth]{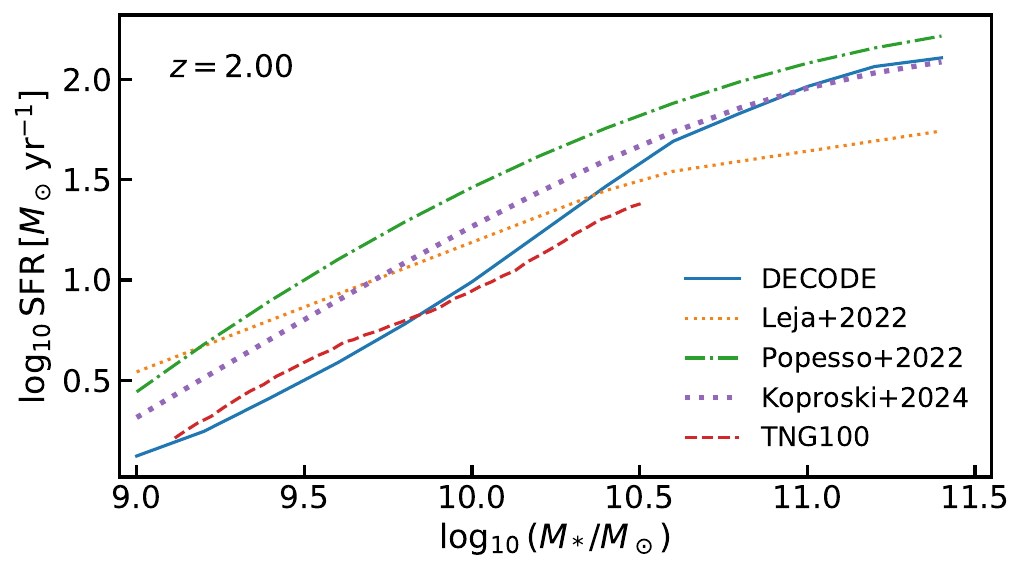}
            \caption{Comparison of the star-forming main sequence at different redshifts when adopting an evolving SMF in input. Upper panel: Main sequence predicted from \decode for $z = 0, 0.6, 1, 1.5 \; {\rm and} \; 2$ (solid lines). The dashed lines show the extreme case where mergers are absent. Central panel: Comparison between the main sequence at $z=1$ from \decode and those from \citet{leja_2022}, \citet{popesso_2023}, \citet{koprowski_2024} and the TNG simulation, as labelled. Lower panel: Same as central panel but for $z=2$.}
            \label{fg_starformingsequence}
        \end{figure}

        Having discussed the mean SFHs of galaxies of different mass along their evolutionary tracks, we now move to the study of the overall SFR-stellar mass relation or main sequence of galaxies as predicted by \decode using our two reference SMF models with and without redshift evolution. In what follows, we focus on the lower mass range, $M_\star \lesssim 10^{11.5} \, M_\odot$, where the fraction of star-forming galaxies dominates the stellar population \citep[e.g.,][]{tomczak_2014, moutard_2016, davidzdon_2017, moster_2018, pillepich_2018, behroozi_2019, donnari_2021, weaver_2023}, guaranteeing a fairly consistent comparison with the observational estimates of the main sequence. Indeed, as also stressed in \paperI, in its present form, \decode generates stochastic realizations of galaxies without distinction between star-forming or quenched galaxies. Nevertheless, we also checked that, weighting the mean SFR-$M_\star$ relation predicted by \decode only by the relative fraction of star-forming galaxies inferred by, e.g., \citet[][]{weaver_2023}, the resulting main sequence relation is changed by less than $0.1$ dex at high redshifts and up to $0.2$ dex at $z=0$ at $M_\star \gtrsim 10^{11} \, M_\odot$, with no impact on its shape, normalization and evolution.
        
        The top panel of Figure \ref{fg_starformingsequence} shows the main sequence as predicted by the evolving SMF model at different redshifts, as labelled. The constant SMF would predict very similar shapes but lower in normalization by a factor of $\sim 1.5-2$ at any given redshift up to $z\sim 3$. We show two realisations of the model, the complete model (solid lines) and the model in which we assume galaxies only grow via star formation at all redshifts and masses (dashed lines), in other words implying that the infalling satellites always have very long dynamical timescales. The latter rendition of the model is clearly possibly an oversimplification, and it would predict too many satellites in the local Universe (see Section \ref{sec_res_sat_abun}), but it is included to provide an overview of the shape of the main sequence in a ``maximal'' model dominated by star formation. We first note from the top panel of Figure \ref{fg_starformingsequence} that the SFR increases at the same pace in redshift at fixed stellar masses, especially below $M_\star \lesssim 3\times 10^{10}\, M_\odot$. At larger stellar masses the increasing rate of the SFR is actually larger, especially when effective mergers are included, which are the main channel that grows the stellar mass in early-type galaxies.

        It is also interesting to note that there is no sign of a downturn in the predicted main sequence at $z>1$ above $M_\star \gtrsim 3\times 10^{10}\, M_\odot$. A break or flattening in the main sequence starts appearing only at $z<1$ (broadly consistent with observation at $z\sim1-2$), and this break is visible even in the extreme model with only SFR. The flattening of the main sequence is a mere consequence of the double power law shape in the input SMHM relation, which implies less stellar mass growth when the central galaxy crosses the knee of the SMHM relation. In the full model inclusive of mergers, the flattening becomes even more evident at $z<1$, creating a downturn in the main sequence, because the majority of massive galaxies is predicted to be quenched with their mass growth being dominated by mergers. In this sense, semi-empirical models represent a powerful tool to probe the effect of the evolution of the SMF, or SMHM relation, on the galaxy SFRs. 
        From the observational point of view, the existence of the break has been widely discussed in the literature with contrasting results, with several works suggesting a single power law shape at all stellar masses and redshifts \citep[e.g.,][]{speagle_2014, rodighiero_2014, pearson_2018}, and others suggesting a break towards high stellar masses \citep[e.g.,][]{whitaker_2014, lee_2015, tomczak_2016, thorne_2021, leja_2022, popesso_2023}.

        The middle and lower panels of Figure \ref{fg_starformingsequence} show a close comparison between the main sequence predicted by the SMF evolving model from \decode and three reference observational results from \citet{leja_2022}, \citet{popesso_2023} and \citet{koprowski_2024}, as well as, for completeness, the predictions from the TNG100 simulation. First off, it is important to note that this comparison is only at a qualitative level as the definitions of stellar masses, despite a tiny deviation due to the choice of the IMF\footnote{The offset in stellar mass corresponds to $0.03$ dex for \citet{kroupa_2001} and \citet{chabrier_2003} IMFs.}, may still be systematically different in \citet{popesso_2023}, \citet{bernardi_2017} and \citet{davidzdon_2017}, which are the ones on which our reference models are based on. At face value, \decode tends to generate a mean SFR that is a factor of $\sim2-3$ systematically lower than in the data. \citet{leja_2022}'s results are based on neural networks to parameterize the galaxy population density in the main sequence plane. With this method, \citet{leja_2022} finds a mean SFR which is also lower than other direct estimates, in better agreement with our models, at least at $z\sim1$. Also \citet{koprowski_2024} and the TNG100 tends to predict on average lower SFRs than \citet{popesso_2023} and in better agreement with our predictions. \citet{leja_2022} also showed that their sSFR time evolution for galaxies of $M_\star = 10^{10}\, M_\odot$ well matches those from numerical simulations, such as EAGLE and IllustrisTNG, even though the main sequence still slightly differs at high redshifts. It has been discussed several times in the literature that some estimates of the SFR-$M_\star$ relations may overproduce the stellar mass density recorded at any given epoch \citep[e.g.,][]{bernardi_2010, rodriguez_puebla_2017, donnari_2019, hashemizadeh_2021, leja_2022}, which is also what we infer here using the evolving SMF model. On the other hand, other groups have found better agreement between the integrated SFR and SMFs \citep[e.g.,][]{bellstedt_2020}. Our results further support the intimate link between observed SFRs and measured SMFs, and how each of these observations can provide valuable constraints on the other one.

    \begin{figure}
        \includegraphics[width=\columnwidth]{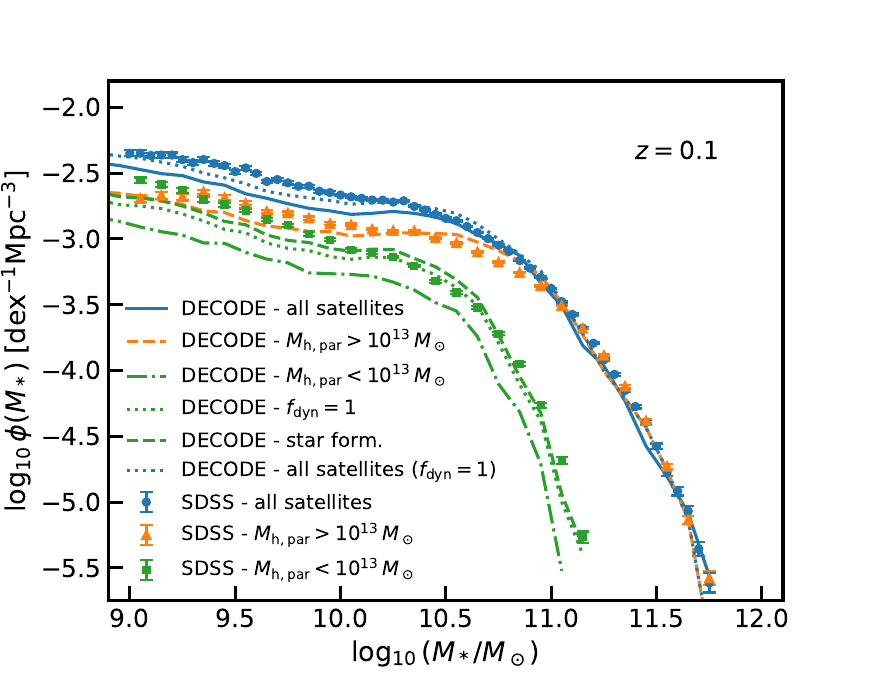}
        \includegraphics[width=\columnwidth]{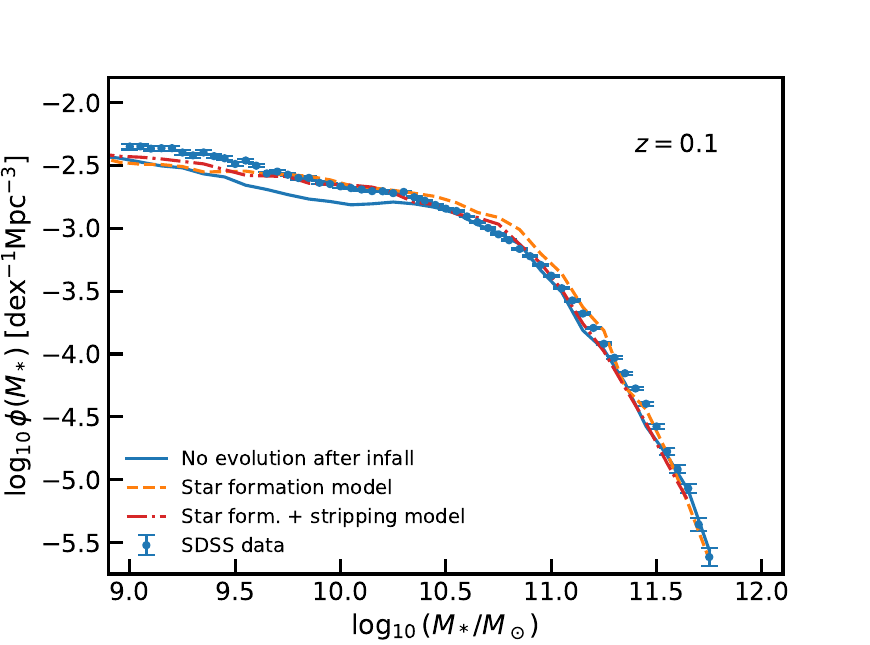}
        \caption{Upper panel: Stellar mass function of surviving satellite galaxies from \decode's evolving SMF model SMHM relation compared to that observed by SDSS. The blue, orange and green solid lines show the SMF for all satellites and parent halo mass greater and smaller than $M_{\rm h,par} = 10^{13} \, M_\odot$, respectively. The green/blue dotted lines show the latter case but with fudge factor $f_{\rm dyn} = 1$, i.e., longer dynamical friction timescales, and the green dashed line with star-forming satellites. Lower panel: Satellites stellar mass function at $z=0.1$ for star-forming and star formation+stripping models. The orange dashed and the red dash-dotted lines show the mass function for star-forming and stripped merging satellites, respectively.}
        \label{fg_smf_sat}
    \end{figure}

    \subsection{Satellite abundances}\label{sec_res_sat_abun}

    As already discussed in other works (e.g., \citealt{moster_2018, behroozi_2019, grylls_paper1}; \citetalias{fu_2022}), the abundances of surviving satellite galaxies provide a complementary and independent test on the shape of the input SMHM relation. Indeed, satellite abundances can be considered as the other side of the coin with respect to mergers, in a dark matter-dominated hierarchical Universe they are effectively ``failed'' mergers at any given epoch, and can thus be used as independent test of the same merger model, and compared against direct observational estimates of the satellite abundances. Any successful cosmological model of galaxy evolution that matches the SFHs and merger histories of galaxies, must also provide the right amount of observed satellites. Any failure in this respect could be, as for any other observable considered so far, attributed to either a shortcoming of the model and/or inconsistencies in the different and independent data sets. Before presenting \decode's predictions on the satellite abundances, we stress once again that the satellite stellar mass function is a genuine prediction of the model, as it depends on both the input SMHM relation, which initialises the subhaloes at infall, and on the prescription for the evolution of the satellites after infall.

    The top panel of Figure \ref{fg_smf_sat} compares \decode's predicted satellite SMF (solid blue line) in the evolving SMF model with frozen satellites against the satellite SDSS galaxy data from \citet{bernardi_2017} as labelled in \citet{yang_2007}'s halo catalogue (blue filled circles). The results with the constant SMF as input do not change appreciably, since the satellites SMF is not very sensible to the input SMHM relation, as already shown in \paperI. It is apparent that the model, without any further fine-tuning, nicely matches the data, at least above $M_{\rm \star,sat} \gtrsim 3\times10^{10}\, M_\odot$, while there is a non-negligible shortfall of $\sim 0.1$ dex at lower masses. When dividing the observed SDSS sample in satellites hosted in parent haloes above and below a chosen host halo mass of $M_{\rm h,par} = 10^{13}\, M_\odot$ (orange triangles and green squares), we see that the major shortfall occurs in lower mass haloes. Allowing for a longer dynamical friction timescale ($f_{\rm dyn}=1$) with respect to \decode's reference timescale (Equation 9 of \paperI), improves the fit at low stellar masses both in lower mass parent haloes as well as in the total satellite SMF (dotted lines). A similar, nearly degenerate, effect of boosting in the number density of satellites is also found by allowing the satellites to continue forming stars after infall as described in Section \ref{sec_decode_sat_sf} (dashed lines). We further show in the bottom panel the effect on the total satellite mass function of including late, post-infall star formation (orange dashed line) and star formation and stellar stripping (red, dash-dotted line), which provides an improved fit to the data with respect to a frozen model at low stellar masses below $M_{\rm \star,sat} \lesssim 3\times10^{10}\, M_\odot$. 

    Although the two model renditions discussed above characterized by a lower dynamical friction timescale or by the inclusion of star formation and stripping, provide both a nearly degenerate match to the local stellar mass function of satellites, we note that some stellar stripping in satellites is expected, and indeed observed, in groups and clusters \citep[e.g.,][]{poggianti_2017, franchetto_2021, akerman_2023}, and can also contribute to the ICL level measured in clusters, as discussed in the next Section.

    \begin{figure}
        \includegraphics[width=\columnwidth]{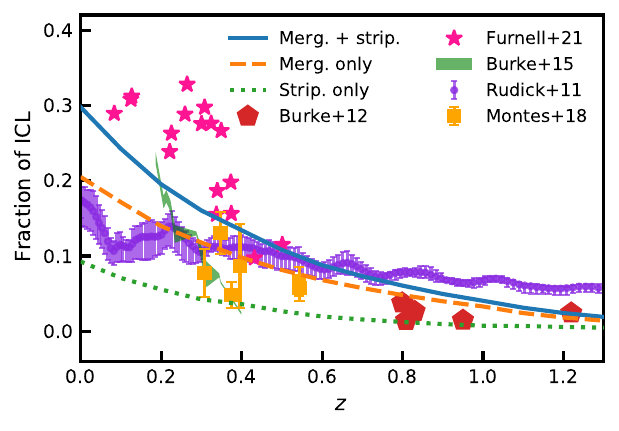}
        \includegraphics[width=\columnwidth]{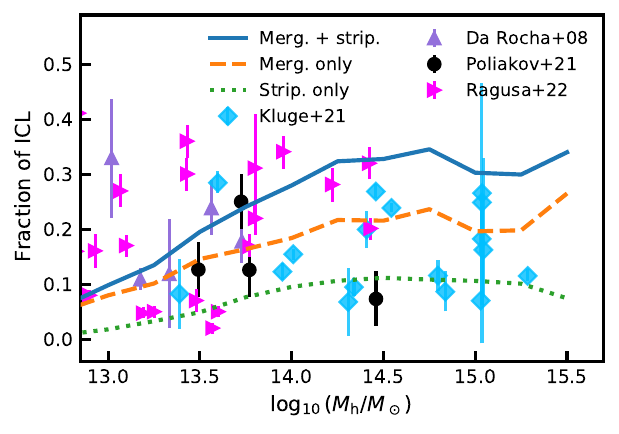}
        \caption{Upper panel: fraction of intracluster light as a function of redshift. The blue solid, orange dashed and green dotted lines show the prediction from \decode's evolving SMF SMHM relation for models with mergers+stripping, mergers only and stripping only, respectively. The points with error bars show  the data from different studies in the literature \citep[][]{rudick_2011, burke_2012, burke_2015, montes_2018, furnell_2021}. Lower panel: same as the upper panel, but as a function of the cluster total mass at redshift $z=0$. The points with error bars show  the data from \citet{da_rocha_2005, da_rocha_2008, kluge_2021, poliakov_2021}.}
        \label{fg_frac_icl}
    \end{figure}
    
    \subsection{Intracluster light}\label{sec_res_icl}

    As a further application of \decode and additional constraint for the assembly history of galaxies, we study the intracluster light (ICL), i.e., the faint diffuse light coming from stars which freely floats within each cluster's gravitational potential and is not bound to any galaxy within the cluster \citep[e.g.,][]{montes_2022}. Many observational works suggested that the origin of the ICL may be due to tidal stripping of satellites and galaxy-galaxy mergers \citep[e.g.,][]{gregg_1998, mihos_2005, contini_2014, contini_2018}. Implementing the production of ICL in our modelling is a valuable addition to constrain the viable models. Regardless of the satellites evolution scenario, a constant fraction of stellar mass is always transferred from mergers to the ICL. Furthermore, when including the stellar stripping in the evolution of satellites, which changes simultaneously the stellar masses of merged and surviving satellites, and merger rates and SFHs of central galaxies, the stellar mass stripped from satellites will contribute to the ICL. Indeed, we compute the ICL in \decode by assuming that it is formed from the stellar mass stripped from the satellite galaxies (Section \ref{sec_decode_strip}) and/or from the stellar mass lost during a galaxy merger. We assume that a fraction of $\sim 20\%$ of the stellar mass during mergers involving the central Brightest Cluster Galaxy (BCG) is lost and transferred to the ICL, similarly to what is assumed in semi-analytic models \citep[see, e.g.,][]{contini_2014}. The amount of ICL might correlate with some cluster properties, such as the redshift and cluster halo mass. We discuss in this Section how the ICL evolves with the latter quantities.
    
    The upper panel of Figure \ref{fg_frac_icl} shows the redshift evolution of the fraction of ICL. Similarly to semi-analytic models, in each cluster, we define the BCG as the most massive central galaxy that lives in the cluster halo \citep[e.g.,][]{tonini_2012}. We define the fraction of ICL as the ratio between the mass of the ICL and the total stellar mass of the cluster (i.e., the sum of the stellar masses of the BCG, surviving satellites and ICL itself). The lines in Figure \ref{fg_frac_icl} show the predictions from \decode for the evolving SMF model, where the ICL is formed by only stripping mechanisms (green dotted, this is an extreme scenario where mergers do not happen), only galaxy merger processes (orange dashed) and both stripping and mergers (blue solid). We compare our results with observational data from \citet{rudick_2011}, \citet{burke_2012}, \citet{burke_2015}, \citet{montes_2018} and \citet{furnell_2021}. We selected haloes with mass between $10^{14} \, M_\odot \lesssim M_{\rm h} \lesssim 10^{15} \, M_\odot$, in order to be as consistent as possible with the data sets we compare to. We stress that the ICL data are often derived in heterogeneous ways, using different methods and calibrations, but they are still all included in the same plot to provide a general term of comparison and broad guide to the models. What is evident from the full collection of the data is that despite the ICL fractions being sparse, they tend to point to averages of $\sim20\%$ at all cluster masses, which is a relevant constraint for models.
    
    Consistently with the observations, we find in the top panel of Figure \ref{fg_frac_icl} that the ICL gradually builds up noticeable mass below $z\lesssim0.7$. If only stellar stripping is considered, then the ICL is only visible at $z<0.3$. However, the predicted ICL in this model would still be limited to $\sim10\%$, which lies below the ensemble average of the data, possibly suggesting that the stellar mass stripped from infalling satellites by itself may not account for all the amount of observed ICL. Figure \ref{fg_frac_icl} shows that the mergers contribution to the ICL is larger with respect to the stripping by a factor of $\sim2$. Overall, both the mergers only and stripping + mergers models better align with the general trend found in simulations and observations, i.e., a negligible amount of ICL at redshift $z \gtrsim 1$ and gradually increasing up to $f_{\rm ICL}\sim 20\%$ in the local Universe \citep[see also][for a detailed discussion]{montes_2022}. The predicted ICL will increase/decrease by few percents and the cumulative mass growth from mergers will decrease/increase by a factor of $\sim 0.05$ dex, by changing the fraction of mass loss in mergers by a factor of $10\%$. Here, once again we can appreciate the power of a semi-empirical model that can gradually build levels of complexity in a transparent way guided by the data.

    The lower panel of Figure \ref{fg_frac_icl} shows the fraction of ICL at redshift $z=0$ as a function of the cluster total mass. The coloured lines show the predictions from \decode, while the data points show the results from \citet{da_rocha_2005, da_rocha_2008, kluge_2021, poliakov_2021}. Exploring the correlation between the ICL and the total halo mass can give an overview on how differently the mass forms in groups and clusters of galaxies. Observations suggest that there is no correlation between the ICL and the halo mass, while simulations found contrasting results, i.e., ICL fraction constant, increasing or decreasing with halo mass \citep[e.g.,][]{murante_2004, lin_2004, purcell_2007, dolag_2010, cui_2014, contini_2018, brough2023, chun_2024}. Interestingly, our results for all the three models suggest an ICL fraction which is overall in agreement with the data, but slightly increasing from the low mass haloes towards high mass haloes. In our semi-empirical framework, this increase in ICL is due to the double power-law shape in the SMHM relation, which implies a higher fraction of major mergers in massive galaxies above the knee of the SMF, which are usually hosted in more massive haloes, as also discussed in Section 4.2 of \paperI. Finally, the recent results of \citet{brough2023}, who computed the ICL from the Horizon-AGN, Hydrangea, IllustrisTNG and Magneticum simulations, also suggests a fraction of ICL of $\sim 10-20\%$ for haloes with mass $14<\log_{10}(M_{\rm h}/M_\odot) <14.5$, broadly in line with our findings.

\section{Discussion}\label{sec_discuss}

In this work, we have highlighted the intimate connection that exists in a hierarchical dark matter-dominated Universe among distinct observables, namely the SMFs, the SFHs, the merger rates, the abundances of satellites, and the level of ICL. All of these probes are all sides of the same coin, and a comprehensive model of galaxy evolution should aim at simultaneously predicting these observations. However, the challenge arises when these independent data sets may suffer from underlying inconsistencies which would prevent a safe comparison between models and data. We proved here that a semi-empirical, data-driven approach is the most suited to shed light on such possible discrepancies in the data, by using a subset of the data in input and others in output. \decode, in particular, uses the SMF as input to predict the SMHM relation, and from there predicts the SFHs, merger rates, and ICL fractions using very minimal additional input assumptions and parameters. 

We first highlighted in Section \ref{sec_res_smhm_models} that the SMF remains poorly known, even at redshift $z=0$, with several works that proposed both very different shapes and redshift evolution for the SMF \citep[see, e.g.,][]{tomczak_2014, bernardi_2017, davidzdon_2017, leja_2020, Kawinwanichakij_2020, weaver_2023}. The analysis in this work points out the strong dependence of the galaxy SFHs on the input SMHM relation, therefore on the SMF, as also found by \citet{grylls_paper2}. Our results suggest that by changing the mapping between stellar mass and halo mass, or the input SMF, the resulting mass growth history can alter significantly by several factors. In this paper, we explored two reference models, one with an evolving and another extreme one with a constant SMF up to $z\sim4$. A slowly evolving SMF at low masses predicts SFHs in good agreement with those retrieved from SED fitting from local data. For galaxies above the knee of the SMF instead, data suggest quite flat SFHs up to $z\sim3$, which would align with the SFHs from a constant SMF at high masses. We found a preference for a SMF characterized by a weak evolution in time at the faint end and by a bright end with flatter slope and significantly evolving. These features allow to predict SFHs for central galaxies with stellar mass today $M_\star \lesssim 10^{11} \, M_\odot$ that well match the observed ones (Section \ref{sec_res_sfhs}) and to produce enough major mergers to simultaneously reproduce the fraction of ellipticals, satellite abundances and B/T distributions in models in which ellipticals are predominantly originating from major mergers (see \paperI and references therein).

Moreover, a SMF with a flatter and evolving high-mass end would be in agreement with the findings of significantly high volume densities of massive star-forming galaxies at $z>3$ from deep ALMA and radio surveys, likely to provide a conspicuous contribution to the high-$z$ SFR density \citep[HST-dark galaxies, e.g.,][]{franco_2018, wang_2019, gruppioni_2020, talia_2021}. The existence of similar numbers of high-$z$ massive star-forming systems are a real challenge for the existing semi-analytical models \citep[e.g.,][]{henriques_2015} and hydrodynamical simulations \citep[e.g.,][]{pillepich_2018}, which underestimate their number density by one to two orders of magnitude \citep[see also][]{wang_2019}. Recent observations with JWST extended these results to even higher redshifts, i.e., $z=8$, finding heavily dust-obscured, massive ($M_\star \sim 10^{10} \, M_\odot$), star-forming sources at $z\sim2-8$ with an surface densities of $\sim 0.8 \, {\rm arcmin}^{-2}$ \citep[][]{barrufet_2023, nelson_2023}. This suggests that an important fraction of massive galaxies may have been missing from our cosmic census at $z>3$ all the way to the Epoch of Reionization.

We also found that the main sequence as observed by several surveys is higher in normalization or flatter in slope at low masses than \decode's predictions, and are not always in line with their observed stellar masses \citep[e.g.,][]{whitaker_2014, speagle_2014, tomczak_2016, popesso_2023}. \citet{leja_2015} extensively discussed this problem and, following a continuity equation approach to evolve galaxy stellar mass growths forwards in time, also found that a main sequence with steeper low-mass end is more suitable to fit the observed SMFs. Similar results are put forward by \citet{leja_2019}, where they evolved the galaxy SMF backwards in time with the SFRs from the 3D-HST catalogues. Similarly, by deriving the SFRs directly from the stellar mass assemblies, which by design fit the mass function, we also found a main sequence which is lower in normalization and steeper in slope than those suggested by some observations cited above, whilst consistent with those from simulations \citep[see, e.g.,][]{donnari_2019}. Moreover, we found that the main sequence presents a break and drops towards high stellar masses, especially at lower redshifts. In our semi-empirical modelling, the shape of the SFR-$M_\star$ relation is a direct by-product of the SMHM relation. As the latter is a double power law, also the main sequence tends to present a similar shape. However, the break in the SFR-$M_\star$ relation changes significantly with redshift, moving from $2\times 10^{11} \, M_\odot$ at $z\geq 2$, to $3\times 10^{10} \, M_\odot$ at $z<1$. Therefore, the main sequence does not present any significant break or flattening up to high stellar masses of a few $10^{11}\,M_\odot$ at high redshifts, with a flattening appearing only below $z<1$, more or less pronounced depending on the shape of the input SMHM relation. This behaviour in the predicted main sequence is visible even in the extreme case where mergers are not included in the model (see Figure \ref{fg_sfhs_highMstar}). When mergers are self-consistently included, they become the dominant process in the galaxy stellar mass assembly leading to a further, more marked drop in the SFR, irrespective of the input SMF model. On the other hand, towards higher redshifts galaxies are still in an active star-forming phase where merger's contribution is less with respect to lower redshifts, and the flattening in the main sequence is barely visible. Our results are in line with the recent works that found a flattening in the slope of the main sequence at high stellar masses, e.g., \citet{leja_2022} and \citet{popesso_2023}, who suggested that such a curvature in the SFR-$M_\star$ relation is most likely due to the suppression of the SFR in galaxies.

Moreover, from the tests on the TNG simulation and from our analysis in Section \ref{sec_res_sat_abun}, we found consistently that the less massive satellite galaxies continue to form stars after their infall before being rapidly quenched \citep[see][for a detailed discussion]{shi_2020}. The addition of this feature allowed us to both reproduce the TNG's and observed satellite abundances. Finally, in this work, we also found evidence for the need of both mergers in addition to the stellar stripping to form the ICL at all stellar masses. This finding is in line with what is found by several observational and theoretical works \citep[see, e.g.,][]{puchwein_2010, rudick_2011, burke_2012, contini_2014, contini_2018, contini_2021, montes_2018, montes_2019}. Indeed, \citet{contini_2014} showed that mergers can form up to $\sim20\%$ of the mass of the ICL in massive clusters. Furthermore, many works, such as \citet{montes_2018} and \citet{contini_2020}, suggested that the stellar stripping forms gradually more mass in the ICL during redshift evolution. The very recent work of \citet{contini_2023} showed that the fraction of ICL also slightly depends on the halo mass, increasing with the latter and then staying steady at $\sim 0.35$, similarly to what we found. These findings are based on the assumption that $20\%$ of the mass of the merging satellites is lost during mergers, which is the value assumed in most of the semi-analytical models in the literature and allows a good compromise between mergers, SFHs and ICL.

\section{Conclusions}\label{sec_conclu}

In this paper, we have presented a holistic perspective encompassing galaxy star formation histories, merger histories, satellite abundances and intracluster light. All of these observables, should in principle be linked together in a hierarchical, dark matter-dominated Universe where galaxies grow via mergers and star formation, and live in multiple environments, from the field to clusters. To probe the connection and self-consistency among these distinct data sets, we made use of \decode in a $\Lambda$CDM hierarchical framework, where dark matter assemblies and merger trees are converted into galaxy mass growth and merger histories via the input SMHM relation. The star formation histories are then derived from the difference between the latter two quantities. We started by testing \decode on a self-consistent and comprehensive model of galaxy formation and evolution, namely the TNG simulation. We found that by using in input TNG's SMHM relation, we were able to simultaneously reproduce the mean star formation and merger histories of TNG's central galaxies, as well as its satellite SMF. We then turned to apply \decode to real data. To this purpose, we used in input two models of the SMHM relation based on the abundance matching between the SMF and the HMF. In one model, we assumed the SMF to be constant up to $z=1.5$, and evolving after that, and in the other we assumed the SMF to evolve significantly from $z=0$. The choice of these two models broadly bracket the most recent observations which are still not agreeing on the exact shape and evolution of the galaxy SMF, as discussed in Section \ref{sec_res_smhm_models}. From these two models we derived via \decode a variety of observables which we compared with a number of independent data sets.

Our main results can be summarised as follows:
\begin{itemize}
    \item A SMHM relation characterized by a weakly evolving low-mass end produces star formation histories in relatively good agreement with those inferred from SED fitting in local surveys such as MaNGA and GAMA, whilst fast evolving low-mass end of the SMF, suggested by some observational groups, are disfavoured (Figure \ref{fg_sfhs_multiMstar}).
    \item For more massive galaxies above $M_\star > 3 \times 10^{10}\, M_\odot$ SED-based estimates of the SFHs do not currently agree with each other. Reproducing the approximately flat stellar mass growth histories inferred from GAMA requires a nearly constant SMF at all stellar masses and up to $z\sim2$, in line with the recent suggestions of an increased star formation efficiency in massive galaxies at $z>2$ from JWST data \citep[e.g.,][]{atek_2023, endsley_2023, labbe_2023, nelson_2023, weibel_2024}.
    \item The evolution of satellites does not affect the star formation histories of central low-mass and intermediate-mass galaxies since the contribution from mergers is small.
    \item The merger histories of massive galaxies above $M_\star > 10^{11} \, M_\odot$ have a relatively weak dependence on the type of post-infall satellite evolution implemented in the model, while the integrated star formation histories and specific star formation rates tend to change by a factor of at most $\sim2-3$ depending on the assumed satellite evolution after infall (Figure \ref{fg_sfhs_highMstar}).
    \item The main sequence SFR-$M_\star$ relation implied by the two SMF models explored here tends to be lower in normalization by a factor $\sim2-3$ with respect to observations. It also shows clear signs of a flattening at $M_\star > 3\times10^{10} \, M_\odot$ but only at $z\lesssim1$, which is independent of the merger rate but a natural byproduct of the break in the SMHM relation. Including mergers in the models tends to further sharpen the drop in the main sequence above $3\times10^{10}-10^{11} \, M_\odot$, as more massive galaxies tend to grow proportionally less via star formation (Figure \ref{fg_starformingsequence}).
    \item The satellites can be considered as failed mergers, and thus their abundances, which depend on the same SMHM in input, represent a valuable complementary probe to test galaxy evolution models. We found that the SDSS local satellites SMF can be nicely reproduced by both the SMHM relations explored here, and including residual star formation in satellites after infall improves the match with the low-mass end of the satellite SMF, especially in more massive parent haloes.
    \item The ICL is another probe of the assembly history of galaxies. Allowing in our reference models for $\sim20\%$ of stellar mass loss during mergers as well as stellar stripping in satellites within clusters, provides a good match to current constraints on the ICL at different redshifts and parent halo masses, irrespective of the input SMHM relation, and with minimal impact on the SFH/merger histories of galaxies (Figure \ref{fg_frac_icl}).
\end{itemize}

In this work, we have put forward a flexible and efficient data-driven approach to probe the self-consistency within a hierarchical, dark matter-dominated Universe, of the SMF with other key independent observational probes such as the star formation rates, the satellite abundance, and the ICL, which will be applied to the imminent data release from extra-galactic missions. The advent of new high-quality data from ongoing surveys, such as JWST \citep[][]{gardner_2006} and Euclid \citep[][]{amiaux_2012}, which will provide self-consistent determinations of the galaxy stellar mass function, will be extremely beneficial for building a more comprehensive and complete understanding of galaxy evolution.

\section*{Acknowledgements}
	
This work received funding from the European Union’s Horizon 2020 research and innovation programme under the Marie Sk\l odowska-Curie grant agreement No. 860744 for the BiD4BESt project (Coordinator: F. Shankar). MA acknowledges funding from the Deutsche Forschungsgemeinschaft (DFG) through an Emmy Noether Research Group (grant number NE 2441/1-1).

\section*{Data Availability}
    
The data underlying this article will be shared on reasonable request to the corresponding author.

\bibliographystyle{mnras}
\bibliography{main}

\begin{thebibliography}{}
\makeatletter
\relax
\def\mn@urlcharsother{\let\do\@makeother \do\$\do\&\do\#\do\^\do\_\do\%\do\~}
\def\mn@doi{\begingroup\mn@urlcharsother \@ifnextchar [ {\mn@doi@}
  {\mn@doi@[]}}
\def\mn@doi@[#1]#2{\def\@tempa{#1}\ifx\@tempa\@empty \href
  {http://dx.doi.org/#2} {doi:#2}\else \href {http://dx.doi.org/#2} {#1}\fi
  \endgroup}
\def\mn@eprint#1#2{\mn@eprint@#1:#2::\@nil}
\def\mn@eprint@arXiv#1{\href {http://arxiv.org/abs/#1} {{\tt arXiv:#1}}}
\def\mn@eprint@dblp#1{\href {http://dblp.uni-trier.de/rec/bibtex/#1.xml}
  {dblp:#1}}
\def\mn@eprint@#1:#2:#3:#4\@nil{\def\@tempa {#1}\def\@tempb {#2}\def\@tempc
  {#3}\ifx \@tempc \@empty \let \@tempc \@tempb \let \@tempb \@tempa \fi \ifx
  \@tempb \@empty \def\@tempb {arXiv}\fi \@ifundefined
  {mn@eprint@\@tempb}{\@tempb:\@tempc}{\expandafter \expandafter \csname
  mn@eprint@\@tempb\endcsname \expandafter{\@tempc}}}

\bibitem[\protect\citeauthoryear{{Abazajian} et~al.,}{{Abazajian}
  et~al.}{2009}]{abazajian_2009}
{Abazajian} K.~N.,  et~al., 2009, \mn@doi [\apjs]
  {10.1088/0067-0049/182/2/543}, \href
  {https://ui.adsabs.harvard.edu/abs/2009ApJS..182..543A} {182, 543}

\bibitem[\protect\citeauthoryear{{Akerman}, {Tonnesen}, {Poggianti}, {Smith},
  {Marasco}, {Kulier}, {M{\"u}ller}  \& {Vulcani}}{{Akerman}
  et~al.}{2023}]{akerman_2023}
{Akerman} N.,  {Tonnesen} S.,  {Poggianti} B.~M.,  {Smith} R.,  {Marasco} A.,
  {Kulier} A.,  {M{\"u}ller} A.,   {Vulcani} B.,  2023, \mn@doi [arXiv
  e-prints] {10.48550/arXiv.2311.04964}, \href
  {https://ui.adsabs.harvard.edu/abs/2023arXiv231104964A} {p. arXiv:2311.04964}

\bibitem[\protect\citeauthoryear{{Amiaux} et~al.,}{{Amiaux}
  et~al.}{2012}]{amiaux_2012}
{Amiaux} J.,  et~al., 2012, in {Clampin} M.~C.,  {Fazio} G.~G.,  {MacEwen}
  H.~A.,   {Oschmann} Jacobus~M. J.,  eds,  Society of Photo-Optical
  Instrumentation Engineers (SPIE) Conference Series Vol. 8442, Space
  Telescopes and Instrumentation 2012: Optical, Infrared, and Millimeter Wave.
  p. 84420Z (\mn@eprint {arXiv} {1209.2228}), \mn@doi{10.1117/12.926513}

\bibitem[\protect\citeauthoryear{{Atek} et~al.,}{{Atek}
  et~al.}{2023}]{atek_2023}
{Atek} H.,  et~al., 2023, \mn@doi [\mnras] {10.1093/mnras/stad1998}, \href
  {https://ui.adsabs.harvard.edu/abs/2023MNRAS.524.5486A} {524, 5486}

\bibitem[\protect\citeauthoryear{{Aversa}, {Lapi}, {de Zotti}, {Shankar}  \&
  {Danese}}{{Aversa} et~al.}{2015}]{aversa_2015}
{Aversa} R.,  {Lapi} A.,  {de Zotti} G.,  {Shankar} F.,   {Danese} L.,  2015,
  \mn@doi [\apj] {10.1088/0004-637X/810/1/74}, \href
  {https://ui.adsabs.harvard.edu/abs/2015ApJ...810...74A} {810, 74}

\bibitem[\protect\citeauthoryear{{Ayromlou}, {Nelson}, {Yates}, {Kauffmann},
  {Renneby}  \& {White}}{{Ayromlou} et~al.}{2021a}]{Ayromlou2021Comparing}
{Ayromlou} M.,  {Nelson} D.,  {Yates} R.~M.,  {Kauffmann} G.,  {Renneby} M.,
  {White} S. D.~M.,  2021a, \mn@doi [\mnras] {10.1093/mnras/staa4011}, \href
  {https://ui.adsabs.harvard.edu/abs/2021MNRAS.502.1051A} {502, 1051}

\bibitem[\protect\citeauthoryear{{Ayromlou}, {Kauffmann}, {Yates}, {Nelson}  \&
  {White}}{{Ayromlou} et~al.}{2021b}]{ayromlou2021}
{Ayromlou} M.,  {Kauffmann} G.,  {Yates} R.~M.,  {Nelson} D.,   {White} S.
  D.~M.,  2021b, \mn@doi [\mnras] {10.1093/mnras/stab1245}, \href
  {https://ui.adsabs.harvard.edu/abs/2021MNRAS.505..492A} {505, 492}

\bibitem[\protect\citeauthoryear{{Barrufet} et~al.,}{{Barrufet}
  et~al.}{2023}]{barrufet_2023}
{Barrufet} L.,  et~al., 2023, \mn@doi [\mnras] {10.1093/mnras/stad947}, \href
  {https://ui.adsabs.harvard.edu/abs/2023MNRAS.522..449B} {522, 449}

\bibitem[\protect\citeauthoryear{{Behroozi}, {Wechsler}, {Hearin}  \&
  {Conroy}}{{Behroozi} et~al.}{2019}]{behroozi_2019}
{Behroozi} P.,  {Wechsler} R.~H.,  {Hearin} A.~P.,   {Conroy} C.,  2019,
  \mn@doi [\mnras] {10.1093/mnras/stz1182}, \href
  {https://ui.adsabs.harvard.edu/abs/2019MNRAS.488.3143B} {488, 3143}

\bibitem[\protect\citeauthoryear{{Bell}}{{Bell}}{2008}]{bell_2008}
{Bell} E.~F.,  2008, \mn@doi [\apj] {10.1086/589551}, \href
  {https://ui.adsabs.harvard.edu/abs/2008ApJ...682..355B} {682, 355}

\bibitem[\protect\citeauthoryear{Bellstedt et~al.,}{Bellstedt
  et~al.}{2020a}]{bellstedt_2020a}
Bellstedt S.,  et~al., 2020a, \mn@doi [MNRAS] {10.1093/mnras/staa1466}, 496,
  3235

\bibitem[\protect\citeauthoryear{{Bellstedt} et~al.,}{{Bellstedt}
  et~al.}{2020b}]{bellstedt_2020}
{Bellstedt} S.,  et~al., 2020b, \mn@doi [\mnras] {10.1093/mnras/staa2620},
  \href {https://ui.adsabs.harvard.edu/abs/2020MNRAS.498.5581B} {498, 5581}

\bibitem[\protect\citeauthoryear{{Bernardi}, {Shankar}, {Hyde}, {Mei},
  {Marulli}  \& {Sheth}}{{Bernardi} et~al.}{2010}]{bernardi_2010}
{Bernardi} M.,  {Shankar} F.,  {Hyde} J.~B.,  {Mei} S.,  {Marulli} F.,
  {Sheth} R.~K.,  2010, \mn@doi [\mnras] {10.1111/j.1365-2966.2010.16425.x},
  \href {https://ui.adsabs.harvard.edu/abs/2010MNRAS.404.2087B} {404, 2087}

\bibitem[\protect\citeauthoryear{{Bernardi}, {Meert}, {Sheth}, {Vikram},
  {Huertas-Company}, {Mei}  \& {Shankar}}{{Bernardi}
  et~al.}{2013}]{bernardi_2013}
{Bernardi} M.,  {Meert} A.,  {Sheth} R.~K.,  {Vikram} V.,  {Huertas-Company}
  M.,  {Mei} S.,   {Shankar} F.,  2013, \mn@doi [\mnras]
  {10.1093/mnras/stt1607}, \href
  {https://ui.adsabs.harvard.edu/abs/2013MNRAS.436..697B} {436, 697}

\bibitem[\protect\citeauthoryear{{Bernardi}, {Meert}, {Sheth},
  {Huertas-Company}, {Maraston}, {Shankar}  \& {Vikram}}{{Bernardi}
  et~al.}{2016}]{bernardi_2016}
{Bernardi} M.,  {Meert} A.,  {Sheth} R.~K.,  {Huertas-Company} M.,  {Maraston}
  C.,  {Shankar} F.,   {Vikram} V.,  2016, \mn@doi [\mnras]
  {10.1093/mnras/stv2487}, \href
  {https://ui.adsabs.harvard.edu/abs/2016MNRAS.455.4122B} {455, 4122}

\bibitem[\protect\citeauthoryear{{Bernardi}, {Meert}, {Sheth}, {Fischer},
  {Huertas-Company}, {Maraston}, {Shankar}  \& {Vikram}}{{Bernardi}
  et~al.}{2017}]{bernardi_2017}
{Bernardi} M.,  {Meert} A.,  {Sheth} R.~K.,  {Fischer} J.~L.,
  {Huertas-Company} M.,  {Maraston} C.,  {Shankar} F.,   {Vikram} V.,  2017,
  \mn@doi [\mnras] {10.1093/mnras/stx176}, \href
  {https://ui.adsabs.harvard.edu/abs/2017MNRAS.467.2217B} {467, 2217}

\bibitem[\protect\citeauthoryear{Bertemes \& Wuyts}{Bertemes \&
  Wuyts}{prep}]{Bertemes2022}
Bertemes C.,  Wuyts S.,  in prep.

\bibitem[\protect\citeauthoryear{{Blanton}, {Kazin}, {Muna}, {Weaver}  \&
  {Price-Whelan}}{{Blanton} et~al.}{2011}]{Blanton2011}
{Blanton} M.~R.,  {Kazin} E.,  {Muna} D.,  {Weaver} B.~A.,   {Price-Whelan} A.,
   2011, \mn@doi [The Astrophysical Journal] {10.1088/0004-6256/142/1/31},
  \href {https://ui.adsabs.harvard.edu/abs/2011AJ....142...31B} {142, 31}

\bibitem[\protect\citeauthoryear{{Blanton} et~al.,}{{Blanton}
  et~al.}{2017}]{Blanton2017_SDSS-IV}
{Blanton} M.~R.,  et~al., 2017, \mn@doi [The Astrophysical Journal]
  {10.3847/1538-3881/aa7567}, \href
  {https://ui.adsabs.harvard.edu/abs/2017AJ....154...28B} {154, 28}

\bibitem[\protect\citeauthoryear{{Bluck}, {Mendel}, {Ellison}, {Moreno},
  {Simard}, {Patton}  \& {Starkenburg}}{{Bluck} et~al.}{2014}]{bluck_2014}
{Bluck} A. F.~L.,  {Mendel} J.~T.,  {Ellison} S.~L.,  {Moreno} J.,  {Simard}
  L.,  {Patton} D.~R.,   {Starkenburg} E.,  2014, \mn@doi [\mnras]
  {10.1093/mnras/stu594}, \href
  {https://ui.adsabs.harvard.edu/abs/2014MNRAS.441..599B} {441, 599}

\bibitem[\protect\citeauthoryear{{Bluck}, {Maiolino}, {Brownson}, {Conselice},
  {Ellison}, {Piotrowska}  \& {Thorp}}{{Bluck} et~al.}{2022}]{bluck_2022}
{Bluck} A. F.~L.,  {Maiolino} R.,  {Brownson} S.,  {Conselice} C.~J.,
  {Ellison} S.~L.,  {Piotrowska} J.~M.,   {Thorp} M.~D.,  2022, \mn@doi [\aap]
  {10.1051/0004-6361/202142643}, \href
  {https://ui.adsabs.harvard.edu/abs/2022A&A...659A.160B} {659, A160}

\bibitem[\protect\citeauthoryear{{Boco}, {Lapi}, {Shankar}, {Fu}, {Gabrielli}
  \& {Sicilia}}{{Boco} et~al.}{2023}]{boco_2023}
{Boco} L.,  {Lapi} A.,  {Shankar} F.,  {Fu} H.,  {Gabrielli} F.,   {Sicilia}
  A.,  2023, \mn@doi [\apj] {10.3847/1538-4357/ace76d}, \href
  {https://ui.adsabs.harvard.edu/abs/2023ApJ...954...97B} {954, 97}

\bibitem[\protect\citeauthoryear{{Bose}, {Eisenstein}, {Hernquist},
  {Pillepich}, {Nelson}, {Marinacci}, {Springel}  \& {Vogelsberger}}{{Bose}
  et~al.}{2019}]{bose_2019}
{Bose} S.,  {Eisenstein} D.~J.,  {Hernquist} L.,  {Pillepich} A.,  {Nelson} D.,
   {Marinacci} F.,  {Springel} V.,   {Vogelsberger} M.,  2019, \mn@doi [\mnras]
  {10.1093/mnras/stz2546}, \href
  {https://ui.adsabs.harvard.edu/abs/2019MNRAS.490.5693B} {490, 5693}

\bibitem[\protect\citeauthoryear{{Brough} et~al.,}{{Brough}
  et~al.}{2023}]{brough2023}
{Brough} S.,  et~al., 2023, \mn@doi [arXiv e-prints]
  {10.48550/arXiv.2311.18016}, \href
  {https://ui.adsabs.harvard.edu/abs/2023arXiv231118016B} {p. arXiv:2311.18016}

\bibitem[\protect\citeauthoryear{Bruzual \& Charlot}{Bruzual \&
  Charlot}{2003}]{Bruzual2003}
Bruzual G.,  Charlot S.,  2003, Monthly Notices of the Royal Astronomical
  Society, 344, 1000

\bibitem[\protect\citeauthoryear{{Buchan} \& {Shankar}}{{Buchan} \&
  {Shankar}}{2016}]{buchan_2016}
{Buchan} S.,  {Shankar} F.,  2016, \mn@doi [\mnras] {10.1093/mnras/stw1771},
  \href {https://ui.adsabs.harvard.edu/abs/2016MNRAS.462.2001B} {462, 2001}

\bibitem[\protect\citeauthoryear{{Bundy} et~al.,}{{Bundy}
  et~al.}{2015}]{Bundy2015}
{Bundy} K.,  et~al., 2015, \mn@doi [The Astrophysical Journal]
  {10.1088/0004-637X/798/1/7}, \href
  {https://ui.adsabs.harvard.edu/abs/2015ApJ...798....7B} {798, 7}

\bibitem[\protect\citeauthoryear{{Burke}, {Collins}, {Stott}  \&
  {Hilton}}{{Burke} et~al.}{2012}]{burke_2012}
{Burke} C.,  {Collins} C.~A.,  {Stott} J.~P.,   {Hilton} M.,  2012, \mn@doi
  [\mnras] {10.1111/j.1365-2966.2012.21555.x}, \href
  {https://ui.adsabs.harvard.edu/abs/2012MNRAS.425.2058B} {425, 2058}

\bibitem[\protect\citeauthoryear{{Burke}, {Hilton}  \& {Collins}}{{Burke}
  et~al.}{2015}]{burke_2015}
{Burke} C.,  {Hilton} M.,   {Collins} C.,  2015, \mn@doi [\mnras]
  {10.1093/mnras/stv450}, \href
  {https://ui.adsabs.harvard.edu/abs/2015MNRAS.449.2353B} {449, 2353}

\bibitem[\protect\citeauthoryear{{Calzetti}, {Armus}, {Bohlin}, {Kinney},
  {Koornneef}  \& {Storchi-Bergmann}}{{Calzetti} et~al.}{2000}]{Calzetti2000}
{Calzetti} D.,  {Armus} L.,  {Bohlin} R.~C.,  {Kinney} A.~L.,  {Koornneef} J.,
   {Storchi-Bergmann} T.,  2000, \mn@doi [The Astrophysical Journal]
  {10.1086/308692}, \href
  {https://ui.adsabs.harvard.edu/abs/2000ApJ...533..682C} {533, 682}

\bibitem[\protect\citeauthoryear{{Cannarozzo} et~al.,}{{Cannarozzo}
  et~al.}{2022}]{connarozzo_2022}
{Cannarozzo} C.,  et~al., 2022, \mn@doi [\mnras] {10.1093/mnras/stac3023},
  \href {https://ui.adsabs.harvard.edu/abs/2022MNRAS.tmp.2860C} {}

\bibitem[\protect\citeauthoryear{{Cappellari}}{{Cappellari}}{2017}]{cappellari_2017}
{Cappellari} M.,  2017, \mn@doi [\mnras] {10.1093/mnras/stw3020}, \href
  {https://ui.adsabs.harvard.edu/abs/2017MNRAS.466..798C} {466, 798}

\bibitem[\protect\citeauthoryear{Carnall, McLure, Dunlop  \&
  Dav{\'{e}}}{Carnall et~al.}{2018}]{Carnall2018}
Carnall A.~C.,  McLure R.~J.,  Dunlop J.~S.,   Dav{\'{e}} R.,  2018, \mn@doi
  [Monthly Notices of the Royal Astronomical Society] {10.1093/MNRAS/STY2169},
  480, 4379

\bibitem[\protect\citeauthoryear{{Cattaneo}, {Mamon}, {Warnick}  \&
  {Knebe}}{{Cattaneo} et~al.}{2011}]{cattaneo_2011}
{Cattaneo} A.,  {Mamon} G.~A.,  {Warnick} K.,   {Knebe} A.,  2011, \mn@doi
  [\aap] {10.1051/0004-6361/201015780}, \href
  {https://ui.adsabs.harvard.edu/abs/2011A&A...533A...5C} {533, A5}

\bibitem[\protect\citeauthoryear{{Cattaneo} et~al.,}{{Cattaneo}
  et~al.}{2017}]{cattaneo_2017}
{Cattaneo} A.,  et~al., 2017, \mn@doi [\mnras] {10.1093/mnras/stx1597}, \href
  {https://ui.adsabs.harvard.edu/abs/2017MNRAS.471.1401C} {471, 1401}

\bibitem[\protect\citeauthoryear{{Cattaneo}, {Koutsouridou}, {Tollet},
  {Devriendt}  \& {Dubois}}{{Cattaneo} et~al.}{2020}]{cattaneo_2020}
{Cattaneo} A.,  {Koutsouridou} I.,  {Tollet} E.,  {Devriendt} J.,   {Dubois}
  Y.,  2020, \mn@doi [\mnras] {10.1093/mnras/staa1832}, \href
  {https://ui.adsabs.harvard.edu/abs/2020MNRAS.497..279C} {497, 279}

\bibitem[\protect\citeauthoryear{{Chabrier}}{{Chabrier}}{2003}]{chabrier_2003}
{Chabrier} G.,  2003, \mn@doi [\pasp] {10.1086/376392}, \href
  {https://ui.adsabs.harvard.edu/abs/2003PASP..115..763C} {115, 763}

\bibitem[\protect\citeauthoryear{{Chen} et~al.,}{{Chen}
  et~al.}{2020}]{chen_2020}
{Chen} Z.,  et~al., 2020, \mn@doi [\apj] {10.3847/1538-4357/ab9633}, \href
  {https://ui.adsabs.harvard.edu/abs/2020ApJ...897..102C} {897, 102}

\bibitem[\protect\citeauthoryear{{Chun}, {Shin}, {Ko}, {Smith}  \&
  {Yoo}}{{Chun} et~al.}{2024}]{chun_2024}
{Chun} K.,  {Shin} J.,  {Ko} J.,  {Smith} R.,   {Yoo} J.,  2024, \mn@doi [arXiv
  e-prints] {10.48550/arXiv.2405.08061}, \href
  {https://ui.adsabs.harvard.edu/abs/2024arXiv240508061C} {p. arXiv:2405.08061}

\bibitem[\protect\citeauthoryear{{Contini}}{{Contini}}{2021}]{contini_2021}
{Contini} E.,  2021, \mn@doi [Galaxies] {10.3390/galaxies9030060}, \href
  {https://ui.adsabs.harvard.edu/abs/2021Galax...9...60C} {9, 60}

\bibitem[\protect\citeauthoryear{{Contini} \& {Gu}}{{Contini} \&
  {Gu}}{2020}]{contini_2020}
{Contini} E.,  {Gu} Q.,  2020, \mn@doi [\apj] {10.3847/1538-4357/abb1aa}, \href
  {https://ui.adsabs.harvard.edu/abs/2020ApJ...901..128C} {901, 128}

\bibitem[\protect\citeauthoryear{{Contini}, {De Lucia}, {Villalobos}  \&
  {Borgani}}{{Contini} et~al.}{2014}]{contini_2014}
{Contini} E.,  {De Lucia} G.,  {Villalobos} {\'A}.,   {Borgani} S.,  2014,
  \mn@doi [\mnras] {10.1093/mnras/stt2174}, \href
  {https://ui.adsabs.harvard.edu/abs/2014MNRAS.437.3787C} {437, 3787}

\bibitem[\protect\citeauthoryear{{Contini}, {Yi}  \& {Kang}}{{Contini}
  et~al.}{2018}]{contini_2018}
{Contini} E.,  {Yi} S.~K.,   {Kang} X.,  2018, \mn@doi [\mnras]
  {10.1093/mnras/sty1518}, \href
  {https://ui.adsabs.harvard.edu/abs/2018MNRAS.479..932C} {479, 932}

\bibitem[\protect\citeauthoryear{{Contini}, {Jeon}, {Rhee}, {Han}  \&
  {Yi}}{{Contini} et~al.}{2023}]{contini_2023}
{Contini} E.,  {Jeon} S.,  {Rhee} J.,  {Han} S.,   {Yi} S.~K.,  2023, arXiv
  e-prints, \href {https://ui.adsabs.harvard.edu/abs/2023arXiv231003263C} {p.
  arXiv:2310.03263}

\bibitem[\protect\citeauthoryear{{Cowie} \& {Songaila}}{{Cowie} \&
  {Songaila}}{1977}]{cowie_1977}
{Cowie} L.~L.,  {Songaila} A.,  1977, \mn@doi [\nat] {10.1038/266501a0}, \href
  {https://ui.adsabs.harvard.edu/abs/1977Natur.266..501C} {266, 501}

\bibitem[\protect\citeauthoryear{{Cui} et~al.,}{{Cui} et~al.}{2014}]{cui_2014}
{Cui} W.,  et~al., 2014, \mn@doi [\mnras] {10.1093/mnras/stt1940}, \href
  {https://ui.adsabs.harvard.edu/abs/2014MNRAS.437..816C} {437, 816}

\bibitem[\protect\citeauthoryear{{Da Rocha} \& {Mendes de Oliveira}}{{Da Rocha}
  \& {Mendes de Oliveira}}{2005}]{da_rocha_2005}
{Da Rocha} C.,  {Mendes de Oliveira} C.,  2005, \mn@doi [\mnras]
  {10.1111/j.1365-2966.2005.09641.x}, \href
  {https://ui.adsabs.harvard.edu/abs/2005MNRAS.364.1069D} {364, 1069}

\bibitem[\protect\citeauthoryear{{Da Rocha}, {Ziegler}  \& {Mendes de
  Oliveira}}{{Da Rocha} et~al.}{2008}]{da_rocha_2008}
{Da Rocha} C.,  {Ziegler} B.~L.,   {Mendes de Oliveira} C.,  2008, \mn@doi
  [\mnras] {10.1111/j.1365-2966.2008.13500.x}, \href
  {https://ui.adsabs.harvard.edu/abs/2008MNRAS.388.1433D} {388, 1433}

\bibitem[\protect\citeauthoryear{{Davidzon} et~al.,}{{Davidzon}
  et~al.}{2017}]{davidzdon_2017}
{Davidzon} I.,  et~al., 2017, \mn@doi [\aap] {10.1051/0004-6361/201730419},
  \href {https://ui.adsabs.harvard.edu/abs/2017A&A...605A..70D} {605, A70}

\bibitem[\protect\citeauthoryear{{Davis}, {Efstathiou}, {Frenk}  \&
  {White}}{{Davis} et~al.}{1985}]{Davis1985TheEvolution}
{Davis} M.,  {Efstathiou} G.,  {Frenk} C.~S.,   {White} S.~D.~M.,  1985,
  \mn@doi [\apj] {10.1086/163168}, \href
  {https://ui.adsabs.harvard.edu/abs/1985ApJ...292..371D} {292, 371}

\bibitem[\protect\citeauthoryear{{Davison}, {Norris}, {Pfeffer}, {Davies}  \&
  {Crain}}{{Davison} et~al.}{2020}]{davison_2020}
{Davison} T.~A.,  {Norris} M.~A.,  {Pfeffer} J.~L.,  {Davies} J.~J.,   {Crain}
  R.~A.,  2020, \mn@doi [\mnras] {10.1093/mnras/staa1816}, \href
  {https://ui.adsabs.harvard.edu/abs/2020MNRAS.497...81D} {497, 81}

\bibitem[\protect\citeauthoryear{{De Lucia}, {Springel}, {White}, {Croton}  \&
  {Kauffmann}}{{De Lucia} et~al.}{2006}]{de_lucia_2006}
{De Lucia} G.,  {Springel} V.,  {White} S. D.~M.,  {Croton} D.,   {Kauffmann}
  G.,  2006, \mn@doi [\mnras] {10.1111/j.1365-2966.2005.09879.x}, \href
  {https://ui.adsabs.harvard.edu/abs/2006MNRAS.366..499D} {366, 499}

\bibitem[\protect\citeauthoryear{{Dekel} \& {Birnboim}}{{Dekel} \&
  {Birnboim}}{2006}]{dekel_2006}
{Dekel} A.,  {Birnboim} Y.,  2006, \mn@doi [\mnras]
  {10.1111/j.1365-2966.2006.10145.x}, \href
  {https://ui.adsabs.harvard.edu/abs/2006MNRAS.368....2D} {368, 2}

\bibitem[\protect\citeauthoryear{{Dekel} \& {Birnboim}}{{Dekel} \&
  {Birnboim}}{2008}]{dekel_2008}
{Dekel} A.,  {Birnboim} Y.,  2008, \mn@doi [\mnras]
  {10.1111/j.1365-2966.2007.12569.x}, \href
  {https://ui.adsabs.harvard.edu/abs/2008MNRAS.383..119D} {383, 119}

\bibitem[\protect\citeauthoryear{{Dimauro} et~al.,}{{Dimauro}
  et~al.}{2022}]{dimauro_2022}
{Dimauro} P.,  et~al., 2022, \mn@doi [\mnras] {10.1093/mnras/stac884}, \href
  {https://ui.adsabs.harvard.edu/abs/2022MNRAS.513..256D} {513, 256}

\bibitem[\protect\citeauthoryear{{Ding}, {Zhu}, {Pillepich}, {van de Ven},
  {Iodice}, {Corsini}  \& {Pinna}}{{Ding} et~al.}{2024}]{ding_2024}
{Ding} Y.,  {Zhu} L.,  {Pillepich} A.,  {van de Ven} G.,  {Iodice} E.,
  {Corsini} E.~M.,   {Pinna} F.,  2024, \mn@doi [arXiv e-prints]
  {10.48550/arXiv.2404.01541}, \href
  {https://ui.adsabs.harvard.edu/abs/2024arXiv240401541D} {p. arXiv:2404.01541}

\bibitem[\protect\citeauthoryear{{Dolag}, {Murante}  \& {Borgani}}{{Dolag}
  et~al.}{2010}]{dolag_2010}
{Dolag} K.,  {Murante} G.,   {Borgani} S.,  2010, \mn@doi [\mnras]
  {10.1111/j.1365-2966.2010.16583.x}, \href
  {https://ui.adsabs.harvard.edu/abs/2010MNRAS.405.1544D} {405, 1544}

\bibitem[\protect\citeauthoryear{{Donnari} et~al.,}{{Donnari}
  et~al.}{2019}]{donnari_2019}
{Donnari} M.,  et~al., 2019, \mn@doi [\mnras] {10.1093/mnras/stz712}, \href
  {https://ui.adsabs.harvard.edu/abs/2019MNRAS.485.4817D} {485, 4817}

\bibitem[\protect\citeauthoryear{{Donnari}, {Pillepich}, {Nelson}, {Marinacci},
  {Vogelsberger}  \& {Hernquist}}{{Donnari} et~al.}{2021}]{donnari_2021}
{Donnari} M.,  {Pillepich} A.,  {Nelson} D.,  {Marinacci} F.,  {Vogelsberger}
  M.,   {Hernquist} L.,  2021, \mn@doi [\mnras] {10.1093/mnras/stab1950}, \href
  {https://ui.adsabs.harvard.edu/abs/2021MNRAS.506.4760D} {506, 4760}

\bibitem[\protect\citeauthoryear{Driver et~al.,}{Driver
  et~al.}{2011}]{driver_2011}
Driver S.~P.,  et~al., 2011, \mn@doi [MNRAS]
  {10.1111/j.1365-2966.2010.18188.x}, 413, 971

\bibitem[\protect\citeauthoryear{Driver et~al.,}{Driver
  et~al.}{2022}]{driver_2022}
Driver S.~P.,  et~al., 2022, \mn@doi [MNRAS] {10.1093/mnras/stac472}, 513, 439

\bibitem[\protect\citeauthoryear{{Eisert}, {Pillepich}, {Nelson}, {Klessen},
  {Huertas-Company}  \& {Rodriguez-Gomez}}{{Eisert} et~al.}{2023}]{eisert_2023}
{Eisert} L.,  {Pillepich} A.,  {Nelson} D.,  {Klessen} R.~S.,
  {Huertas-Company} M.,   {Rodriguez-Gomez} V.,  2023, \mn@doi [\mnras]
  {10.1093/mnras/stac3295}, \href
  {https://ui.adsabs.harvard.edu/abs/2023MNRAS.519.2199E} {519, 2199}

\bibitem[\protect\citeauthoryear{{Endsley}, {Stark}, {Whitler}, {Topping},
  {Chen}, {Plat}, {Chisholm}  \& {Charlot}}{{Endsley}
  et~al.}{2023}]{endsley_2023}
{Endsley} R.,  {Stark} D.~P.,  {Whitler} L.,  {Topping} M.~W.,  {Chen} Z.,
  {Plat} A.,  {Chisholm} J.,   {Charlot} S.,  2023, \mn@doi [\mnras]
  {10.1093/mnras/stad1919}, \href
  {https://ui.adsabs.harvard.edu/abs/2023MNRAS.524.2312E} {524, 2312}

\bibitem[\protect\citeauthoryear{{Engler}, {Pillepich}, {Joshi}, {Pasquali},
  {Nelson}  \& {Grebel}}{{Engler} et~al.}{2023}]{engler_2023}
{Engler} C.,  {Pillepich} A.,  {Joshi} G.~D.,  {Pasquali} A.,  {Nelson} D.,
  {Grebel} E.~K.,  2023, \mn@doi [\mnras] {10.1093/mnras/stad1357}, \href
  {https://ui.adsabs.harvard.edu/abs/2023MNRAS.tmp.1353E} {}

\bibitem[\protect\citeauthoryear{{Fillingham}, {Cooper}, {Pace},
  {Boylan-Kolchin}, {Bullock}, {Garrison-Kimmel}  \& {Wheeler}}{{Fillingham}
  et~al.}{2016}]{fillingham_2016}
{Fillingham} S.~P.,  {Cooper} M.~C.,  {Pace} A.~B.,  {Boylan-Kolchin} M.,
  {Bullock} J.~S.,  {Garrison-Kimmel} S.,   {Wheeler} C.,  2016, \mn@doi
  [\mnras] {10.1093/mnras/stw2131}, \href
  {https://ui.adsabs.harvard.edu/abs/2016MNRAS.463.1916F} {463, 1916}

\bibitem[\protect\citeauthoryear{{Fischer}, {Bernardi}  \& {Meert}}{{Fischer}
  et~al.}{2017}]{fischer_2017}
{Fischer} J.~L.,  {Bernardi} M.,   {Meert} A.,  2017, \mn@doi [\mnras]
  {10.1093/mnras/stx136}, \href
  {https://ui.adsabs.harvard.edu/abs/2017MNRAS.467..490F} {467, 490}

\bibitem[\protect\citeauthoryear{{Fontanot}, {Macci{\`o}}, {Hirschmann}, {De
  Lucia}, {Kannan}, {Somerville}  \& {Wilman}}{{Fontanot}
  et~al.}{2015}]{fontanot_2015}
{Fontanot} F.,  {Macci{\`o}} A.~V.,  {Hirschmann} M.,  {De Lucia} G.,  {Kannan}
  R.,  {Somerville} R.~S.,   {Wilman} D.,  2015, \mn@doi [\mnras]
  {10.1093/mnras/stv1119}, \href
  {https://ui.adsabs.harvard.edu/abs/2015MNRAS.451.2968F} {451, 2968}

\bibitem[\protect\citeauthoryear{{Franchetto} et~al.,}{{Franchetto}
  et~al.}{2021}]{franchetto_2021}
{Franchetto} A.,  et~al., 2021, \mn@doi [VizieR Online Data Catalog]
  {10.26093/cds/vizier.18950106}, \href
  {https://ui.adsabs.harvard.edu/abs/2021yCat..18950106F} {p. J/ApJ/895/106}

\bibitem[\protect\citeauthoryear{{Franco} et~al.,}{{Franco}
  et~al.}{2018}]{franco_2018}
{Franco} M.,  et~al., 2018, \mn@doi [\aap] {10.1051/0004-6361/201832928}, \href
  {https://ui.adsabs.harvard.edu/abs/2018A&A...620A.152F} {620, A152}

\bibitem[\protect\citeauthoryear{{Fu} et~al.,}{{Fu} et~al.}{2022}]{fu_2022}
{Fu} H.,  et~al., 2022, \mn@doi [\mnras] {10.1093/mnras/stac2205}, \href
  {https://ui.adsabs.harvard.edu/abs/2022MNRAS.516.3206F} {516, 3206}

\bibitem[\protect\citeauthoryear{{Furnell} et~al.,}{{Furnell}
  et~al.}{2021}]{furnell_2021}
{Furnell} K.~E.,  et~al., 2021, \mn@doi [\mnras] {10.1093/mnras/stab065}, \href
  {https://ui.adsabs.harvard.edu/abs/2021MNRAS.502.2419F} {502, 2419}

\bibitem[\protect\citeauthoryear{{Gardner} et~al.,}{{Gardner}
  et~al.}{2006}]{gardner_2006}
{Gardner} J.~P.,  et~al., 2006, \mn@doi [\ssr] {10.1007/s11214-006-8315-7},
  \href {https://ui.adsabs.harvard.edu/abs/2006SSRv..123..485G} {123, 485}

\bibitem[\protect\citeauthoryear{{Gonz{\'a}lez}, {Labb{\'e}}, {Bouwens},
  {Illingworth}, {Franx}  \& {Kriek}}{{Gonz{\'a}lez}
  et~al.}{2011}]{gonzalez_2011}
{Gonz{\'a}lez} V.,  {Labb{\'e}} I.,  {Bouwens} R.~J.,  {Illingworth} G.,
  {Franx} M.,   {Kriek} M.,  2011, \mn@doi [\apjl]
  {10.1088/2041-8205/735/2/L34}, \href
  {https://ui.adsabs.harvard.edu/abs/2011ApJ...735L..34G} {735, L34}

\bibitem[\protect\citeauthoryear{{Granato}, {De Zotti}, {Silva}, {Bressan}  \&
  {Danese}}{{Granato} et~al.}{2004}]{granato_2004}
{Granato} G.~L.,  {De Zotti} G.,  {Silva} L.,  {Bressan} A.,   {Danese} L.,
  2004, \mn@doi [\apj] {10.1086/379875}, \href
  {https://ui.adsabs.harvard.edu/abs/2004ApJ...600..580G} {600, 580}

\bibitem[\protect\citeauthoryear{{Gregg} \& {West}}{{Gregg} \&
  {West}}{1998}]{gregg_1998}
{Gregg} M.~D.,  {West} M.~J.,  1998, \mn@doi [\nat] {10.1038/25078}, \href
  {https://ui.adsabs.harvard.edu/abs/1998Natur.396..549G} {396, 549}

\bibitem[\protect\citeauthoryear{{Groenewald}, {Skelton}, {Gilbank}  \&
  {Loubser}}{{Groenewald} et~al.}{2017}]{groenewald_2017}
{Groenewald} D.~N.,  {Skelton} R.~E.,  {Gilbank} D.~G.,   {Loubser} S.~I.,
  2017, \mn@doi [\mnras] {10.1093/mnras/stx340}, \href
  {https://ui.adsabs.harvard.edu/abs/2017MNRAS.467.4101G} {467, 4101}

\bibitem[\protect\citeauthoryear{{Gruppioni} et~al.,}{{Gruppioni}
  et~al.}{2020}]{gruppioni_2020}
{Gruppioni} C.,  et~al., 2020, \mn@doi [\aap] {10.1051/0004-6361/202038487},
  \href {https://ui.adsabs.harvard.edu/abs/2020A&A...643A...8G} {643, A8}

\bibitem[\protect\citeauthoryear{{Grylls}, {Shankar}, {Zanisi}  \&
  {Bernardi}}{{Grylls} et~al.}{2019}]{grylls_paper1}
{Grylls} P.~J.,  {Shankar} F.,  {Zanisi} L.,   {Bernardi} M.,  2019, \mn@doi
  [\mnras] {10.1093/mnras/sty3281}, \href
  {https://ui.adsabs.harvard.edu/abs/2019MNRAS.483.2506G} {483, 2506}

\bibitem[\protect\citeauthoryear{{Grylls}, {Shankar}, {Leja}, {Menci},
  {Moster}, {Behroozi}  \& {Zanisi}}{{Grylls} et~al.}{2020}]{grylls_paper2}
{Grylls} P.~J.,  {Shankar} F.,  {Leja} J.,  {Menci} N.,  {Moster} B.,
  {Behroozi} P.,   {Zanisi} L.,  2020, \mn@doi [\mnras]
  {10.1093/mnras/stz2956}, \href
  {https://ui.adsabs.harvard.edu/abs/2020MNRAS.491..634G} {491, 634}

\bibitem[\protect\citeauthoryear{{Gunn} \& {Gott}}{{Gunn} \&
  {Gott}}{1972}]{gunn_1972}
{Gunn} J.~E.,  {Gott} J.~Richard I.,  1972, \mn@doi [\apj] {10.1086/151605},
  \href {https://ui.adsabs.harvard.edu/abs/1972ApJ...176....1G} {176, 1}

\bibitem[\protect\citeauthoryear{{Guo} \& {White}}{{Guo} \&
  {White}}{2008}]{guo_2008}
{Guo} Q.,  {White} S.~D.~M.,  2008, \mn@doi [\mnras]
  {10.1111/j.1365-2966.2007.12619.x}, \href
  {https://ui.adsabs.harvard.edu/abs/2008MNRAS.384....2G} {384, 2}

\bibitem[\protect\citeauthoryear{{Guo} et~al.,}{{Guo} et~al.}{2011}]{guo_2011}
{Guo} Q.,  et~al., 2011, \mn@doi [\mnras] {10.1111/j.1365-2966.2010.18114.x},
  \href {https://ui.adsabs.harvard.edu/abs/2011MNRAS.413..101G} {413, 101}

\bibitem[\protect\citeauthoryear{{Hashemizadeh} et~al.,}{{Hashemizadeh}
  et~al.}{2021}]{hashemizadeh_2021}
{Hashemizadeh} A.,  et~al., 2021, \mn@doi [\mnras] {10.1093/mnras/stab600},
  \href {https://ui.adsabs.harvard.edu/abs/2021MNRAS.505..136H} {505, 136}

\bibitem[\protect\citeauthoryear{{Hatton}, {Devriendt}, {Ninin}, {Bouchet},
  {Guiderdoni}  \& {Vibert}}{{Hatton} et~al.}{2003}]{hatton_2003}
{Hatton} S.,  {Devriendt} J. E.~G.,  {Ninin} S.,  {Bouchet} F.~R.,
  {Guiderdoni} B.,   {Vibert} D.,  2003, \mn@doi [\mnras]
  {10.1046/j.1365-8711.2003.05589.x}, \href
  {https://ui.adsabs.harvard.edu/abs/2003MNRAS.343...75H} {343, 75}

\bibitem[\protect\citeauthoryear{{Henriques}, {White}, {Thomas}, {Angulo},
  {Guo}, {Lemson}, {Springel}  \& {Overzier}}{{Henriques}
  et~al.}{2015}]{henriques_2015}
{Henriques} B. M.~B.,  {White} S. D.~M.,  {Thomas} P.~A.,  {Angulo} R.,  {Guo}
  Q.,  {Lemson} G.,  {Springel} V.,   {Overzier} R.,  2015, \mn@doi [\mnras]
  {10.1093/mnras/stv705}, \href
  {https://ui.adsabs.harvard.edu/abs/2015MNRAS.451.2663H} {451, 2663}

\bibitem[\protect\citeauthoryear{{Hopkins}, {Hernquist}, {Cox}  \&
  {Kere{\v{s}}}}{{Hopkins} et~al.}{2008}]{hopkins_2008}
{Hopkins} P.~F.,  {Hernquist} L.,  {Cox} T.~J.,   {Kere{\v{s}}} D.,  2008,
  \mn@doi [\apjs] {10.1086/524362}, \href
  {https://ui.adsabs.harvard.edu/abs/2008ApJS..175..356H} {175, 356}

\bibitem[\protect\citeauthoryear{{Hopkins} et~al.,}{{Hopkins}
  et~al.}{2010a}]{hopkins_2010a}
{Hopkins} P.~F.,  et~al., 2010a, \mn@doi [\apj] {10.1088/0004-637X/715/1/202},
  \href {https://ui.adsabs.harvard.edu/abs/2010ApJ...715..202H} {715, 202}

\bibitem[\protect\citeauthoryear{{Hopkins} et~al.,}{{Hopkins}
  et~al.}{2010b}]{hopkins_2010b}
{Hopkins} P.~F.,  et~al., 2010b, \mn@doi [\apj] {10.1088/0004-637X/724/2/915},
  \href {https://ui.adsabs.harvard.edu/abs/2010ApJ...724..915H} {724, 915}

\bibitem[\protect\citeauthoryear{{Hu{\v{s}}ko}, {Lacey}  \&
  {Baugh}}{{Hu{\v{s}}ko} et~al.}{2022}]{husko_2022}
{Hu{\v{s}}ko} F.,  {Lacey} C.~G.,   {Baugh} C.~M.,  2022, \mn@doi [\mnras]
  {10.1093/mnras/stab3324}, \href
  {https://ui.adsabs.harvard.edu/abs/2022MNRAS.509.5918H} {509, 5918}

\bibitem[\protect\citeauthoryear{{Jiang}, {Jing}, {Faltenbacher}, {Lin}  \&
  {Li}}{{Jiang} et~al.}{2008}]{jiang_2008}
{Jiang} C.~Y.,  {Jing} Y.~P.,  {Faltenbacher} A.,  {Lin} W.~P.,   {Li} C.,
  2008, \mn@doi [\apj] {10.1086/526412}, \href
  {https://ui.adsabs.harvard.edu/abs/2008ApJ...675.1095J} {675, 1095}

\bibitem[\protect\citeauthoryear{{Jiang}, {Dekel}, {Freundlich}, {van den
  Bosch}, {Green}, {Hopkins}, {Benson}  \& {Du}}{{Jiang}
  et~al.}{2021}]{jiang_2021}
{Jiang} F.,  {Dekel} A.,  {Freundlich} J.,  {van den Bosch} F.~C.,  {Green}
  S.~B.,  {Hopkins} P.~F.,  {Benson} A.,   {Du} X.,  2021, \mn@doi [\mnras]
  {10.1093/mnras/staa4034}, \href
  {https://ui.adsabs.harvard.edu/abs/2021MNRAS.502..621J} {502, 621}

\bibitem[\protect\citeauthoryear{{Kannan}, {Macci{\`o}}, {Fontanot}, {Moster},
  {Karman}  \& {Somerville}}{{Kannan} et~al.}{2015}]{kannan_2015}
{Kannan} R.,  {Macci{\`o}} A.~V.,  {Fontanot} F.,  {Moster} B.~P.,  {Karman}
  W.,   {Somerville} R.~S.,  2015, \mn@doi [\mnras] {10.1093/mnras/stv1633},
  \href {https://ui.adsabs.harvard.edu/abs/2015MNRAS.452.4347K} {452, 4347}

\bibitem[\protect\citeauthoryear{{Kawinwanichakij} et~al.,}{{Kawinwanichakij}
  et~al.}{2020}]{Kawinwanichakij_2020}
{Kawinwanichakij} L.,  et~al., 2020, \mn@doi [\apj] {10.3847/1538-4357/ab75c4},
  \href {https://ui.adsabs.harvard.edu/abs/2020ApJ...892....7K} {892, 7}

\bibitem[\protect\citeauthoryear{{Kluge}, {Bender}, {Riffeser}, {Goessl},
  {Hopp}, {Schmidt}  \& {Ries}}{{Kluge} et~al.}{2021}]{kluge_2021}
{Kluge} M.,  {Bender} R.,  {Riffeser} A.,  {Goessl} C.,  {Hopp} U.,  {Schmidt}
  M.,   {Ries} C.,  2021, \mn@doi [\apjs] {10.3847/1538-4365/abcda6}, \href
  {https://ui.adsabs.harvard.edu/abs/2021ApJS..252...27K} {252, 27}

\bibitem[\protect\citeauthoryear{{Koprowski}, {Wijesekera}, {Dunlop}, {McLeod},
  {Micha{\l}owski}, {Lisiecki}  \& {McLure}}{{Koprowski}
  et~al.}{2024}]{koprowski_2024}
{Koprowski} M.~P.,  {Wijesekera} J.~V.,  {Dunlop} J.~S.,  {McLeod} D.~J.,
  {Micha{\l}owski} M.~J.,  {Lisiecki} K.,   {McLure} R.~J.,  2024, \mn@doi
  [arXiv e-prints] {10.48550/arXiv.2403.06575}, \href
  {https://ui.adsabs.harvard.edu/abs/2024arXiv240306575K} {p. arXiv:2403.06575}

\bibitem[\protect\citeauthoryear{{Koutsouridou} \& {Cattaneo}}{{Koutsouridou}
  \& {Cattaneo}}{2019}]{koutsouridou_2019}
{Koutsouridou} I.,  {Cattaneo} A.,  2019, \mn@doi [\mnras]
  {10.1093/mnras/stz2916}, \href
  {https://ui.adsabs.harvard.edu/abs/2019MNRAS.490.5375K} {490, 5375}

\bibitem[\protect\citeauthoryear{{Koutsouridou} \& {Cattaneo}}{{Koutsouridou}
  \& {Cattaneo}}{2022}]{koutsouridou_2022}
{Koutsouridou} I.,  {Cattaneo} A.,  2022, \mn@doi [\mnras]
  {10.1093/mnras/stac2240}, \href
  {https://ui.adsabs.harvard.edu/abs/2022MNRAS.516.4194K} {516, 4194}

\bibitem[\protect\citeauthoryear{{Kroupa}}{{Kroupa}}{2001}]{kroupa_2001}
{Kroupa} P.,  2001, \mn@doi [\mnras] {10.1046/j.1365-8711.2001.04022.x}, \href
  {https://ui.adsabs.harvard.edu/abs/2001MNRAS.322..231K} {322, 231}

\bibitem[\protect\citeauthoryear{{Kroupa} \& {Boily}}{{Kroupa} \&
  {Boily}}{2002}]{Kroupa2002}
{Kroupa} P.,  {Boily} C.~M.,  2002, \mn@doi [Monthly Notices of the Royal
  Astronomical Society] {10.1046/j.1365-8711.2002.05848.x}, \href
  {https://ui.adsabs.harvard.edu/abs/2002MNRAS.336.1188K} {336, 1188}

\bibitem[\protect\citeauthoryear{{Labbe} et~al.,}{{Labbe}
  et~al.}{2023}]{labbe_2023}
{Labbe} I.,  et~al., 2023, \mn@doi [arXiv e-prints]
  {10.48550/arXiv.2306.07320}, \href
  {https://ui.adsabs.harvard.edu/abs/2023arXiv230607320L} {p. arXiv:2306.07320}

\bibitem[\protect\citeauthoryear{{Lackner}, {Cen}, {Ostriker}  \&
  {Joung}}{{Lackner} et~al.}{2012}]{lackner_2012}
{Lackner} C.~N.,  {Cen} R.,  {Ostriker} J.~P.,   {Joung} M.~R.,  2012, \mn@doi
  [\mnras] {10.1111/j.1365-2966.2012.21525.x}, \href
  {https://ui.adsabs.harvard.edu/abs/2012MNRAS.425..641L} {425, 641}

\bibitem[\protect\citeauthoryear{{Lang} et~al.,}{{Lang}
  et~al.}{2014}]{lang_2014}
{Lang} P.,  et~al., 2014, \mn@doi [\apj] {10.1088/0004-637X/788/1/11}, \href
  {https://ui.adsabs.harvard.edu/abs/2014ApJ...788...11L} {788, 11}

\bibitem[\protect\citeauthoryear{{Lapi}, {Mancuso}, {Bressan}  \&
  {Danese}}{{Lapi} et~al.}{2017}]{lapi_2017}
{Lapi} A.,  {Mancuso} C.,  {Bressan} A.,   {Danese} L.,  2017, \mn@doi [\apj]
  {10.3847/1538-4357/aa88c9}, \href
  {https://ui.adsabs.harvard.edu/abs/2017ApJ...847...13L} {847, 13}

\bibitem[\protect\citeauthoryear{{Lapi} et~al.,}{{Lapi}
  et~al.}{2018}]{lapi_2018}
{Lapi} A.,  et~al., 2018, \mn@doi [\apj] {10.3847/1538-4357/aab6af}, \href
  {https://ui.adsabs.harvard.edu/abs/2018ApJ...857...22L} {857, 22}

\bibitem[\protect\citeauthoryear{{Larson}, {Tinsley}  \& {Caldwell}}{{Larson}
  et~al.}{1980}]{larson_1980}
{Larson} R.~B.,  {Tinsley} B.~M.,   {Caldwell} C.~N.,  1980, \mn@doi [\apj]
  {10.1086/157917}, \href
  {https://ui.adsabs.harvard.edu/abs/1980ApJ...237..692L} {237, 692}

\bibitem[\protect\citeauthoryear{{Lee} et~al.,}{{Lee} et~al.}{2015}]{lee_2015}
{Lee} N.,  et~al., 2015, \mn@doi [\apj] {10.1088/0004-637X/801/2/80}, \href
  {https://ui.adsabs.harvard.edu/abs/2015ApJ...801...80L} {801, 80}

\bibitem[\protect\citeauthoryear{{Leitner}}{{Leitner}}{2012}]{leitner_2012}
{Leitner} S.~N.,  2012, \mn@doi [\apj] {10.1088/0004-637X/745/2/149}, \href
  {https://ui.adsabs.harvard.edu/abs/2012ApJ...745..149L} {745, 149}

\bibitem[\protect\citeauthoryear{{Leitner} \& {Kravtsov}}{{Leitner} \&
  {Kravtsov}}{2011}]{leitner_2011}
{Leitner} S.~N.,  {Kravtsov} A.~V.,  2011, \mn@doi [\apj]
  {10.1088/0004-637X/734/1/48}, \href
  {https://ui.adsabs.harvard.edu/abs/2011ApJ...734...48L} {734, 48}

\bibitem[\protect\citeauthoryear{{Leja}, {van Dokkum}, {Franx}  \&
  {Whitaker}}{{Leja} et~al.}{2015}]{leja_2015}
{Leja} J.,  {van Dokkum} P.~G.,  {Franx} M.,   {Whitaker} K.~E.,  2015, \mn@doi
  [\apj] {10.1088/0004-637X/798/2/115}, \href
  {https://ui.adsabs.harvard.edu/abs/2015ApJ...798..115L} {798, 115}

\bibitem[\protect\citeauthoryear{Leja, Carnall, Johnson, Conroy  \&
  Speagle}{Leja et~al.}{2019a}]{Leja2019a}
Leja J.,  Carnall A.~C.,  Johnson B.~D.,  Conroy C.,   Speagle J.~S.,  2019a,
  \mn@doi [The Astrophysical Journal] {10.3847/1538-4357/ab133c}, 876, 3

\bibitem[\protect\citeauthoryear{{Leja} et~al.,}{{Leja}
  et~al.}{2019b}]{leja_2019}
{Leja} J.,  et~al., 2019b, \mn@doi [\apj] {10.3847/1538-4357/ab1d5a}, \href
  {https://ui.adsabs.harvard.edu/abs/2019ApJ...877..140L} {877, 140}

\bibitem[\protect\citeauthoryear{{Leja}, {Speagle}, {Johnson}, {Conroy}, {van
  Dokkum}  \& {Franx}}{{Leja} et~al.}{2020}]{leja_2020}
{Leja} J.,  {Speagle} J.~S.,  {Johnson} B.~D.,  {Conroy} C.,  {van Dokkum} P.,
   {Franx} M.,  2020, \mn@doi [\apj] {10.3847/1538-4357/ab7e27}, \href
  {https://ui.adsabs.harvard.edu/abs/2020ApJ...893..111L} {893, 111}

\bibitem[\protect\citeauthoryear{{Leja} et~al.,}{{Leja}
  et~al.}{2022}]{leja_2022}
{Leja} J.,  et~al., 2022, \mn@doi [\apj] {10.3847/1538-4357/ac887d}, \href
  {https://ui.adsabs.harvard.edu/abs/2022ApJ...936..165L} {936, 165}

\bibitem[\protect\citeauthoryear{{Lilly}, {Carollo}, {Pipino}, {Renzini}  \&
  {Peng}}{{Lilly} et~al.}{2013}]{lilly_2013}
{Lilly} S.~J.,  {Carollo} C.~M.,  {Pipino} A.,  {Renzini} A.,   {Peng} Y.,
  2013, \mn@doi [\apj] {10.1088/0004-637X/772/2/119}, \href
  {https://ui.adsabs.harvard.edu/abs/2013ApJ...772..119L} {772, 119}

\bibitem[\protect\citeauthoryear{{Lin} \& {Mohr}}{{Lin} \&
  {Mohr}}{2004}]{lin_2004}
{Lin} Y.-T.,  {Mohr} J.~J.,  2004, \mn@doi [\apj] {10.1086/425412}, \href
  {https://ui.adsabs.harvard.edu/abs/2004ApJ...617..879L} {617, 879}

\bibitem[\protect\citeauthoryear{Liske et~al.,}{Liske
  et~al.}{2015}]{liske_2015}
Liske J.,  et~al., 2015, \mn@doi [MNRAS] {10.1093/mnras/stv1436}, 452, 2087

\bibitem[\protect\citeauthoryear{{Lyu} et~al.,}{{Lyu} et~al.}{2023}]{lyu_2023}
{Lyu} C.,  et~al., 2023, \mn@doi [\apj] {10.3847/1538-4357/ad036b}, \href
  {https://ui.adsabs.harvard.edu/abs/2023ApJ...959....5L} {959, 5}

\bibitem[\protect\citeauthoryear{{Madau} \& {Dickinson}}{{Madau} \&
  {Dickinson}}{2014}]{madau_2014}
{Madau} P.,  {Dickinson} M.,  2014, \mn@doi [\araa]
  {10.1146/annurev-astro-081811-125615}, \href
  {https://ui.adsabs.harvard.edu/abs/2014ARA&A..52..415M} {52, 415}

\bibitem[\protect\citeauthoryear{{Marinacci} et~al.,}{{Marinacci}
  et~al.}{2018}]{marinacci2018first}
{Marinacci} F.,  et~al., 2018, \mn@doi [\mnras] {10.1093/mnras/sty2206}, \href
  {https://ui.adsabs.harvard.edu/abs/2018MNRAS.480.5113M} {480, 5113}

\bibitem[\protect\citeauthoryear{{Martig}, {Bournaud}, {Teyssier}  \&
  {Dekel}}{{Martig} et~al.}{2009}]{martig_2009}
{Martig} M.,  {Bournaud} F.,  {Teyssier} R.,   {Dekel} A.,  2009, \mn@doi
  [\apj] {10.1088/0004-637X/707/1/250}, \href
  {https://ui.adsabs.harvard.edu/abs/2009ApJ...707..250M} {707, 250}

\bibitem[\protect\citeauthoryear{{Matharu} et~al.,}{{Matharu}
  et~al.}{2019}]{matharu_2019}
{Matharu} J.,  et~al., 2019, \mn@doi [\mnras] {10.1093/mnras/sty3465}, \href
  {https://ui.adsabs.harvard.edu/abs/2019MNRAS.484..595M} {484, 595}

\bibitem[\protect\citeauthoryear{{Meert}, {Vikram}  \& {Bernardi}}{{Meert}
  et~al.}{2015}]{meert_2015}
{Meert} A.,  {Vikram} V.,   {Bernardi} M.,  2015, \mn@doi [\mnras]
  {10.1093/mnras/stu2333}, \href
  {https://ui.adsabs.harvard.edu/abs/2015MNRAS.446.3943M} {446, 3943}

\bibitem[\protect\citeauthoryear{{Meert}, {Vikram}  \& {Bernardi}}{{Meert}
  et~al.}{2016}]{meert_2016}
{Meert} A.,  {Vikram} V.,   {Bernardi} M.,  2016, \mn@doi [\mnras]
  {10.1093/mnras/stv2475}, \href
  {https://ui.adsabs.harvard.edu/abs/2016MNRAS.455.2440M} {455, 2440}

\bibitem[\protect\citeauthoryear{{Mendel}, {Simard}, {Palmer}, {Ellison}  \&
  {Patton}}{{Mendel} et~al.}{2014}]{mendel_2014}
{Mendel} J.~T.,  {Simard} L.,  {Palmer} M.,  {Ellison} S.~L.,   {Patton} D.~R.,
   2014, \mn@doi [\apjs] {10.1088/0067-0049/210/1/3}, \href
  {https://ui.adsabs.harvard.edu/abs/2014ApJS..210....3M} {210, 3}

\bibitem[\protect\citeauthoryear{{Mihos}, {Harding}, {Feldmeier}  \&
  {Morrison}}{{Mihos} et~al.}{2005}]{mihos_2005}
{Mihos} J.~C.,  {Harding} P.,  {Feldmeier} J.,   {Morrison} H.,  2005, \mn@doi
  [\apjl] {10.1086/497030}, \href
  {https://ui.adsabs.harvard.edu/abs/2005ApJ...631L..41M} {631, L41}

\bibitem[\protect\citeauthoryear{{Monachesi} et~al.,}{{Monachesi}
  et~al.}{2019}]{monachesi_2019}
{Monachesi} A.,  et~al., 2019, \mn@doi [\mnras] {10.1093/mnras/stz538}, \href
  {https://ui.adsabs.harvard.edu/abs/2019MNRAS.485.2589M} {485, 2589}

\bibitem[\protect\citeauthoryear{{Montes}}{{Montes}}{2022}]{montes_2022}
{Montes} M.,  2022, \mn@doi [Nature Astronomy] {10.1038/s41550-022-01616-z},
  \href {https://ui.adsabs.harvard.edu/abs/2022NatAs...6..308M} {6, 308}

\bibitem[\protect\citeauthoryear{{Montes} \& {Trujillo}}{{Montes} \&
  {Trujillo}}{2018}]{montes_2018}
{Montes} M.,  {Trujillo} I.,  2018, \mn@doi [\mnras] {10.1093/mnras/stx2847},
  \href {https://ui.adsabs.harvard.edu/abs/2018MNRAS.474..917M} {474, 917}

\bibitem[\protect\citeauthoryear{{Montes} \& {Trujillo}}{{Montes} \&
  {Trujillo}}{2019}]{montes_2019}
{Montes} M.,  {Trujillo} I.,  2019, \mn@doi [\mnras] {10.1093/mnras/sty2858},
  \href {https://ui.adsabs.harvard.edu/abs/2019MNRAS.482.2838M} {482, 2838}

\bibitem[\protect\citeauthoryear{{Moster}, {Somerville}, {Maulbetsch}, {van den
  Bosch}, {Macci{\`o}}, {Naab}  \& {Oser}}{{Moster} et~al.}{2010}]{moster_2010}
{Moster} B.~P.,  {Somerville} R.~S.,  {Maulbetsch} C.,  {van den Bosch} F.~C.,
  {Macci{\`o}} A.~V.,  {Naab} T.,   {Oser} L.,  2010, \mn@doi [\apj]
  {10.1088/0004-637X/710/2/903}, \href
  {https://ui.adsabs.harvard.edu/abs/2010ApJ...710..903M} {710, 903}

\bibitem[\protect\citeauthoryear{{Moster}, {Naab}  \& {White}}{{Moster}
  et~al.}{2013}]{moster_2013}
{Moster} B.~P.,  {Naab} T.,   {White} S. D.~M.,  2013, \mn@doi [\mnras]
  {10.1093/mnras/sts261}, \href
  {https://ui.adsabs.harvard.edu/abs/2013MNRAS.428.3121M} {428, 3121}

\bibitem[\protect\citeauthoryear{{Moster}, {Naab}  \& {White}}{{Moster}
  et~al.}{2018}]{moster_2018}
{Moster} B.~P.,  {Naab} T.,   {White} S. D.~M.,  2018, \mn@doi [\mnras]
  {10.1093/mnras/sty655}, \href
  {https://ui.adsabs.harvard.edu/abs/2018MNRAS.477.1822M} {477, 1822}

\bibitem[\protect\citeauthoryear{{Moustakas} et~al.,}{{Moustakas}
  et~al.}{2013}]{moustakas_2013}
{Moustakas} J.,  et~al., 2013, \mn@doi [\apj] {10.1088/0004-637X/767/1/50},
  \href {https://ui.adsabs.harvard.edu/abs/2013ApJ...767...50M} {767, 50}

\bibitem[\protect\citeauthoryear{{Moutard} et~al.,}{{Moutard}
  et~al.}{2016}]{moutard_2016}
{Moutard} T.,  et~al., 2016, \mn@doi [\aap] {10.1051/0004-6361/201527294},
  \href {https://ui.adsabs.harvard.edu/abs/2016A&A...590A.103M} {590, A103}

\bibitem[\protect\citeauthoryear{{Murante} et~al.,}{{Murante}
  et~al.}{2004}]{murante_2004}
{Murante} G.,  et~al., 2004, \mn@doi [\apjl] {10.1086/421348}, \href
  {https://ui.adsabs.harvard.edu/abs/2004ApJ...607L..83M} {607, L83}

\bibitem[\protect\citeauthoryear{{Naiman} et~al.,}{{Naiman}
  et~al.}{2018}]{naiman2018first}
{Naiman} J.~P.,  et~al., 2018, \mn@doi [\mnras] {10.1093/mnras/sty618}, \href
  {https://ui.adsabs.harvard.edu/abs/2018MNRAS.477.1206N} {477, 1206}

\bibitem[\protect\citeauthoryear{{Neistein} \& {Weinmann}}{{Neistein} \&
  {Weinmann}}{2010}]{neistein_2010}
{Neistein} E.,  {Weinmann} S.~M.,  2010, \mn@doi [\mnras]
  {10.1111/j.1365-2966.2010.16656.x}, \href
  {https://ui.adsabs.harvard.edu/abs/2010MNRAS.405.2717N} {405, 2717}

\bibitem[\protect\citeauthoryear{{Nelson} et~al.,}{{Nelson}
  et~al.}{2018}]{nelson18a}
{Nelson} D.,  et~al., 2018, \mn@doi [\mnras] {10.1093/mnras/stx3040}, \href
  {http://adsabs.harvard.edu/abs/2018MNRAS.475..624N} {475, 624}

\bibitem[\protect\citeauthoryear{{Nelson} et~al.,}{{Nelson}
  et~al.}{2019}]{nelson_2019}
{Nelson} D.,  et~al., 2019, \mn@doi [Computational Astrophysics and Cosmology]
  {10.1186/s40668-019-0028-x}, \href
  {https://ui.adsabs.harvard.edu/abs/2019ComAC...6....2N} {6, 2}

\bibitem[\protect\citeauthoryear{{Nelson} et~al.,}{{Nelson}
  et~al.}{2023}]{nelson_2023}
{Nelson} E.~J.,  et~al., 2023, \mn@doi [\apjl] {10.3847/2041-8213/acc1e1},
  \href {https://ui.adsabs.harvard.edu/abs/2023ApJ...948L..18N} {948, L18}

\bibitem[\protect\citeauthoryear{{O'Leary}, {Moster}, {Naab}  \&
  {Somerville}}{{O'Leary} et~al.}{2021}]{oleary_2021}
{O'Leary} J.~A.,  {Moster} B.~P.,  {Naab} T.,   {Somerville} R.~S.,  2021,
  \mn@doi [\mnras] {10.1093/mnras/staa3746}, \href
  {https://ui.adsabs.harvard.edu/abs/2021MNRAS.501.3215O} {501, 3215}

\bibitem[\protect\citeauthoryear{{Oser}, {Ostriker}, {Naab}, {Johansson}  \&
  {Burkert}}{{Oser} et~al.}{2010}]{oser_2010}
{Oser} L.,  {Ostriker} J.~P.,  {Naab} T.,  {Johansson} P.~H.,   {Burkert} A.,
  2010, \mn@doi [\apj] {10.1088/0004-637X/725/2/2312}, \href
  {https://ui.adsabs.harvard.edu/abs/2010ApJ...725.2312O} {725, 2312}

\bibitem[\protect\citeauthoryear{{Pakmor} \& {Springel}}{{Pakmor} \&
  {Springel}}{2013}]{pakmor2013simulations}
{Pakmor} R.,  {Springel} V.,  2013, \mn@doi [\mnras] {10.1093/mnras/stt428},
  \href {https://ui.adsabs.harvard.edu/abs/2013MNRAS.432..176P} {432, 176}

\bibitem[\protect\citeauthoryear{{Pakmor}, {Bauer}  \& {Springel}}{{Pakmor}
  et~al.}{2011}]{pakmor2011magnetohydrodynamics}
{Pakmor} R.,  {Bauer} A.,   {Springel} V.,  2011, \mn@doi [\mnras]
  {10.1111/j.1365-2966.2011.19591.x}, \href
  {https://ui.adsabs.harvard.edu/abs/2011MNRAS.418.1392P} {418, 1392}

\bibitem[\protect\citeauthoryear{{Pearson} et~al.,}{{Pearson}
  et~al.}{2018}]{pearson_2018}
{Pearson} W.~J.,  et~al., 2018, \mn@doi [\aap] {10.1051/0004-6361/201832821},
  \href {https://ui.adsabs.harvard.edu/abs/2018A&A...615A.146P} {615, A146}

\bibitem[\protect\citeauthoryear{{Pietrinferni}, {Cassisi}, {Salaris}  \&
  {Castelli}}{{Pietrinferni} et~al.}{2004}]{pietrinferni_2004}
{Pietrinferni} A.,  {Cassisi} S.,  {Salaris} M.,   {Castelli} F.,  2004,
  \mn@doi [\apj] {10.1086/422498}, \href
  {https://ui.adsabs.harvard.edu/abs/2004ApJ...612..168P} {612, 168}

\bibitem[\protect\citeauthoryear{{Pietrinferni}, {Cassisi}, {Salaris}  \&
  {Castelli}}{{Pietrinferni} et~al.}{2006}]{pietrinferni_2006}
{Pietrinferni} A.,  {Cassisi} S.,  {Salaris} M.,   {Castelli} F.,  2006,
  \mn@doi [\apj] {10.1086/501344}, \href
  {https://ui.adsabs.harvard.edu/abs/2006ApJ...642..797P} {642, 797}

\bibitem[\protect\citeauthoryear{{Pillepich} et~al.,}{{Pillepich}
  et~al.}{2018a}]{pillepich2018Simulating}
{Pillepich} A.,  et~al., 2018a, \mn@doi [\mnras] {10.1093/mnras/stx2656}, \href
  {https://ui.adsabs.harvard.edu/abs/2018MNRAS.473.4077P} {473, 4077}

\bibitem[\protect\citeauthoryear{{Pillepich} et~al.,}{{Pillepich}
  et~al.}{2018b}]{pillepich_2018}
{Pillepich} A.,  et~al., 2018b, \mn@doi [\mnras] {10.1093/mnras/stx3112}, \href
  {https://ui.adsabs.harvard.edu/abs/2018MNRAS.475..648P} {475, 648}

\bibitem[\protect\citeauthoryear{{Planck Collaboration} et~al.,}{{Planck
  Collaboration} et~al.}{2018}]{planck_2018_cosmo_params}
{Planck Collaboration} et~al., 2018, arXiv e-prints, \href
  {https://ui.adsabs.harvard.edu/abs/2018arXiv180706209P} {p. arXiv:1807.06209}

\bibitem[\protect\citeauthoryear{{Poggianti} et~al.,}{{Poggianti}
  et~al.}{2017}]{poggianti_2017}
{Poggianti} B.~M.,  et~al., 2017, \mn@doi [\apj] {10.3847/1538-4357/aa78ed},
  \href {https://ui.adsabs.harvard.edu/abs/2017ApJ...844...48P} {844, 48}

\bibitem[\protect\citeauthoryear{{Poliakov}, {Mosenkov}, {Brosch}, {Koriski}
  \& {Rich}}{{Poliakov} et~al.}{2021}]{poliakov_2021}
{Poliakov} D.,  {Mosenkov} A.~V.,  {Brosch} N.,  {Koriski} S.,   {Rich} R.~M.,
  2021, \mn@doi [\mnras] {10.1093/mnras/stab853}, \href
  {https://ui.adsabs.harvard.edu/abs/2021MNRAS.503.6059P} {503, 6059}

\bibitem[\protect\citeauthoryear{{Popesso} et~al.,}{{Popesso}
  et~al.}{2023}]{popesso_2023}
{Popesso} P.,  et~al., 2023, \mn@doi [\mnras] {10.1093/mnras/stac3214}, \href
  {https://ui.adsabs.harvard.edu/abs/2023MNRAS.519.1526P} {519, 1526}

\bibitem[\protect\citeauthoryear{{Puchwein}, {Springel}, {Sijacki}  \&
  {Dolag}}{{Puchwein} et~al.}{2010}]{puchwein_2010}
{Puchwein} E.,  {Springel} V.,  {Sijacki} D.,   {Dolag} K.,  2010, \mn@doi
  [\mnras] {10.1111/j.1365-2966.2010.16786.x}, \href
  {https://ui.adsabs.harvard.edu/abs/2010MNRAS.406..936P} {406, 936}

\bibitem[\protect\citeauthoryear{{Purcell}, {Bullock}  \& {Zentner}}{{Purcell}
  et~al.}{2007}]{purcell_2007}
{Purcell} C.~W.,  {Bullock} J.~S.,   {Zentner} A.~R.,  2007, \mn@doi [\apj]
  {10.1086/519787}, \href
  {https://ui.adsabs.harvard.edu/abs/2007ApJ...666...20P} {666, 20}

\bibitem[\protect\citeauthoryear{Robotham, Davies, Driver, Koushan, Taranu,
  Casura  \& Liske}{Robotham et~al.}{2018}]{robotham_2018}
Robotham A. S.~G.,  Davies L. J.~M.,  Driver S.~P.,  Koushan S.,  Taranu D.~S.,
   Casura S.,   Liske J.,  2018, \mn@doi [MNRAS] {10.1093/mnras/sty440}, 476,
  3137

\bibitem[\protect\citeauthoryear{Robotham, Bellstedt, Lagos, Thorne, Davies,
  Driver  \& Bravo}{Robotham et~al.}{2020}]{robotham_2020}
Robotham A. S.~G.,  Bellstedt S.,  Lagos C. d.~P.,  Thorne J.~E.,  Davies
  L.~J.,  Driver S.~P.,   Bravo M.,  2020, \mn@doi [MNRAS]
  {10.1093/mnras/staa1116}, 495, 905

\bibitem[\protect\citeauthoryear{{Rodighiero} et~al.,}{{Rodighiero}
  et~al.}{2014}]{rodighiero_2014}
{Rodighiero} G.,  et~al., 2014, \mn@doi [\mnras] {10.1093/mnras/stu1110}, \href
  {https://ui.adsabs.harvard.edu/abs/2014MNRAS.443...19R} {443, 19}

\bibitem[\protect\citeauthoryear{{Rodriguez-Gomez} et~al.,}{{Rodriguez-Gomez}
  et~al.}{2015}]{rodriguez_gomez_2015}
{Rodriguez-Gomez} V.,  et~al., 2015, \mn@doi [\mnras] {10.1093/mnras/stv264},
  \href {https://ui.adsabs.harvard.edu/abs/2015MNRAS.449...49R} {449, 49}

\bibitem[\protect\citeauthoryear{{Rodriguez-Gomez} et~al.,}{{Rodriguez-Gomez}
  et~al.}{2016}]{rodriguez_gomez_2016}
{Rodriguez-Gomez} V.,  et~al., 2016, \mn@doi [\mnras] {10.1093/mnras/stw456},
  \href {https://ui.adsabs.harvard.edu/abs/2016MNRAS.458.2371R} {458, 2371}

\bibitem[\protect\citeauthoryear{{Rodr{\'\i}guez-Puebla}, {Primack},
  {Avila-Reese}  \& {Faber}}{{Rodr{\'\i}guez-Puebla}
  et~al.}{2017}]{rodriguez_puebla_2017}
{Rodr{\'\i}guez-Puebla} A.,  {Primack} J.~R.,  {Avila-Reese} V.,   {Faber}
  S.~M.,  2017, \mn@doi [\mnras] {10.1093/mnras/stx1172}, \href
  {https://ui.adsabs.harvard.edu/abs/2017MNRAS.470..651R} {470, 651}

\bibitem[\protect\citeauthoryear{{Rohr}, {Pillepich}, {Nelson}, {Zinger},
  {Joshi}  \& {Ayromlou}}{{Rohr} et~al.}{2023}]{Rohr2023Jellyfish}
{Rohr} E.,  {Pillepich} A.,  {Nelson} D.,  {Zinger} E.,  {Joshi} G.,
  {Ayromlou} M.,  2023, \mn@doi [arXiv e-prints] {10.48550/arXiv.2304.09196},
  \href {https://ui.adsabs.harvard.edu/abs/2023arXiv230409196R} {p.
  arXiv:2304.09196}

\bibitem[\protect\citeauthoryear{{Rudick}, {Mihos}  \& {McBride}}{{Rudick}
  et~al.}{2011}]{rudick_2011}
{Rudick} C.~S.,  {Mihos} J.~C.,   {McBride} C.~K.,  2011, \mn@doi [\apj]
  {10.1088/0004-637X/732/1/48}, \href
  {https://ui.adsabs.harvard.edu/abs/2011ApJ...732...48R} {732, 48}

\bibitem[\protect\citeauthoryear{{S{\'a}nchez} et~al.,}{{S{\'a}nchez}
  et~al.}{2019}]{sanchez_2019}
{S{\'a}nchez} S.~F.,  et~al., 2019, \mn@doi [\mnras] {10.1093/mnras/sty2730},
  \href {https://ui.adsabs.harvard.edu/abs/2019MNRAS.482.1557S} {482, 1557}

\bibitem[\protect\citeauthoryear{{Schawinski} et~al.,}{{Schawinski}
  et~al.}{2014}]{schawinski_2014}
{Schawinski} K.,  et~al., 2014, \mn@doi [\mnras] {10.1093/mnras/stu327}, \href
  {https://ui.adsabs.harvard.edu/abs/2014MNRAS.440..889S} {440, 889}

\bibitem[\protect\citeauthoryear{{Shankar}, {Lapi}, {Salucci}, {De Zotti}  \&
  {Danese}}{{Shankar} et~al.}{2006}]{shankar_2006}
{Shankar} F.,  {Lapi} A.,  {Salucci} P.,  {De Zotti} G.,   {Danese} L.,  2006,
  \mn@doi [\apj] {10.1086/502794}, \href
  {https://ui.adsabs.harvard.edu/abs/2006ApJ...643...14S} {643, 14}

\bibitem[\protect\citeauthoryear{{Shankar} et~al.,}{{Shankar}
  et~al.}{2014}]{shankar_2014}
{Shankar} F.,  et~al., 2014, \mn@doi [\mnras] {10.1093/mnras/stt2470}, \href
  {https://ui.adsabs.harvard.edu/abs/2014MNRAS.439.3189S} {439, 3189}

\bibitem[\protect\citeauthoryear{{Shankar} et~al.,}{{Shankar}
  et~al.}{2015}]{shankar_2015}
{Shankar} F.,  et~al., 2015, \mn@doi [\apj] {10.1088/0004-637X/802/2/73}, \href
  {https://ui.adsabs.harvard.edu/abs/2015ApJ...802...73S} {802, 73}

\bibitem[\protect\citeauthoryear{{Shi} et~al.,}{{Shi} et~al.}{2020}]{shi_2020}
{Shi} J.,  et~al., 2020, \mn@doi [\apj] {10.3847/1538-4357/ab8464}, \href
  {https://ui.adsabs.harvard.edu/abs/2020ApJ...893..139S} {893, 139}

\bibitem[\protect\citeauthoryear{{Smith}, {Choi}, {Lee}, {Rhee},
  {Sanchez-Janssen}  \& {Yi}}{{Smith} et~al.}{2016}]{smith_2016}
{Smith} R.,  {Choi} H.,  {Lee} J.,  {Rhee} J.,  {Sanchez-Janssen} R.,   {Yi}
  S.~K.,  2016, \mn@doi [\apj] {10.3847/1538-4357/833/1/109}, \href
  {https://ui.adsabs.harvard.edu/abs/2016ApJ...833..109S} {833, 109}

\bibitem[\protect\citeauthoryear{{Speagle}, {Steinhardt}, {Capak}  \&
  {Silverman}}{{Speagle} et~al.}{2014}]{speagle_2014}
{Speagle} J.~S.,  {Steinhardt} C.~L.,  {Capak} P.~L.,   {Silverman} J.~D.,
  2014, \mn@doi [\apjs] {10.1088/0067-0049/214/2/15}, \href
  {https://ui.adsabs.harvard.edu/abs/2014ApJS..214...15S} {214, 15}

\bibitem[\protect\citeauthoryear{{Springel}}{{Springel}}{2010}]{springel2010pur}
{Springel} V.,  2010, \mn@doi [\mnras] {10.1111/j.1365-2966.2009.15715.x},
  \href {https://ui.adsabs.harvard.edu/abs/2010MNRAS.401..791S} {401, 791}

\bibitem[\protect\citeauthoryear{{Springel}, {White}, {Tormen}  \&
  {Kauffmann}}{{Springel} et~al.}{2001}]{springel2001populating}
{Springel} V.,  {White} S. D.~M.,  {Tormen} G.,   {Kauffmann} G.,  2001,
  \mn@doi [\mnras] {10.1046/j.1365-8711.2001.04912.x}, \href
  {https://ui.adsabs.harvard.edu/abs/2001MNRAS.328..726S} {328, 726}

\bibitem[\protect\citeauthoryear{{Springel} et~al.,}{{Springel}
  et~al.}{2018}]{springel2018first}
{Springel} V.,  et~al., 2018, \mn@doi [\mnras] {10.1093/mnras/stx3304}, \href
  {https://ui.adsabs.harvard.edu/abs/2018MNRAS.475..676S} {475, 676}

\bibitem[\protect\citeauthoryear{{Talia}, {Cimatti}, {Giulietti}, {Zamorani},
  {Bethermin}, {Faisst}, {Le F{\`e}vre}  \& {Smol{\c{c}}i{\'c}}}{{Talia}
  et~al.}{2021}]{talia_2021}
{Talia} M.,  {Cimatti} A.,  {Giulietti} M.,  {Zamorani} G.,  {Bethermin} M.,
  {Faisst} A.,  {Le F{\`e}vre} O.,   {Smol{\c{c}}i{\'c}} V.,  2021, \mn@doi
  [\apj] {10.3847/1538-4357/abd6e3}, \href
  {https://ui.adsabs.harvard.edu/abs/2021ApJ...909...23T} {909, 23}

\bibitem[\protect\citeauthoryear{{Thomas} et~al.,}{{Thomas}
  et~al.}{2019}]{thomas_2019}
{Thomas} R.,  et~al., 2019, \mn@doi [\aap] {10.1051/0004-6361/201935813}, \href
  {https://ui.adsabs.harvard.edu/abs/2019A&A...630A.145T} {630, A145}

\bibitem[\protect\citeauthoryear{{Thorne} et~al.,}{{Thorne}
  et~al.}{2021}]{thorne_2021}
{Thorne} J.~E.,  et~al., 2021, \mn@doi [\mnras] {10.1093/mnras/stab1294}, \href
  {https://ui.adsabs.harvard.edu/abs/2021MNRAS.505..540T} {505, 540}

\bibitem[\protect\citeauthoryear{{Tollet}, {Cattaneo}, {Mamon}, {Moutard}  \&
  {van den Bosch}}{{Tollet} et~al.}{2017}]{tollet_2017}
{Tollet} {\'E}.,  {Cattaneo} A.,  {Mamon} G.~A.,  {Moutard} T.,   {van den
  Bosch} F.~C.,  2017, \mn@doi [\mnras] {10.1093/mnras/stx1840}, \href
  {https://ui.adsabs.harvard.edu/abs/2017MNRAS.471.4170T} {471, 4170}

\bibitem[\protect\citeauthoryear{{Tomczak} et~al.,}{{Tomczak}
  et~al.}{2014}]{tomczak_2014}
{Tomczak} A.~R.,  et~al., 2014, \mn@doi [\apj] {10.1088/0004-637X/783/2/85},
  \href {https://ui.adsabs.harvard.edu/abs/2014ApJ...783...85T} {783, 85}

\bibitem[\protect\citeauthoryear{{Tomczak} et~al.,}{{Tomczak}
  et~al.}{2016}]{tomczak_2016}
{Tomczak} A.~R.,  et~al., 2016, \mn@doi [\apj] {10.3847/0004-637X/817/2/118},
  \href {https://ui.adsabs.harvard.edu/abs/2016ApJ...817..118T} {817, 118}

\bibitem[\protect\citeauthoryear{{Tonini}, {Bernyk}, {Croton}, {Maraston}  \&
  {Thomas}}{{Tonini} et~al.}{2012}]{tonini_2012}
{Tonini} C.,  {Bernyk} M.,  {Croton} D.,  {Maraston} C.,   {Thomas} D.,  2012,
  \mn@doi [\apj] {10.1088/0004-637X/759/1/43}, \href
  {https://ui.adsabs.harvard.edu/abs/2012ApJ...759...43T} {759, 43}

\bibitem[\protect\citeauthoryear{{Trujillo-Gomez}, {Kruijssen}, {Pfeffer},
  {Reina-Campos}, {Crain}, {Bastian}  \& {Cabrera-Ziri}}{{Trujillo-Gomez}
  et~al.}{2023}]{trujillo_gomez_2023}
{Trujillo-Gomez} S.,  {Kruijssen} J.~M.~D.,  {Pfeffer} J.,  {Reina-Campos} M.,
  {Crain} R.~A.,  {Bastian} N.,   {Cabrera-Ziri} I.,  2023, arXiv e-prints,
  \href {https://ui.adsabs.harvard.edu/abs/2023arXiv230105716T} {p.
  arXiv:2301.05716}

\bibitem[\protect\citeauthoryear{{Vazdekis}, {S{\'a}nchez-Bl{\'a}zquez},
  {Falc{\'o}n-Barroso}, {Cenarro}, {Beasley}, {Cardiel}, {Gorgas}  \&
  {Peletier}}{{Vazdekis} et~al.}{2010}]{vazdekis_2010}
{Vazdekis} A.,  {S{\'a}nchez-Bl{\'a}zquez} P.,  {Falc{\'o}n-Barroso} J.,
  {Cenarro} A.~J.,  {Beasley} M.~A.,  {Cardiel} N.,  {Gorgas} J.,   {Peletier}
  R.~F.,  2010, \mn@doi [\mnras] {10.1111/j.1365-2966.2010.16407.x}, \href
  {https://ui.adsabs.harvard.edu/abs/2010MNRAS.404.1639V} {404, 1639}

\bibitem[\protect\citeauthoryear{{Vogelsberger} et~al.,}{{Vogelsberger}
  et~al.}{2014}]{vogelsberger_2014}
{Vogelsberger} M.,  et~al., 2014, \mn@doi [\mnras] {10.1093/mnras/stu1536},
  \href {https://ui.adsabs.harvard.edu/abs/2014MNRAS.444.1518V} {444, 1518}

\bibitem[\protect\citeauthoryear{{Wang}}{{Wang}}{2019}]{wang_2019}
{Wang} T.,  2019, in ALMA2019: Science Results and Cross-Facility Synergies.
  p.~6, \mn@doi{10.5281/zenodo.3585174}

\bibitem[\protect\citeauthoryear{{Weaver} et~al.,}{{Weaver}
  et~al.}{2023}]{weaver_2023}
{Weaver} J.~R.,  et~al., 2023, \mn@doi [\aap] {10.1051/0004-6361/202245581},
  \href {https://ui.adsabs.harvard.edu/abs/2023A&A...677A.184W} {677, A184}

\bibitem[\protect\citeauthoryear{{Wechsler} \& {Tinker}}{{Wechsler} \&
  {Tinker}}{2018}]{wechsler_2018}
{Wechsler} R.~H.,  {Tinker} J.~L.,  2018, \mn@doi [\araa]
  {10.1146/annurev-astro-081817-051756}, \href
  {https://ui.adsabs.harvard.edu/abs/2018ARA&A..56..435W} {56, 435}

\bibitem[\protect\citeauthoryear{{Weibel} et~al.,}{{Weibel}
  et~al.}{2024}]{weibel_2024}
{Weibel} A.,  et~al., 2024, \mn@doi [arXiv e-prints]
  {10.48550/arXiv.2403.08872}, \href
  {https://ui.adsabs.harvard.edu/abs/2024arXiv240308872W} {p. arXiv:2403.08872}

\bibitem[\protect\citeauthoryear{{Weinberger} et~al.,}{{Weinberger}
  et~al.}{2017}]{weinberger17}
{Weinberger} R.,  et~al., 2017, \mn@doi [\mnras] {10.1093/mnras/stw2944}, \href
  {https://ui.adsabs.harvard.edu/abs/2017MNRAS.465.3291W} {465, 3291}

\bibitem[\protect\citeauthoryear{{Westfall} et~al.,}{{Westfall}
  et~al.}{2019}]{Westfall2019}
{Westfall} K.~B.,  et~al., 2019, \mn@doi [\aj] {10.3847/1538-3881/ab44a2},
  \href {https://ui.adsabs.harvard.edu/abs/2019AJ....158..231W} {158, 231}

\bibitem[\protect\citeauthoryear{{Wetzel}, {Tinker}, {Conroy}  \& {van den
  Bosch}}{{Wetzel} et~al.}{2013}]{wetzel_2013}
{Wetzel} A.~R.,  {Tinker} J.~L.,  {Conroy} C.,   {van den Bosch} F.~C.,  2013,
  \mn@doi [\mnras] {10.1093/mnras/stt46910.48550/arXiv.1206.3571}, \href
  {https://ui.adsabs.harvard.edu/abs/2013MNRAS.432..336W} {432, 336}

\bibitem[\protect\citeauthoryear{{Whitaker} et~al.,}{{Whitaker}
  et~al.}{2014}]{whitaker_2014}
{Whitaker} K.~E.,  et~al., 2014, \mn@doi [\apj] {10.1088/0004-637X/795/2/104},
  \href {https://ui.adsabs.harvard.edu/abs/2014ApJ...795..104W} {795, 104}

\bibitem[\protect\citeauthoryear{{Wright}, {Lagos}, {Power}, {Stevens},
  {Cortese}  \& {Poulton}}{{Wright} et~al.}{2022}]{wright_2022}
{Wright} R.~J.,  {Lagos} C. d.~P.,  {Power} C.,  {Stevens} A. R.~H.,  {Cortese}
  L.,   {Poulton} R. J.~J.,  2022, \mn@doi [\mnras] {10.1093/mnras/stac2042},
  \href {https://ui.adsabs.harvard.edu/abs/2022MNRAS.516.2891W} {516, 2891}

\bibitem[\protect\citeauthoryear{{Xu} \& {Peng}}{{Xu} \&
  {Peng}}{2021}]{xu_2021}
{Xu} B.,  {Peng} Y.,  2021, \mn@doi [\apjl] {10.3847/2041-8213/ac3a01}, \href
  {https://ui.adsabs.harvard.edu/abs/2021ApJ...923L..29X} {923, L29}

\bibitem[\protect\citeauthoryear{{Yang}, {Mo}, {van den Bosch}, {Pasquali},
  {Li}  \& {Barden}}{{Yang} et~al.}{2007}]{yang_2007}
{Yang} X.,  {Mo} H.~J.,  {van den Bosch} F.~C.,  {Pasquali} A.,  {Li} C.,
  {Barden} M.,  2007, \mn@doi [\apj] {10.1086/522027}, \href
  {https://ui.adsabs.harvard.edu/abs/2007ApJ...671..153Y} {671, 153}

\bibitem[\protect\citeauthoryear{{Yang}, {Mo}, {van den Bosch}, {Zhang}  \&
  {Han}}{{Yang} et~al.}{2012}]{yang_2012}
{Yang} X.,  {Mo} H.~J.,  {van den Bosch} F.~C.,  {Zhang} Y.,   {Han} J.,  2012,
  \mn@doi [\apj] {10.1088/0004-637X/752/1/41}, \href
  {https://ui.adsabs.harvard.edu/abs/2012ApJ...752...41Y} {752, 41}

\bibitem[\protect\citeauthoryear{{van Dokkum} et~al.,}{{van Dokkum}
  et~al.}{2010}]{van_dokkum_2010}
{van Dokkum} P.~G.,  et~al., 2010, \mn@doi [\apj]
  {10.1088/0004-637X/709/2/1018}, \href
  {https://ui.adsabs.harvard.edu/abs/2010ApJ...709.1018V} {709, 1018}

\makeatother
\end{thebibliography}



\appendix

\section{Testing the abundance matching}\label{app_abun_match_check}

\begin{figure*}
    \includegraphics[width=0.9\textwidth]{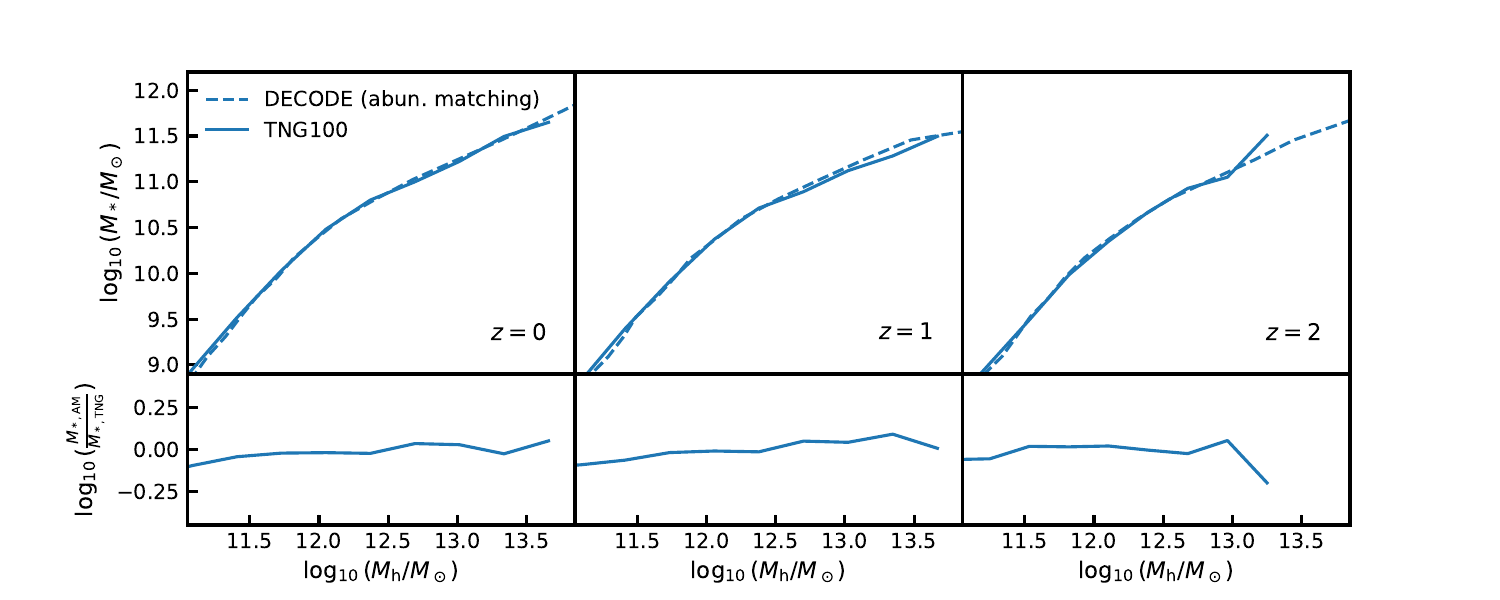}
    \caption{Upper panels: SMHM relation from the TNG simulation (solid lines) at $z=0, 1 \, {\rm and} \, 2$, compared to those computed via \decode’s abundance matching taking the TNG’s SMF and HMF as input (dashed lines). Lower panels: residuals, computed as logarithmic difference, between the SMHM relations shown in the upper panels at the same redshifts.}
    \label{fg_smhm_TNG_AM_comp}
\end{figure*}

Figure \ref{fg_smhm_TNG_AM_comp} shows the SMHM relation from our abundance matching using TNG's SMF input, compared to that directly derived from the simulation (see discussion in Section \ref{sec_decode_valid_tng}).

\section{Effect of satellites star formation on central galaxies growth}\label{app_sats_evo_effect}

\begin{figure}
    \includegraphics[width=\columnwidth]{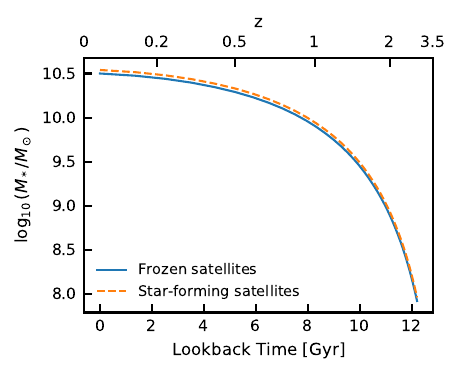}
    \caption{Example of mean star formation history of galaxies with stellar mass $M_\star 10^{10.5}\, M_\odot$ at $z=0$ for the frozen satellites (blue solid line) and star-forming satellites (orange dashed line) scenarios.}
    \label{fg_example_mstar_growth_10.5}
\end{figure}

Figure \ref{fg_example_mstar_growth_10.5} shows the mean SFH in the stellar mass bin of $M_\star 10^{10.5}\, M_\odot$ at redshift $z=0$ for the frozen and star-forming satellites scenarios, showing that the effect of the star formation in satellites on the SFH of the central galaxies is minimal.

\bsp	
\label{lastpage}
\end{document}